\documentclass[manuscript,numberedappendix,appendixfloats,twocolappendix]{emulateapj}
\usepackage{longtable,amsmath}

\shorttitle{Kinematics of M31 dSphs}
\shortauthors{Collins et al.}
\def\ltsima{$\; \buildrel < \over \sim \;$}
\def\lta{\lower.5ex\hbox{\ltsima}}
\def\gtsima{$\; \buildrel > \over \sim \;$}
\def\simgt{\lower.5ex\hbox{\gtsima}}
\def\kms{{\rm\,km\,s^{-1}}}

\def\kpc{{\rm\,kpc}}

\def\msun{{\rm\,M_\odot}}
\def\lsun{{\rm\,L_\odot}}

\newcommand{\Lagr}{\mathcal{L}}

\def\AA{$\; \buildrel \circ \over {\mathrm A}$}

\def\deg{^\circ}

\def\s{\ifmmode \widetilde \else \~\fi}
\def\={\overline}

\def\spose#1{\hbox to 0pt{#1\hss}}

\def\lta{\mathrel{\spose{\lower 3pt\hbox{$\mathchar"218$}}
     \raise 2.0pt\hbox{$\mathchar"13C$}}}
\def\gta{\mathrel{\spose{\lower 3pt\hbox{$\mathchar"218$}}
     \raise 2.0pt\hbox{$\mathchar"13E$}}}
\def\Dt{\spose{\raise 1.5ex\hbox{\hskip3pt$\mathchar"201$}}}    
\def\dt{\spose{\raise 1.0ex\hbox{\hskip2pt$\mathchar"201$}}}    

\def\feh{{\rm[Fe/H]}}

\shorttitle{The kinematics of Andromeda's dSphs}
\shortauthors{Collins et al.}

\begin{document}

\title{A kinematic study of the Andromeda dwarf spheroidal system}

\author{Michelle L. M. Collins\altaffilmark{1,2}, Scott C. Chapman\altaffilmark{2,3}, R. Michael Rich\altaffilmark{4}, Rodrigo A. Ibata\altaffilmark{5}, Nicolas F. Martin\altaffilmark{5,1}, Michael J. Irwin\altaffilmark{2}, Nicholas F. Bate\altaffilmark{6}, Geraint F. Lewis\altaffilmark{6}, Jorge Pe\~narrubia\altaffilmark{7,8}, Nobuo Arimoto\altaffilmark{9,10}, Caitlin M. Casey\altaffilmark{11}, Annette M. N. Ferguson\altaffilmark{8}, Andreas Koch\altaffilmark{12}, Alan W. McConnachie\altaffilmark{13}, Nial Tanvir\altaffilmark{14}}
\altaffiltext{1}{Max-Planck-Institut
  f\"ur Astronomie, K\"onigstuhl 17, D-69117 Heidelberg, Germany}
\altaffiltext{2}{Institute of
  Astronomy,Madingley Rise, Cambridge, CB3 0HA ,UK}
\altaffiltext{3}{Dalhousie University Dept. of Physics and Atmospheric Science
  Coburg Road Halifax, B3H1A6, Canada}
\altaffiltext{4}{Department of Physics and Astronomy, University of
  California, Los Angeles, CA 90095-1547}
\altaffiltext{5}{Observatoire astronomique de Strasbourg, Université de Strasbourg, CNRS, UMR 7550, 11 rue de l’Université, F-67000 Strasbourg, France}
\altaffiltext{6}{Sydney Institute
  for Astronomy, School of Physics, A28, University of Sydney, NSW 2006,
  Australia}
\altaffiltext{7}{Ram\'on y Cajal Fellow, Instituto de
  Astrof\'isica de Andalucia-CSIC, Glorieta de la Astronom\'ia s/n, 18008,
  Granada, Spain}
\altaffiltext{8}{Institute for Astronomy, University of Edinburgh, Royal
  Observatory, Blackford Hill, Edinburgh, EH9 3HJ, UK}
\altaffiltext{9}{Subaru Telescope, National Astronomical Observatory of Japan 650 North
  A'ohoku Place, Hilo, Hawaii 96720, U.S.A.}
\altaffiltext{10}{Graduate University for Advanced Studies 2-21-1 Osawa, Mitaka,
  Tokyo 181-8588, Japan}
\altaffiltext{11}{Institute for Astronomy, 2680 Woodlawn Drive Honolulu, HI 96822-1839
  USA }
\altaffiltext{12}{Zentrum f\"ur Astronomie der Universit\"at Heidelberg, Landessternwarte,
  K\"onigstuhl 12, 69117 Heidelberg, Germany}
\altaffiltext{13}{NRC Herzberg Institute of Astrophysics, 5071 West Saanich
  Road, British Columbia, Victoria V9E 2E7, Canada}
\altaffiltext{14}{Department of
  Physics \& Astronomy, University of Leicester, University Road, Leicester
  LE1 7RH, UK}

\begin{abstract}
  We present a homogeneous kinematic analysis of red giant branch stars within
  18 of the 28 Andromeda dwarf spheroidal (dSph) galaxies, obtained using the
  Keck I LRIS and Keck II DEIMOS spectrographs. Based on their $g-i$ colors
  (taken with the CFHT MegaCam imager), physical positions on the sky, and
  radial velocities, we assign probabilities of dSph membership to each
  observed star. Using this information, the velocity dispersions, central
  masses and central densities of the dark matter halos are calculated for
  these objects, and compared with the properties of the Milky Way dSph
  population. We also measure the average metallicity ([Fe/H]) from the
  co-added spectra of member stars for each M31 dSph and find that they are
  consistent with the trend of decreasing [Fe/H] with luminosity observed in
  the Milky Way population. We find that three of our studied M31 dSphs appear
  as significant outliers in terms of their central velocity dispersion, And
  XIX, XXI and XXV, all of which have large half-light radii ($\gta700$pc) and
  low velocity dispersions ($\sigma_v<5\kms$). In addition, And XXV has a
  mass-to-light ratio within its half-light radius of just $[M/L]_{\rm
    half}=10.3^{+7.0}_{-6.7}$, making it consistent with a simple stellar
  system with no appreciable dark matter component within its $1\sigma$
  uncertainties. We suggest that the structure of the dark matter halos of
  these outliers have been significantly altered by tides.

\end{abstract}


\section{Introduction}

The underlying nature of the dark matter halos of dwarf spheroidal galaxies
(dSphs) has garnered significant attention from the scientific community over
the past decade. The goal of recent observational studies of these objects has
been to make critical tests of structure formation scenarios, particularly
focusing on the viability of the canonical $\Lambda$CDM model. There is the
long standing issue of the relative dearth of these faintest of galaxies
observed surrounding nearby galaxies when compared with the number of dark
matter subhalos produced in $N-$body simulations, which is referred to as the
``missing satellite'' problem \citep{klypin99,moore99}. The extent to which
this mismatch is considered problematic has decreased over recent years as
both theorists and observers have sought to reconcile the simulated and
observable Universe. From a modelling point of view, one does not expect stars
to be able to form within all dark matter subhalos seen in simulations, and at
a certain mass limit ($V_{\rm max}\lta15\kms$, see
\citealt{penarrubia08a,koposov09}), star formation is unable to proceed. Thus,
there is a lower limit placed on galaxy formation. This mass limit is also
tied to feedback processes that can remove the baryonic reservoirs required
for star formation
(e.g., \citealt{bullock00,somerville02,kravtsov10,bullock10,nickerson11,kazantzidis11}). This
would imply that only the most massive subhalos seen in simulations are able
to form and retain luminous populations. Observers have also attempted to
quantify current survey completeness and radial selection effects to account
for the number of satellites we are not currently able to detect
(e.g., \citealt{koposov08,tollerud08,walsh09}). These studies suggest that
there are of order a few hundred satellites within the Milky Way's virial
radius that we have yet to detect.

The high dark matter dominance of dSph galaxies also singles them out as
objects of interest. With total dynamic mass-to-light ratios of
$[M/L]\sim10-1000$s and half-light radii of $r_{\rm half}\sim10-1000$~pc, they
are ideal systems with which to probe the inner density profiles of dark
matter halos. Recent imaging and spectroscopic observations of these objects
within the Local Group have shown that, despite spanning approximately 5
decades in luminosity, the dSphs of the MW share a common mass scale and a
universal density profile \citep{strigari08,walker09b,wolf10}. With a
kinematic resolution of a few 10s of parsecs, these objects allow us to start
addressing the question of whether the central regions of these halos follow
cuspy density profiles as predicted by cosmological simulations
\citep{navarro97}, or constant density cores, similar to what is observed in
low surface brightness galaxies
\citep{blaise01,deblok02,deblok03,deblok05,swaters03,kassin06,spano08}. From
studies of brighter dSphs, such as Sculptor and Fornax
\citep{walker11,amorisco12,jardel12} it appears that their halo density
profiles are also inconsistent with hosting central cusps. It is possible that
these objects originally formed with cuspy density profiles, and that these
have been subsequently modified by baryonic feedback. If this is truly the
case, one should be able to observe cuspy profiles in the fainter dSph
population (\citealt{zolotov12}) as these do not contain enough baryons to
drive this change in the dark matter density profile. Perhaps the only way to
gain further insight into this contentious issue is by measuring the
kinematics for large numbers of stellar tracers within these objects and
analysing them with detailed models that do not make {\it ab initio}
assumptions about the underlying density profiles, or the velocity anisotropy
of both the dark matter and stars, such as those employed by \citet{walker11}
and \citet{jardel12}.

To date, the majority of studies involving the detailed kinematics of dSphs
have revolved largely around those belonging to the Milky Way, as these are
nearby enough that we can measure the velocities of their member stars to a
high degree of accuracy. However, there are currently only $\sim25$ known MW
dSphs, with luminosities ranging from $10^2-10^7\lsun$. For the very faintest,
some controversy remains as to whether they are massively dark-matter dominated
(see e.g., \citealt{niederste09,simon11}), but almost all of them have been
shown to be consistent with the universal mass profiles of \citet{walker09a}
and \citet{wolf10}. One notable exception to this is the Hercules object
\citep{aden09}, which some have argued is currently undergoing significant
tidal disruption \citep{martin10}. Andromeda represents the only other system
for which comparable kinematic analyses can be performed. M31 now has 28 dSph
companions known, whose luminosities range from $\sim10^4 - 10^8\lsun$, the
majority of which have been discovered by the CFHT Pan-Andromeda
Archaeological Survey (PAndAS
\citealt{martin06,ibata07,irwin08,mcconnachie08,martin09,richardson11}). The
relatively brighter lower bound for the luminosities of M31 dSphs compared to
the MW is a detection limit issue, rather than a sign of differing stellar
populations (Martin et al. 2013, in prep). It has been noted by a number of
authors (e.g., \citealt{mcconnachie06a,tollerud12,mcconnachie12}) that for the
brighter dSphs ($M_V<-8$), those belonging to M31 are 2--3 times more extended
in terms of their half-light ($r_{\rm half}$) and tidal ($r_t$) radii compared
with the MW. In these papers, the underlying cause of this discrepancy was not
identified, but it has been argued that it could be an effect of environment,
with the mass distribution of the host playing an important role. Subsequent
work by \citet{brasseur11b}, who included the fainter, non-classic M31 dSphs
for the first time, showed that statistically, the relationship between size
and luminosity for dSphs in the MW and Andromeda are actually largely
consistent with one another, however there remain a number of significantly
extended outliers within the Andromedean system (e.g., And II, And XIX, $r_{\rm
  half}\sim1.2\kpc$ and $1.5\kpc$ respectively), and the scatter in this
relationship is large (up to an order of magnitude at a given luminosity,
\citealt{mcconnachie12}). 

Working from the \citet{mcconnachie06a} results,
\citet{penarrubia08a} modelled the expected velocity dispersions for the M31
dSphs, assuming that all dSph galaxies are embedded within similar mass dark
matter halos. A robust prediction of their modeling was that, given the larger
radial extents, the dSphs of M31 should be {\it kinematically hotter} than
their MW counterparts by a factor of $\sim2$. At the time of writing, they had
only 2 measured velocity dispersions for the M31 dSphs, those of And II and
And IX \citep{cote99,chapman05}. New studies of the kinematics of M31 dSphs
\citep{collins10,collins11b,kalirai10,tollerud12,chapman12} have 
dramatically increased the number of systems with a measured velocity
dispersion, and have shown that instead of being kinematically hotter, these
systems are either very similar to, or in a number of cases (e.g., And II, And
XII, And XIV, And XV and And XXII), {\it colder} than their MW
counterparts. In particular, a significant recent kinematic study of 15 M31
dSph companions using the Keck II DEIMOS spectrograph by the Spectroscopic and
Photometric Landscape of Andromeda's Stellar Halo (SPLASH,
\citealt{tollerud12}) concluded that the M31 dSph system largely obeys very
similar mass-size-luminosity scalings as those of the MW.  However, they also
identified 3 outliers (And XIV, XV and XVI) that appear to possess much lower
velocity dispersions, and hence maximum circular velocities, than would be
expected for these systems. Such a result suggests that there are significant
differences in the formation and/or evolution of the M31 and MW dSph systems.

To investigate this further, our group has been systematically surveying the
known dSphs of M31 with the Keck I LRIS and Keck II DEIMOS spectrographs, and
have obtained kinematic data for 18 of the 28 galaxies. In this paper, we
present new spectroscopic analysis for the 11 dSphs, Andromeda (And) XVII, And
XVIII, And XIX, And XX, And XXI, And XXIII, And XXIV, And XXV, And XXVI, the
tidally disrupting And XXVII, And XXVIII and And XXX (Cassiopeia II) using an
algorithm we have developed that implements a probabilistic method of
determining membership for each galaxy. In addition we re-analyze the
kinematics of 6 dSphs that our group has previously observed (And V, VI, XI,
XII, XIII, XXII) using this method with the aim of providing a homogeneous
analysis of all dSphs observed by our group to date. We also provide the
individual stellar velocities and properties for every star observed in our
dSphs survey, allowing us to present a large catalog of stellar kinematics
that will be of interest to those studying dSph systems and Milky Way-like
galaxies, whether observationally or theoretically.

The outline of this paper is as follows. In \S~2 we discuss the relevant
observations, data reduction techniques. In \S~3 we outline our algorithm for
the classification of member stars. In \S~4 we present an analysis of our new
kinematic datasets. In \S~5 we report on the masses and densities of the dark
matter halos of the M31 dSphs, comparing them to those of the MW dSphs. In \S6
we report on the metallicities, [Fe/H], of the M31 dSphs as measured from the
co-added spectra of their member stars. Finally, we conclude in \S~7.

\section{Observations}
\label{sect:obs}

\subsection{Photometry and target selection}
\label{sect:photobs}

The PAndAS survey \citep{mcconnachie09}, conducted using the 3.6 metre Canada
France Hawaii Telescope (CFHT), maps out the stellar density of the disc and
halo regions of the M31--M33 system over a projected area of
$\sim350\rm{deg}^2$ ($\sim55,000\kpc^2$), resolving individual stars to depths
of $g=26.5$ and $i=25.5$ with a signal to noise ratio of 10, making this
survey the deepest, highest resolution, contiguous map of the majority of the
extended stellar halo of an L$_*$ galaxy to date. Each of the $411$ fields in
this survey ($0.96\times0.94$ deg$^2$) has been observed for at least 1350s in
both MegaCam $g$ and $i$ filters, in $<0.8^{\prime\prime}$ seeing. This survey
was initiated following two precursor surveys of the M31 system, the first of
which surveyed the central $\sim40$~deg$^2$ conducted with the 2.5 metre Isaac
Newton Telescope \citep{ferguson02,irwin05}, and revealed a wealth of
substructure in the Andromeda stellar halo, including the giant southern
stream \citep{ibata01c}. To better understand this feature, and to probe
deeper into the M31 (and M33) stellar halo, a survey of the south west
quadrant of the M31 halo was initiated using the CFHT \citep{ibata07}, and
revealed yet more substructure, including the arc like stream Cp and Cr
\citep{chapman08} and a number of dwarf spheroidal satellites
\citep{martin06}. This CFHT survey was then extended into the full PAndAS
project. For details of the processing and reduction of these data, see
\citet{richardson11}. This survey has introduced us to a wealth of stellar
substructure, debris and globular clusters within the Andromeda--Triangulum
system. In addition, it has led to the discovery of 17 dSphs. These objects
were detected in the PAndAS survey maps as over-densities in matched-filter
surface density maps of metal poor red giant branch (RGB) stars and were
presented in \citet{martin06,ibata07,irwin08,mcconnachie08,martin09} and
\citet{richardson11}. We briefly summarise the photometric properties of all
dSphs discussed within this paper in Table~\ref{tab:photobs}.

\begin{deluxetable*}{lccccc}
  \tabletypesize{\footnotesize} \tablecolumns{6} \tablewidth{0pt}
  \tablecaption{Details of the structural properties for each dwarf, as
    derived from CFHT photometry by
    \citet{ibata07,letarte09,irwin08,mcconnachie08,martin09,collins10,collins11b,brasseur11b,richardson11}
    and \citet{mcconnachie12}, updated for distances presented in
    \citet{conn12}. The photometry for And VI is derived from Subaru
    SuprimeCam photometry \citep{collins11b}, and the values for And XXVIII
    come from SDSS photometry \citep{slater11}.
\label{tab:photobs}}
\tablehead{
\colhead{Property} & \colhead{$\alpha_{0,J2000}$}  &  \colhead{$\delta_{0,J2000}$}  & \colhead{ $M_V$}  & \colhead{$r_{\rm half}$ (pc) }  & \colhead{$D$ (kpc)}}
\startdata
 And V       & 01:10:17.1 & +47:37:41.0 & -10.2 & 302$\pm44$ & 742$^{+21}_{-22}$  \\
 And VI      & 23:51:39.0 & +24:35:42.0 & -10.6 & 524$\pm49$ & 783$\pm28$  \\
 And XI      & 00:46:20.0 & +33:48:05.0 & -6.9  & 158$^{+9}_{-23}$ & 763$^{+29}_{-106}$  \\
 And XII     & 00:47:27.0 & +34:22:29.0 & -6.4  & 324$^{+56}_{-72}$ & 928$^{+40}_{-136}$  \\
 And XIII    & 00:51:51.0 & +33:00:16.0 & -6.7  & 172$^{+34}_{-39}$ & 760$^{+126}_{-154}$  \\
 And XVII    & 00:37:07.0 & +44:19:20.0 & -8.5  & 262$^{+53}_{-46}$ & 727$^{+39}_{-25}$  \\
 And XVIII   & 00:02:14.5 & +45:05:20.0 & -9.7  & 325$\pm24$ & 1214$^{+40}_{-43}$ \\
 And XIX     & 00:19:32.1 & +35:02:37.1 & -9.6  & 1481$^{+62}_{-268}$& 821$^{+32}_{-148}$  \\
 And XX      & 00:07:30.7 & +35:07:56.4 & -6.3  & 114$^{+31}_{-12}$ & 741$^{+42}_{-52}$  \\
 And XXI     & 23:54:47.7 & +42:28:15.0 & -9.8  & 842$\pm77$ & 827$^{+23}_{-25}$  \\
 And XXII    & 01:27:40.0 & +28:05:25.0 & -6.5  & 252$^{+28}_{-47}$ & 920$^{+32}_{-139}$  \\
 And XXIII   & 01:29:21.8 & +38:43:08.0 & -10.2 & 1001$^{+53}_{-52}$& 748$^{+31}_{-21}$  \\  
 And XXIV    & 01:18:30.0 & +46:21:58.0 & -7.6  & 548$^{+31}_{-37}$ & 898$^{+28}_{-42}$  \\
 And XXV     & 00:30:08.9 & +46:51:07.0 & -9.7  & 642$^{+47}_{-74}$ & 736$^{+23}_{-69}$  \\
 And XXVI    & 00:23:45.6 & +47:54:58.0 & -7.1  & 219$^{+67}_{-52}$ & 754$^{+218}_{-164}$   \\
 And XXVII   & 00:37:27.2 & +45:23:13.0 & -7.9  & 657$^{+112}_{-271}$ &1255$^{+42}_{-474}$\\
 And XXVIII  & 22:32:41.2 & +31:12:51.2 & -8.5  & 210$^{+60}_{-50}$ &650$^{+150}_{-80}$\\
 And XXX (Cass II)     & 00:36:34.9 & +49:38:48.0 & -8.0  & 267$^{+23}_{-36}$ & 681$^{+32}_{-78}$  \\
\enddata
\end{deluxetable*}

For the majority of these objects, the
PAndAS dataset formed the basis for our spectroscopic target selection. Using
the color selection boxes presented in \citet{mcconnachie08,martin09} and
\citet{richardson11}, we isolated the RGBs of each dSph, then prioritised each
star on this sequence depending on their color, $i$-band magnitude, and
distance from the centre of the dSph. Stars lying directly on the RGB, with
$20.3<i_0<22.5$ and distance, $d< 4r_{\rm half}$ (where $r_{\rm half}$ is the
half-light radius, measured on the semi-major axis of the dSph) were highly
prioritised (priority A), followed by stars on the RGB within the same
distance from the centre with $22.5<i_0<23.5$ (priority B). The remainder of
the mask was then filled with stars in the field with $20.3<i_0<23.5$ and
$0.5<g-i< 4$ (priority C). In general, it is the brighter, higher priority A
stars that ultimately show the highest probability of membership. We also use
the PAndAS photometry to help us determine membership of the dSph (discussed
in \S~\ref{sect:membership}). 

And XXVIII is the one dSph in our sample that is not covered by the PAndAS
survey. For this object, target selection followed an identical methodology
tot hat detailed above, but used photometry from the 8th data release of the
Sloan Digital Sky Survey (SDSS-III). Details of the observations and analysis
of this photometry can be found in \citet{slater11}.

\subsection{Keck Spectroscopic Observations}
\label{sect:specobs}

The DEep-Imaging Multi-Object Spectrograph (DEIMOS), situated on the Nasmyth
focus of the Keck II telescope is an ideal instrument for obtaining medium
resolution (R$\sim1.4$\AA) spectra of multiple, faint stellar targets in the
M31 dSphs. The data for the dSphs within this work were taken between Sept
2004 and Sept 2012 in photometric conditions, with typical seeing between
$0.5-1^{\prime\prime}$.  Our chosen instrumental setting covered the observed
wavelength range from 5600--9800\AA\ and employed exposure times of 3x20
minute integrations. The majority of observations employed the 1200 line/mm
grating, although for 4 dSphs (And XI, XII, XIII and XXIV) the lower
resolution 600 line/mm grating ($R\sim3.8$\AA\ FWHM) was used. The spectra
from both setups typically possess signal-to-noise (S:N) ratios of
$>3$\AA$^{-1}$ for our bright targets ($i\lta22.0$), though some of our
fainter targets fall below this level. Information regarding the spectroscopic
setup and observations for each dSph are displayed in Table~\ref{tab:specobs}.

The resulting science spectra are reduced using a custom built pipeline, as
described in \citet{ibata11}. Briefly, the pipeline identifies and removes
cosmic rays, corrects for scattered light, performs flat-fielding to correct
for pixel-to-pixel variations, corrects for illumination, slit function and
fringing, wavelength calibrates each pixel using arc-lamp exposures, performs
a 2-dimensional sky subtraction, and finally extracts each spectra  -- without
resampling -- in a small spatial region around the target. This results in a
large set of pixels for each target, each of which carries a flux and
wavelength (with associated uncertainties) plus the value of the target
spatial profile at that pixel.  We then derive velocities for all our stars
with a Bayesian approach, using the Ca II Triplet absorption feature. Located
at rest wavelengths of 8498, 8542 and 8662 \AA, these strong features are
ideal for determining the velocities of our observed stars. We determine the
velocities by using an Markov Chain Monte Carlo procedure where a template Ca
II spectrum was cross-correlated with the non-resampled data, generating a
most-likely velocity for each star, and a likely uncertainty based on the
posterior distribution that incorporates all the uncertainties for each
pixel. Typically our velocity uncertainties lie in the range of
$5-15\kms$. Finally, we also correct these velocities to the heliocentric
frame.

\begin{deluxetable*}{lcccccccccc}
\tabletypesize{\footnotesize}
\tablecolumns{11} 
\tablewidth{0pt}
\tablecaption{Details of spectroscopic observations
\label{tab:specobs}}
\tablehead{
\colhead{Object}  & \colhead{Date}  & \colhead{Instrument} & \colhead{Grating} & \colhead{Res.} & \colhead{$\alpha_{0,J2000}$}& \colhead{$\delta_{0,J2000}$} &  \colhead{P.A. }& \colhead{Exp. (s)}
& \colhead{$N_{Targ}$} & \colhead{$N_{Mem}$}}
\startdata
And V      & 16 Aug 2009     & LRIS   &831/8200&3.0\AA& 01:10:18.21 & +47:37:53.3 & 0$\deg$ & 3600 & 50&15 \\
And VI     & 17--19 Sept 2009& DEIMOS & 1200 & 1.3\AA & 23:51:51.49 & +24:34:57.0 & 0$\deg$ & 5400 & 113& 43\\
And XI     & 23 Sept 2006    & DEIMOS & 600  & 3.5\AA & 00:46:28.08 & +33:46:28.8 & 0$\deg$ & 3600 &33 &5 \\
And XII    & 21--23 Sept 2006& DEIMOS & 600  & 3.5\AA & 00:47:32.89 & +34:22:28.6 & 0$\deg$ & 3600 &49 &8 \\
And XIII   & 23 Sept 2006    & DEIMOS & 600  & 3.5\AA & 00:52:00.22 &+32:59:16.2 & 0$\deg$ & 3600 & 46& 4\\
And XVII   & 26 Sept 2011    & DEIMOS & 1200 & 1.3\AA & 00:37:51.09 & +44:17:51.9 & 280$\deg$ & 3600 &149& 8 \\
And XVIII  &  9 Sept 2010    & DEIMOS & 1200 & 1.3\AA & 00:02:14.50 &+45:05:20.0&90$\deg$& 3600 & 73 & 4 \\
And XIXa   & 26 Sept 2011    & DEIMOS & 600  & 3.5\AA & 00:19:45.04 & +35:05:28.8 & 270$\deg$& 3600 & 107&15\\
And XIXb   & 26 Sept 2011    & DEIMOS & 1200 & 1.3\AA & 00:19:30.88 & +35:07:34.1 & 0$\deg$ & 3600 & 108& 9\\
And XX     &  9 Sept 2010    & DEIMOS & 1200 & 1.3\AA & 00:07:30.69 & +35:08:02.4 & 90$\deg$ &3600 & 85  &4\\
And XXI    & 26 Sept 2011    & DEIMOS & 1200 & 1.3\AA & 23:54:47.70 & +42:28:33.6 & 180$\deg$ &3600& 157&32 \\
And XXIIa  & 23 Sept 2009    & DEIMOS & 1200 & 1.3\AA & 01:27:52.37 &+28:05:22.3 & 90$\deg$&1200 &93 &4\\
And XXIIb  & 10 Sept 2010    & DEIMOS & 1200 & 1.3\AA & 01:27:52.37 & +28:05:22.3 & 0$\deg$&3600& 73&6\\
And XXIIIa &  9 Sept 2010    & DEIMOS & 1200 & 1.3\AA & 01:29:18.18 & +38:43:50.4 & 315$\deg$ & 3600 &196 & 24\\
And XXIIIb & 10 Sept 2010    & DEIMOS & 1200 & 1.3\AA & 01:29:21.87 & +38:44:58.7 & 245$\deg$ & 3600 & 189&18\\
And XXIVa  &  9 Sept 2010    & DEIMOS & 1200 & 1.3\AA & 01:18:32.90 & +46:22:50.0 & 30$\deg$ &3600 & 192 & 1\\
And XXIVb  &  31 May 2011    & DEIMOS & 600  & 3.5\AA & 01:18:32.90 & +46:22:50.0 & 0$\deg$ &2700 & 115 & 11\\
And XXV    & 10 Sept 2010    & DEIMOS & 1200 & 1.3\AA & 00:30:01.88 &+46:50:31.0 & 90$\deg$ &3600& 183& 26\\
And XXVI   & 10 Sept 2010    & DEIMOS & 1200 & 1.3\AA & 00:23:41.42 & +47:54:56.8 & 90$\deg$ &3600 & 179 &6
\\
And XXVII & 10 Sept 2010    & DEIMOS & 1200 & 1.3\AA & 00:37:31.93 & +45:23:55.4 & 45$\deg$ & 3600 & 131&8\\
And XXVIII& 21 Sept 2012    & DEIMOS & 1200 & 1.3\AA & 22:32:32.71 & +31:13:13.2 & 140$\deg$ & 3600 & 102&17\\
Cass II    & 28 Sept 2011    & DEIMOS & 1200 & 1.3\AA & 00:37:00.84 & +49:39:12.0 & 270$\deg$ & 3600 & 156& 8\\ 
\enddata
\vspace{-0.2cm}
\end{deluxetable*}

\subsubsection{Telluric velocity corrections}
\label{sect:telluric}

\begin{figure*}
\begin{center}
\includegraphics[angle=0,width=0.45\hsize]{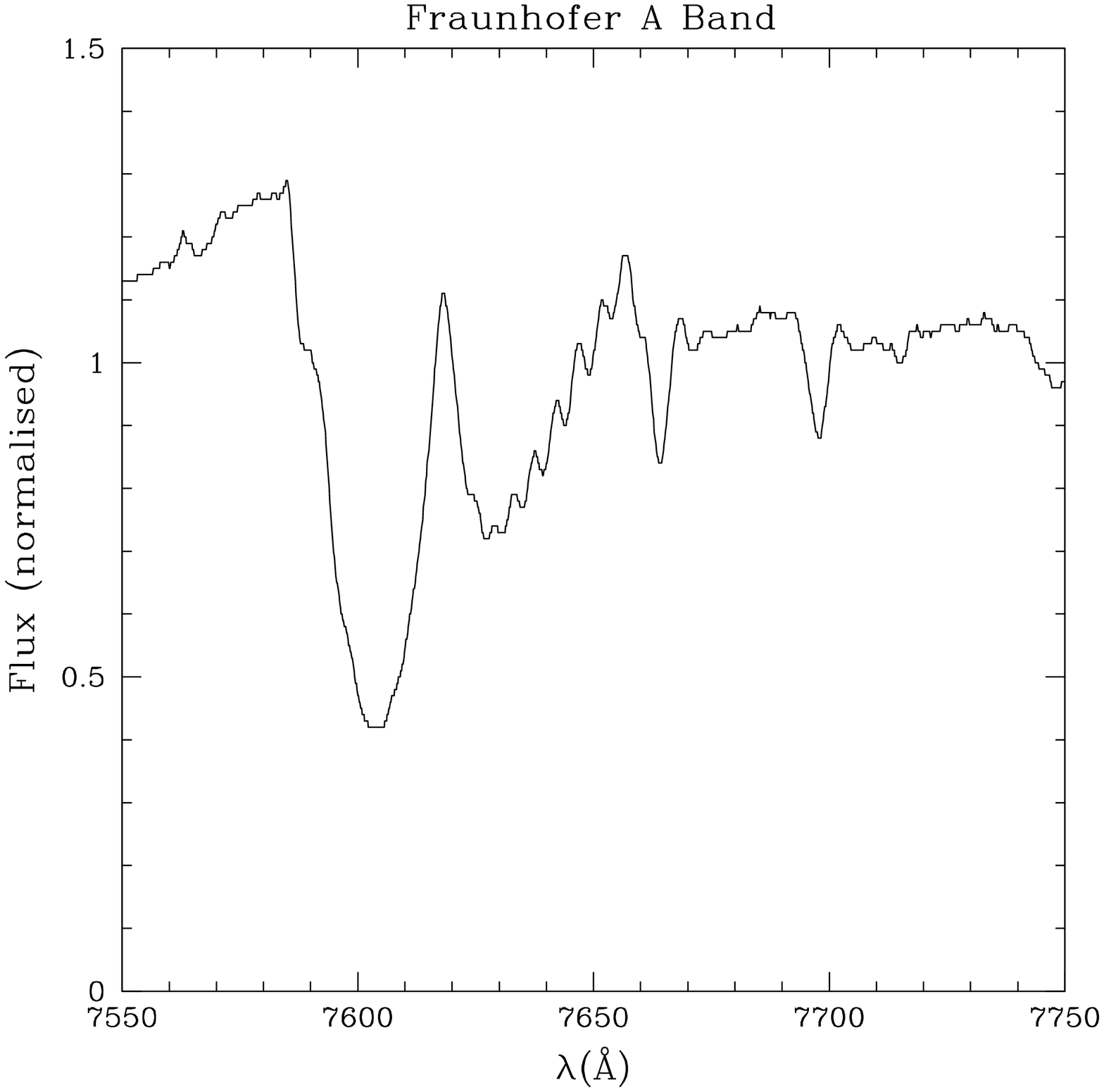}
\includegraphics[angle=0,width=0.45\hsize]{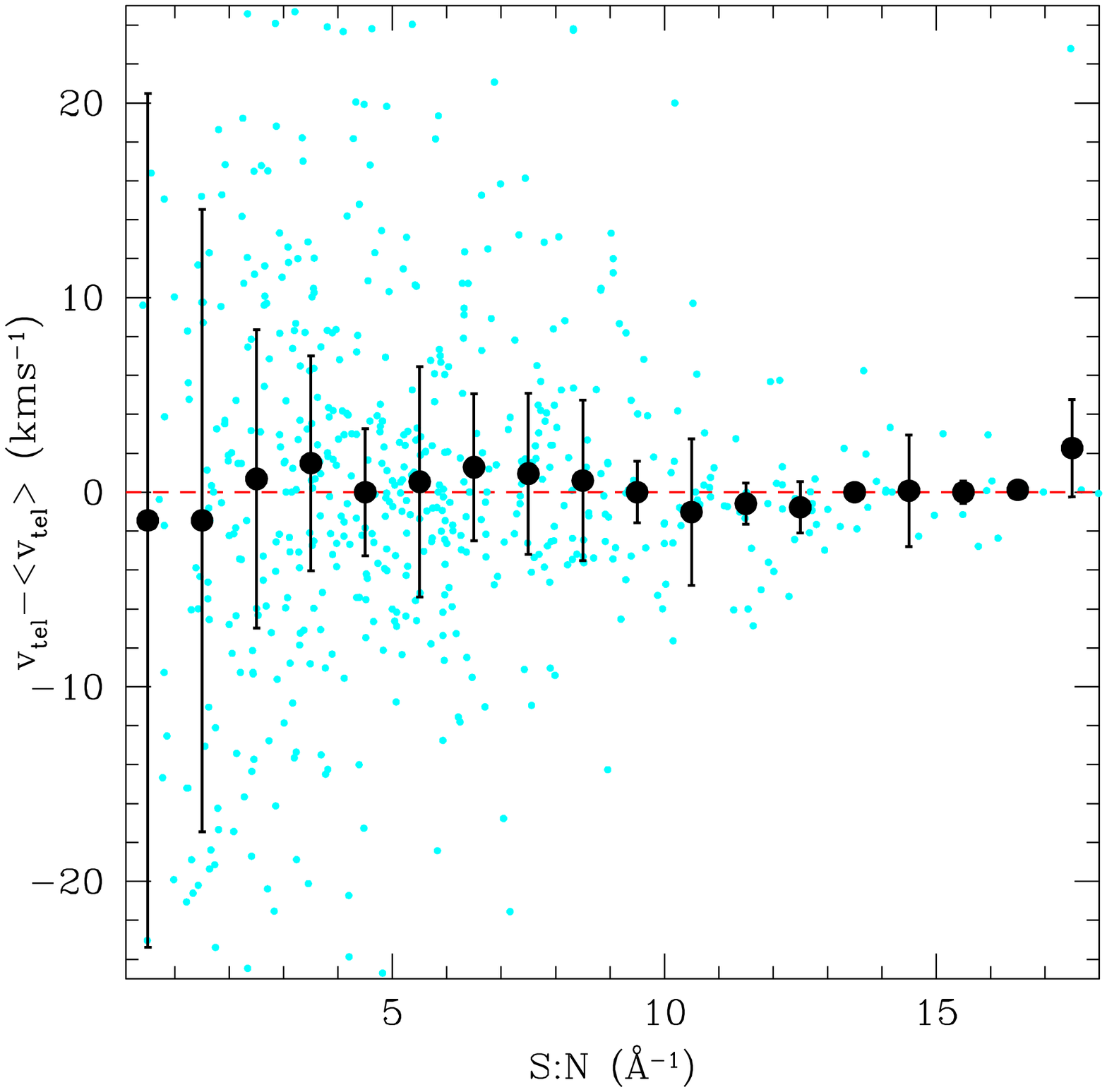}
\caption{{\bf Left: } Keck II DEIMOS spectrum of a bright foreground star,
  centered on the region of the strong telluric absorption feature, the
  Fraunhofer A band. {\bf Right:} The deviation of the telluric correction for
  every observed star within our sample from the average correction determined
  for the spectroscopic mask ($v_{tel}-<v_{tel}>$) as a function of S:N. Cyan
  points represent the individual data, and the large black points represent
  the median of all points within 1\AA\ bins in S:N. Error bars show the
  standard deviation within the bin. The median value for each bin is
  consistent with zero (i.e., the median), and the dispersion increases with
  decreasing S:N, arguing against using this correction in the low S:N
  regime.}
\label{fig:telluric}
\end{center}
\end{figure*}

With slit-spectroscopy, systematic velocity errors can be introduced if stars
are not well aligned within the centre of their slits. Such misalignments can
result from astrometric uncertainties or a slight offset in the position angle
of the mask on the sky. For our astrometry, we take the positions of stars
from PAndAS photometry, which have an internal accuracy of
$\sim0.1^{\prime\prime}$ and a global accuracy of $\sim0.^{\prime\prime}25$
\citep{segall07}. This can translate to velocity uncertainties of up to
$\sim15\kms$ for our DEIMOS setup. In previous studies, authors have tried to
correct for this effect by cross correlating their observed spectra with
telluric absorption features
(e.g., \citealt{sohn07,simon07,kalirai10,collins10,tollerud12}). These
atmospheric absorption lines are superimposed on each science spectrum, and
should always be observed at their rest-frame wavelengths. Thus, if one is
able to determine the offset of these features, shifts caused by misalignment
of the science star within the slit can be corrected for, and this can be
applied on a slit by slit basis. The strongest of these features is the
Fraunhofer A-band, located between 7595--7630\AA. An example of this feature
is shown in the left panel of Fig.~\ref{fig:telluric}.

While we believe this correction is robust in the high S:N regime, we argue
against applying this correction in studies of Andromedean satellites, where
S:N is often quite low (typically less than 8--10\AA$^{-1})$ for 1 hour observations
of faint ($i\gta21$) RGB stars. We find that when we compute this offset for
all stars within our sample, those with high S:N tend to cluster within a few
$\kms$ of the average telluric correction found for the mask. As the S:N
decreases below about 10\AA$^{-1}$, the scatter about this mean value increases
dramatically, as do the uncertainties computed for each individual correction. This
is because the telluric feature is a single, very broad and asymmetric
feature. It is therefore easy in the noisy regime for the cross correlation
routine to misalign the template and science spectrum whilst still producing a
high confidence cross-correlation maximum. We show this effect explicitly in
the right hand panel of Fig.~\ref{fig:telluric} where we plot the deviation of
the telluric correction for every star within our sample from the average
correction determined for the spectroscopic mask it was observed with
($v_{tel}-<v_{tel}>$) as a function of S:N. The cyan points represent the
individual data, and the large black points represent the median of all points
within 1\AA\ bins in S:N. The error bars represent the dispersion within each
bin. It is plainly seen that the median value for each bin is consistent with
zero (i.e., the median), and that the dispersion increases with decreasing
S:N. If we were to apply these velocity corrections to all our stars, it is
probable that we would merely increase the velocity uncertainties rather
than reducing them. 

For this reason, we take a different approach. Using solely the telluric velocity
corrections of stars from each observed mask whose spectra have S:N$>7$, we
measure (a) the average telluric correction for the mask and (b) the evolution
of the telluric correction as a function of mask position. In this way, we can
track any gradient in our measurements that could be caused by e.g., rotational
offsets in our mask. In general, we find these corrections to be slight. The
average measured offset across all our masks is $3.8\kms$ (ranging from
between $-3.4\kms$ and $+10.6\kms$). The measured gradients are very slight,
resulting in an average end-to-end mask difference of 2.6$\kms$, with a range of
$0.1-7.2\kms$, typically within our measured velocity uncertainties.

\section{A probabilistic determination of membership}
\label{sect:membership}

Determining membership for Andromedean dwarf spheroidals is notoriously
difficult in the best of cases. We only possess information about the
velocity, CMD position, distance from the centre of the dSph and spectroscopic
metallicity (although this carries large uncertainties of $>0.3$ dex for
individual stars). Depending on the systemic velocity of the dSph, we must try
to use these properties to distinguish the likely members from either Milky
Way halo K-dwarfs ($v_{hel}\ga-150\kms$) or M31 halo giants
($v_{halo}\approx-300\kms$, $\sigma_{v,halo}\approx90\kms$,
\citealt{chapman06}). In the case of Galactic contamination, our spectra also
cover the region of the Na I doublets ($\sim 8100$\AA). As this feature is
dependent on the stellar surface gravity, it is typically stronger in dwarf
stars than in giants. However, there is a significant overlap between the two,
especially in the CMD color region of interest for Andromedean RGB stars. In
the past, groups have focused on making hard cuts on likely members in an
attempt to weed out likely contaminants (e.g., based on their distance from the
centre of the galaxy or their velocity,
\citealt{chapman05,collins10,collins11b,kalirai10}), but such `by eye'
techniques are not particularly robust. \citet{tollerud12} recently presented
an analysis of a number of M31 dSphs where they used a more statistical method
to ascertain likely membership, using the distance from the centre of, and
position in the CMD of stars targeted within their DEIMOS masks. Here we
employ a similar technique that will assess the probability of stars being
members of a dSph based on (i) their position on the CMD; (ii) their distance
from the centre of the dSph and, in addition; (iii) their position in velocity
space, giving the likelihood of membership as:

\begin{equation}
\label{eqn:probtotal}
P_i\propto P_{CMD}\times P_{dist}\times P_{vel}
\end{equation}

\noindent Below, we fully outline our method, and implement a series of tests
to check it can robustly recover the kinematics of the M31 dSphs.

\subsection{Probability based on CMD position, $P_{CMD}$}

For the first term, $P_{CMD}$, we are interested in where a given
star observed in our mask falls with respect to the RGB of the
observed dSph. \citet{tollerud12} use the distance of a given star
from a fiducial isochrone fit to the dwarf photometry to measure this
probability. Here, we determine this value from the data itself,
rather than using isochrones. Using the PAndAS CFHT photometry, we
construct a normalised Hess diagram for the central region
(i.e., within $2\times r_{\rm half}$) of the dSph, and one of a
surrounding `field' comparison region. By combining these two Hess
diagrams, we can then map both the color distribution of the dSph and
that of our contaminating populations.  We use both directly as
probability maps, where the densest region would have a value of
1.

\begin{figure*}
  \begin{center}
    \includegraphics[angle=0,width=0.45\hsize]{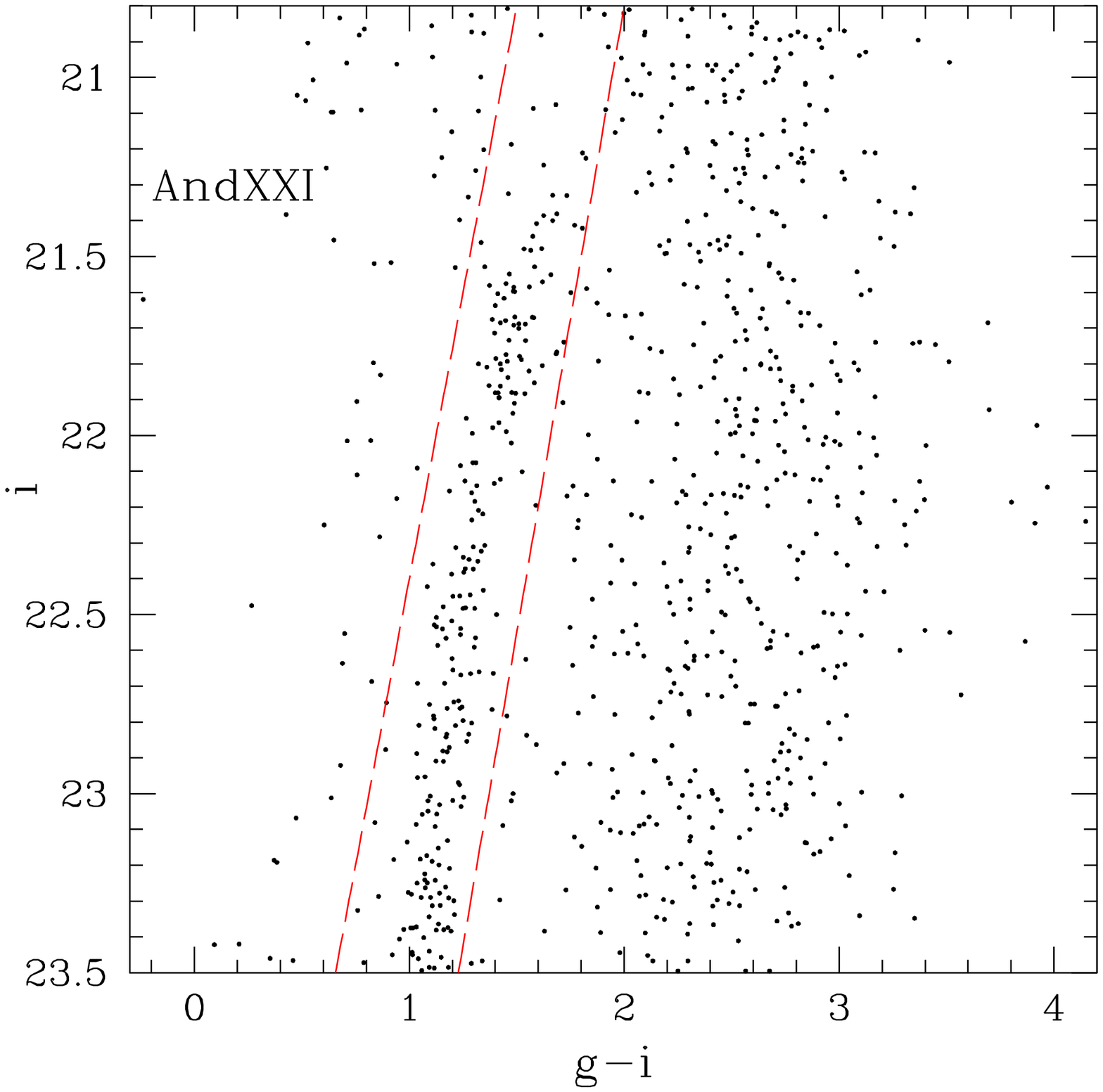}
    \includegraphics[angle=0,width=0.45\hsize]{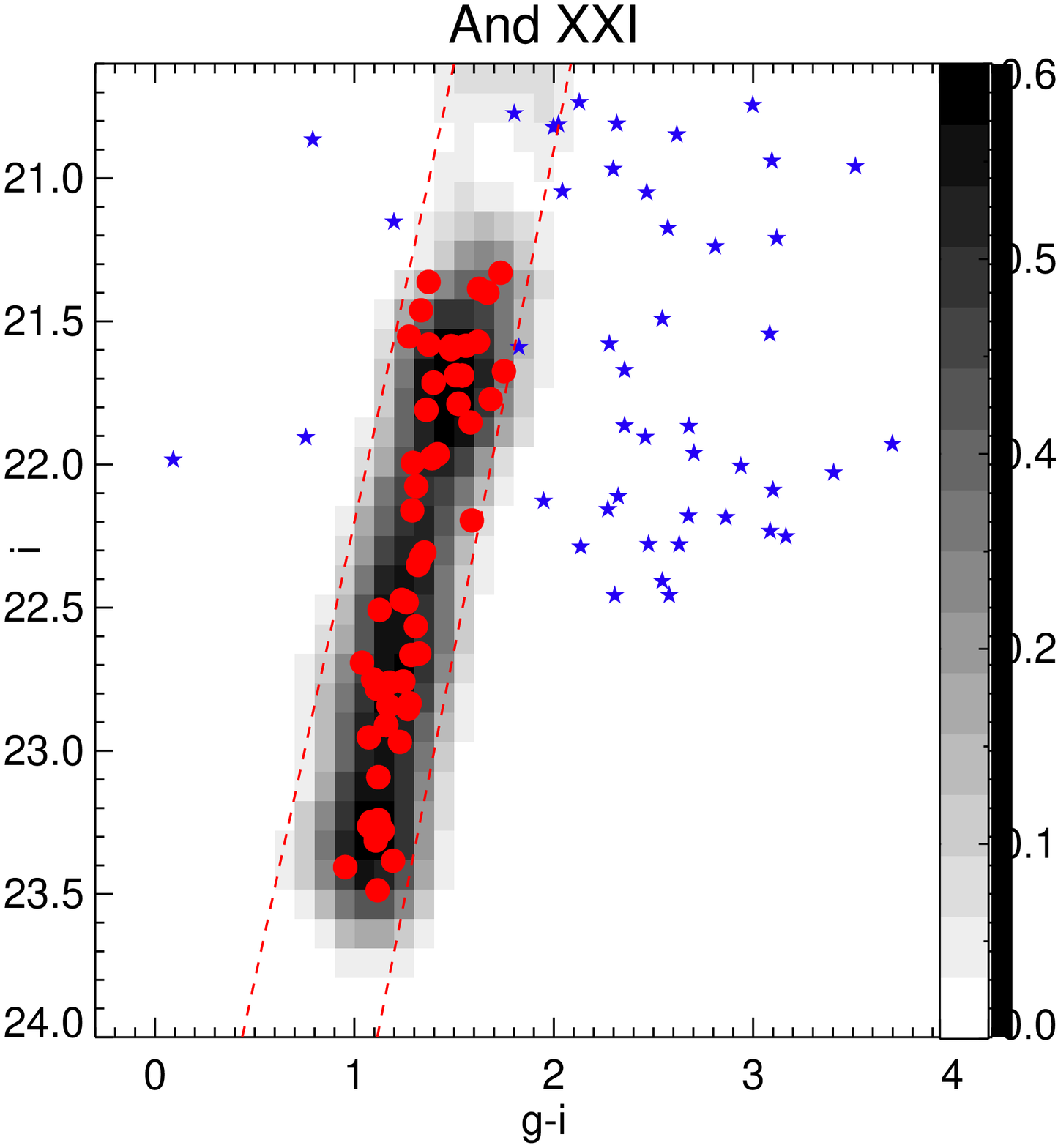}
    \caption{{\bf Left:} The PAndAS CMD for And XXI, showing all stars
    within $2\times r_{\rm half}$ of the dwarf centre. The red dashed
    lines show the bounding box we use to construct our probability
    map for the RGB of the dSph. {\bf Right:} A background normalized
    Hess diagram for the RGB region of And XXI. The color represents
    the probability, $P_{CMD}$ in each cell for a star in that
    position belonging to And XXI. This diagram is used as a
    probability map for all our spectroscopically observed stars. Red
    points represent stars DEIMOS stars for which $P_{CMD}>10^{-6}$, while
    the blue stars represent objects for which $P_{CMD}\leq10^{-6}$.}
\label{hess}
\end{center}
\end{figure*}

So we are not dominated by shot noise of sparsely populated regions of
the CMD, we use only the region of the RGB. We do this by assigning a
generous bounding box around the RGB as seen in the left hand panel of
Fig.~\ref{hess} where we display the PAndAS CMD for the well populated
And XXI RGB. We have zoomed in on the region for which DEIMOS
observations with reliable velocities can be obtained,
e.g., $i<23.5$. The bounding box is shown with red dashed
lines. Anything that falls outside this region is therefore assigned a
probability of $P_{CMD}=0$. The resulting probability map for And XXI
is shown in the right hand panel. Red points represent all DEIMOS
stars that have $P_{CMD}>10^{-6}$, while the blue stars show stars from the
DEIMOS mask that are far removed from the And XXI RGB, and thus not
considered to be members.

\subsection{Probability based on distance position, $P_{dist}$}

The second term in our probability function, $P_{dist}$ can be easily
determined from the known radial profile of the dSphs. The half-light radii of
all these objects are known and can be found in
\citet{mcconnachie06b,zucker04,zucker07,mcconnachie08,martin09,collins10,collins11b,richardson11}.
We also know that their density profiles are well represented by a Plummer
profile with a scale radius of $r_p\equiv r_{\rm half}$. Therefore, we can
define the probability function as a normalised Plummer profile \citep{plummer11}, i.e.,:

\begin{equation}
P_{dist}=\frac{1}{\pi r_p^2[1+(r/r_p)^{2}]^2}
\end{equation}

The above equation assumes that the systems we are studying are perfectly
spherical. While the majority of these systems are not significantly
elliptical, it is important to consider the effect of any observed deviations
from sphericity. We therefore modify $r_p$ based on a given stars angular
position with respect to the dwarfs major axis, $\theta_i$, such that:

\begin{equation}
r_p=\frac{r_{\rm half}(1-\epsilon)}{1+\epsilon{\rm cos}\theta_i}
\end{equation}

\noindent where $\epsilon$ is the measured ellipticity of the dSph as taken
from \citet{mcconnachie12}.

\subsection{Probability based on velocity, $P_{vel}$}
\label{sect:velprob}
The final term, $P_{vel}$, contains information about the likelihood of a
given star belonging to a kinematic substructure that is not well described by
the velocity profiles of either the MW halo dwarfs or the Andromeda halo
giants, both of which are determined empirically from our DEIMOS database of
$>20000$ stars. From an analysis of these stars (selecting out any obvious
substructures and the Andromedean disc) we find that the M31 halo is well
approximated by a single Gaussian with a systemic velocity of
$v_{r,halo}=-308.8\kms$ and $\sigma_{v,halo}=96.3\kms$, giving a probability
density function for a given star with a velocity $v_i$ and velocity
uncertainty of $v_{err,i}$ of:

\begin{equation} 
\begin{aligned}
  P_{halo}=\frac{1}{\sqrt{2\pi(\sigma_{v,halo}^2+v_{err,i}^2)}}\times\\{\rm exp}\Big[-\frac{1}{2}\left(\frac{v_{r,halo}-v_{r,i}}{\sqrt{\sigma_{v,halo}^2+v_{err,i}^2}}\right)^2\Big] \end{aligned}
\end{equation}

\noindent The MW halo population is well approximated by 2 Gaussians with
$v_{r,MW 1}=-81.2\kms$, $\sigma_{v,MW 1}=36.5\kms$ and $v_{r,MW 2}=-40.2\kms$ and
$\sigma_{v,MW 2}=48.5\kms$, resulting in a probability density function for a
given star with a velocity $v_i$ and velocity uncertainty of $v_{err,i}$ of:
\begin{equation} 
\begin{aligned}
  P_{MW}=\frac{R}{\sqrt{2\pi(\sigma_{v,MW
      1}^2+v_{err,i}^2)}}\times\\{\rm exp}\Big[-\frac{1}{2}\left(\frac{v_{r,MW
        1}-v_{r,i}}{\sqrt{\sigma_{v,MW1}^2+v_{err,i}^2}}\right)^2\Big] \\+ \frac{(1-R)}{\sqrt{2\pi(\sigma_{v,MW
      2}^2+v_{err,i}^2)}}\times\\{\rm exp}\Big[-\frac{1}{2}\left(\frac{v_{r,MW 2}-v_{r,i}}{\sqrt{\sigma_{v,MW
        2}^2+v_{err,i}^2}}\right)^2\Big] 
\end{aligned}
\end{equation}

\noindent where $R$ is the fraction of stars in the first MW peak, and
$(1-R)$ is the fraction of stars in the second peak. The value of $R$
is determined empirically from our DEIMOS data set. 

A strong kinematic peak outside of these two populations can then be
searched for using a maximum likelihood technique, based on the
approach of \citet{martin07}. We search for the maximum in the
likelihood function that incorporates the two contamination
populations plus an additional Gaussian structure with systemic
velocity $v_{r,substr}$ and a dispersion of $\sigma_{v,substr}$,
defined as:

\begin{equation} 
\begin{aligned}
\label{eq:mlpeta}
P_{substr}=\frac{1}{\sqrt{2\pi([\eta\sigma_{v,substr}]^2+v_{err,i}^2)}}\times\\{\rm exp}\Big[-\frac{1}{2}\left(\frac{v_{r,substr}-v_{r,i}}{\sqrt{[\eta\sigma_{v,substr}]^2+v_{err,i}^2}})]\right)^2\Big]
\end{aligned}
\end{equation}

\begin{figure*}
\begin{center}
    \includegraphics[angle=0,width=0.45\hsize]{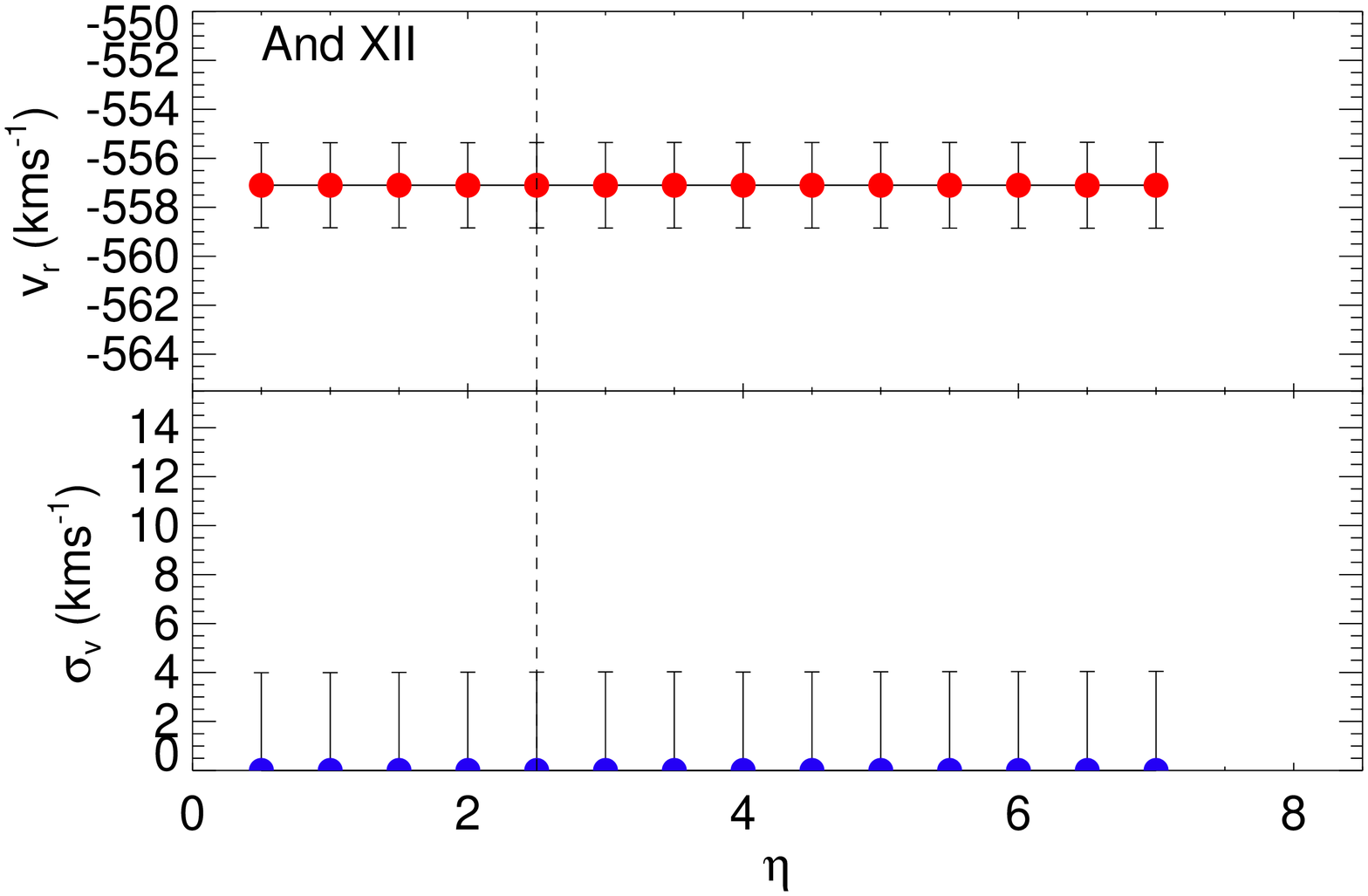}
    \includegraphics[angle=0,width=0.45\hsize]{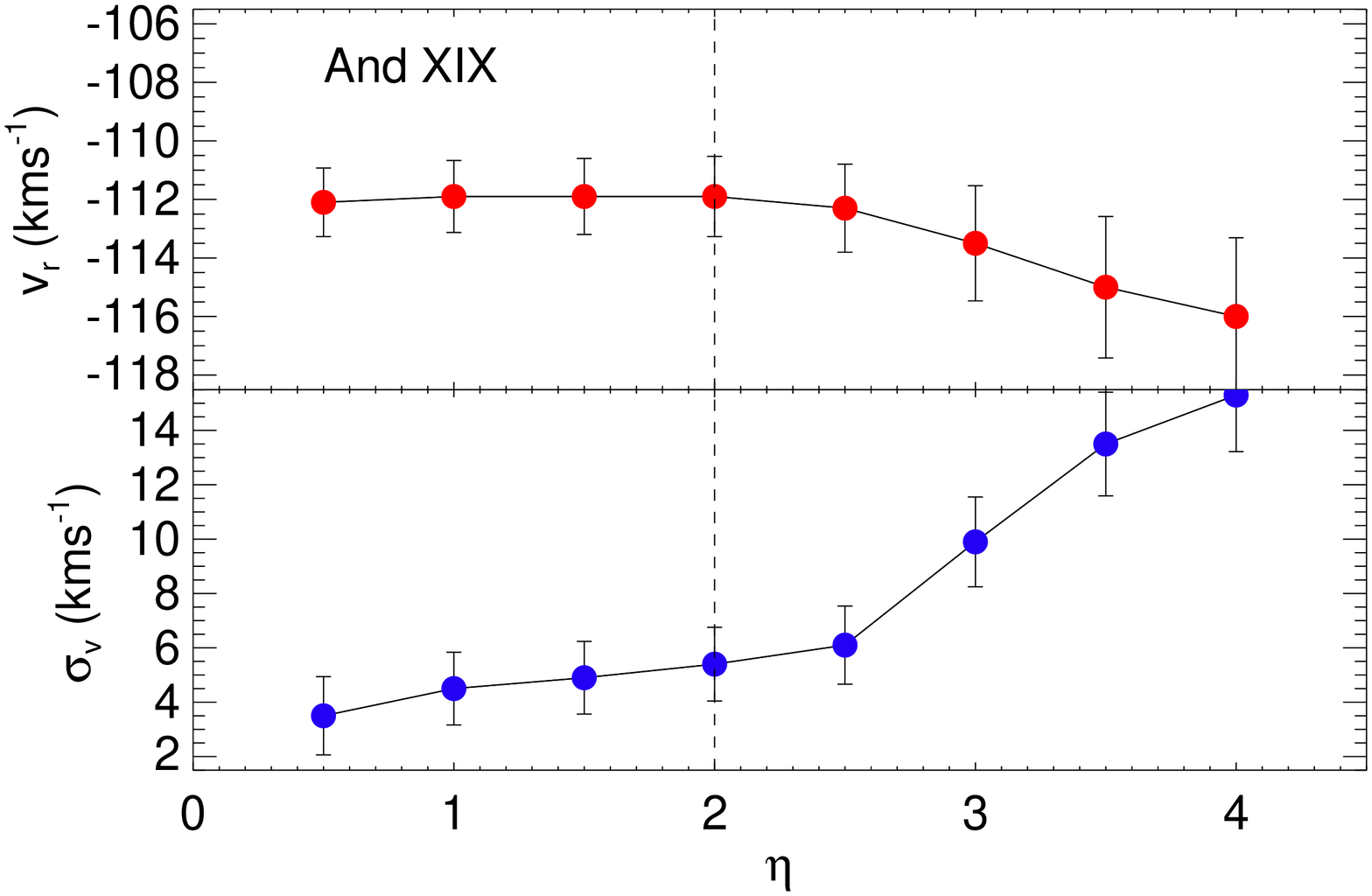}
    \includegraphics[angle=0,width=0.45\hsize]{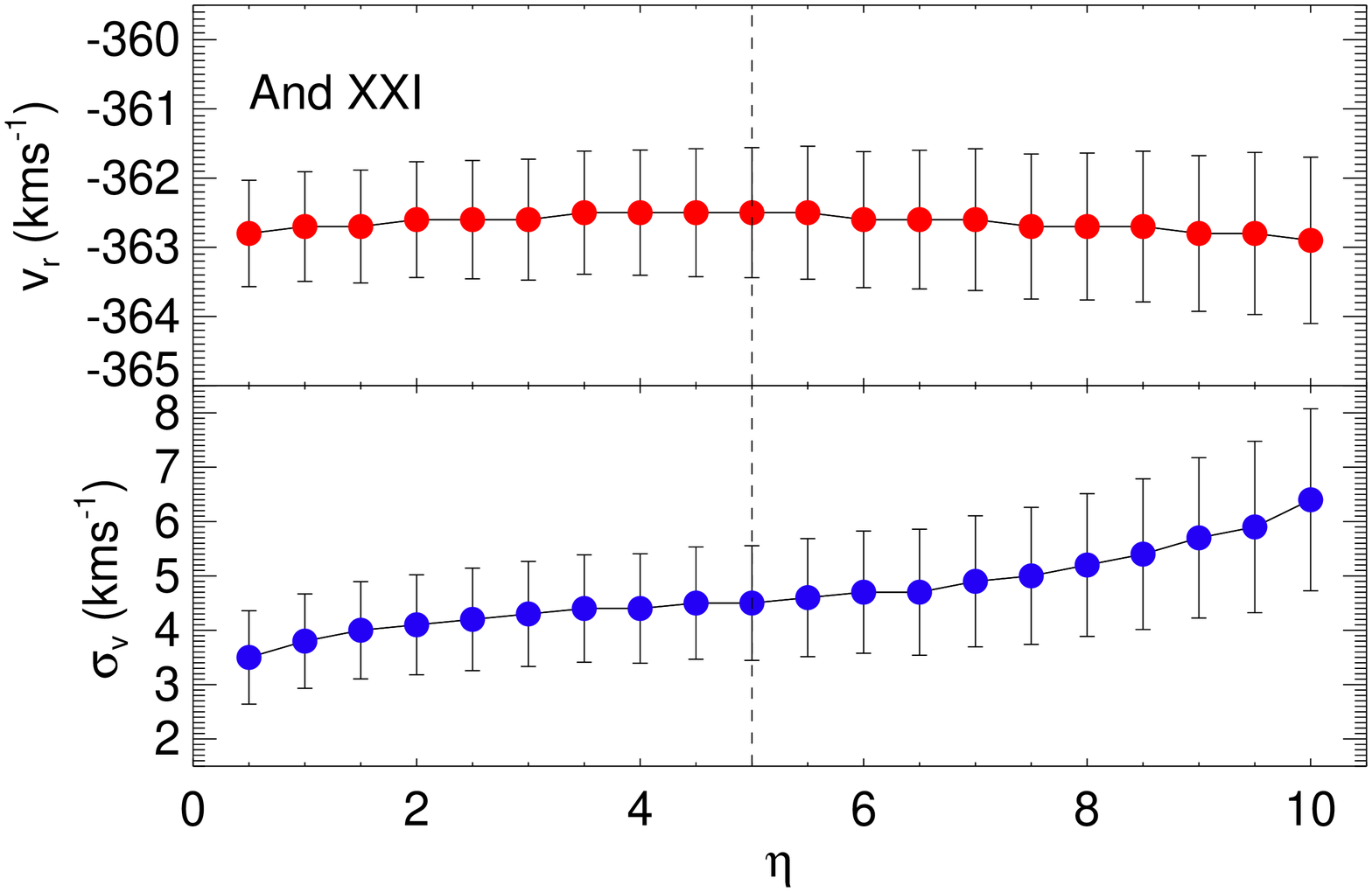}
    \includegraphics[angle=0,width=0.45\hsize]{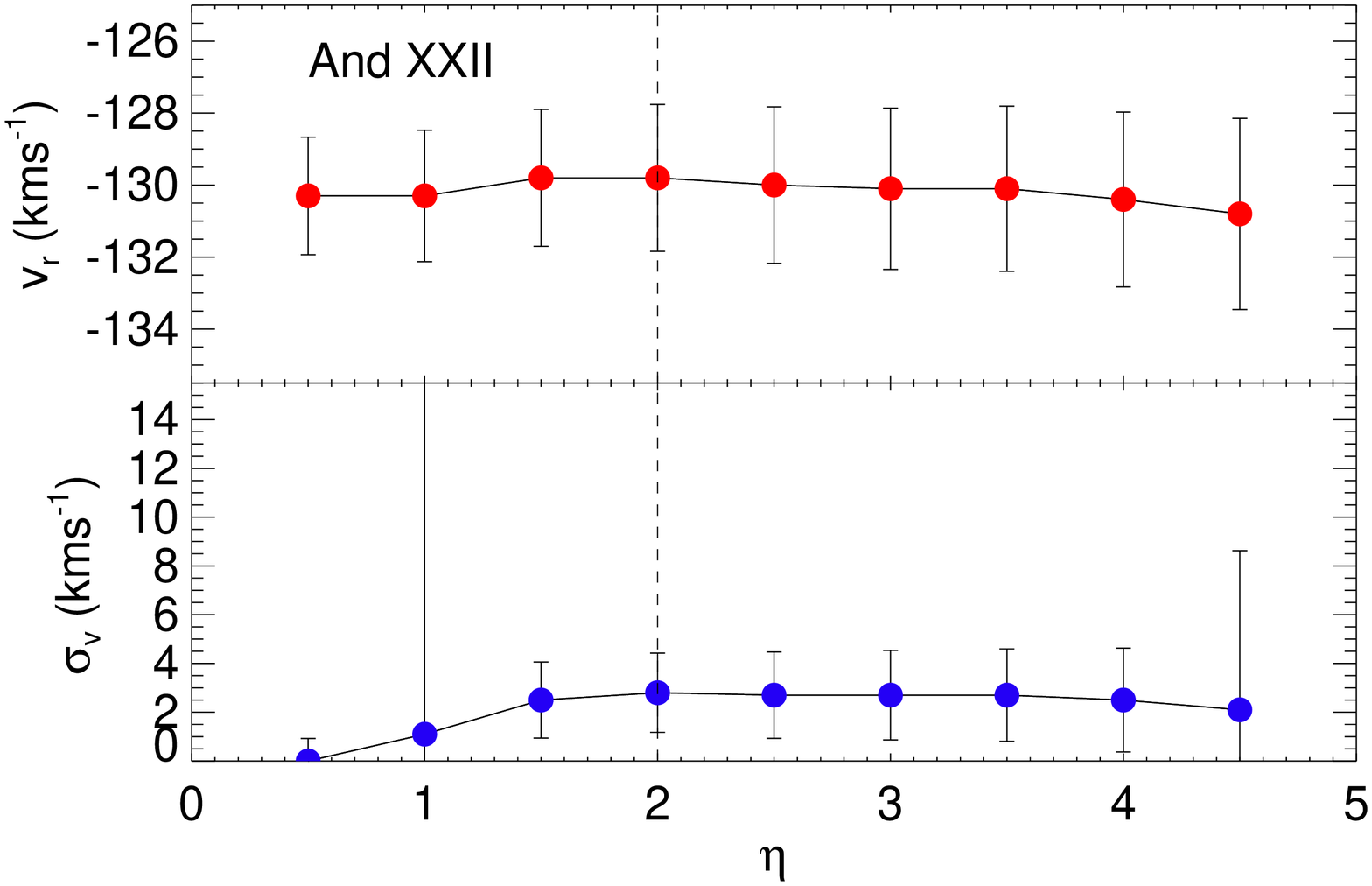}
    \includegraphics[angle=0,width=0.45\hsize]{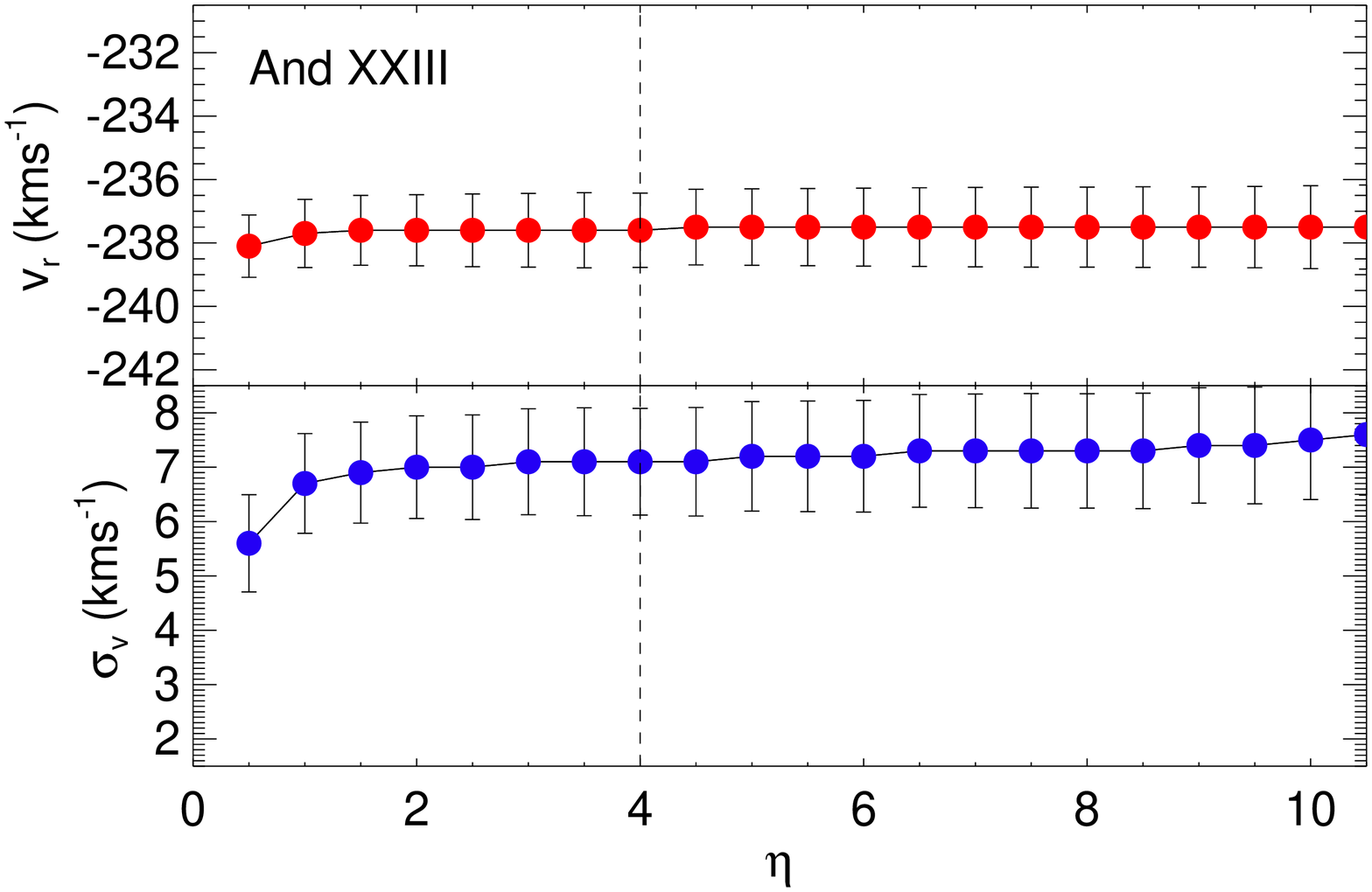}
    \includegraphics[angle=0,width=0.45\hsize]{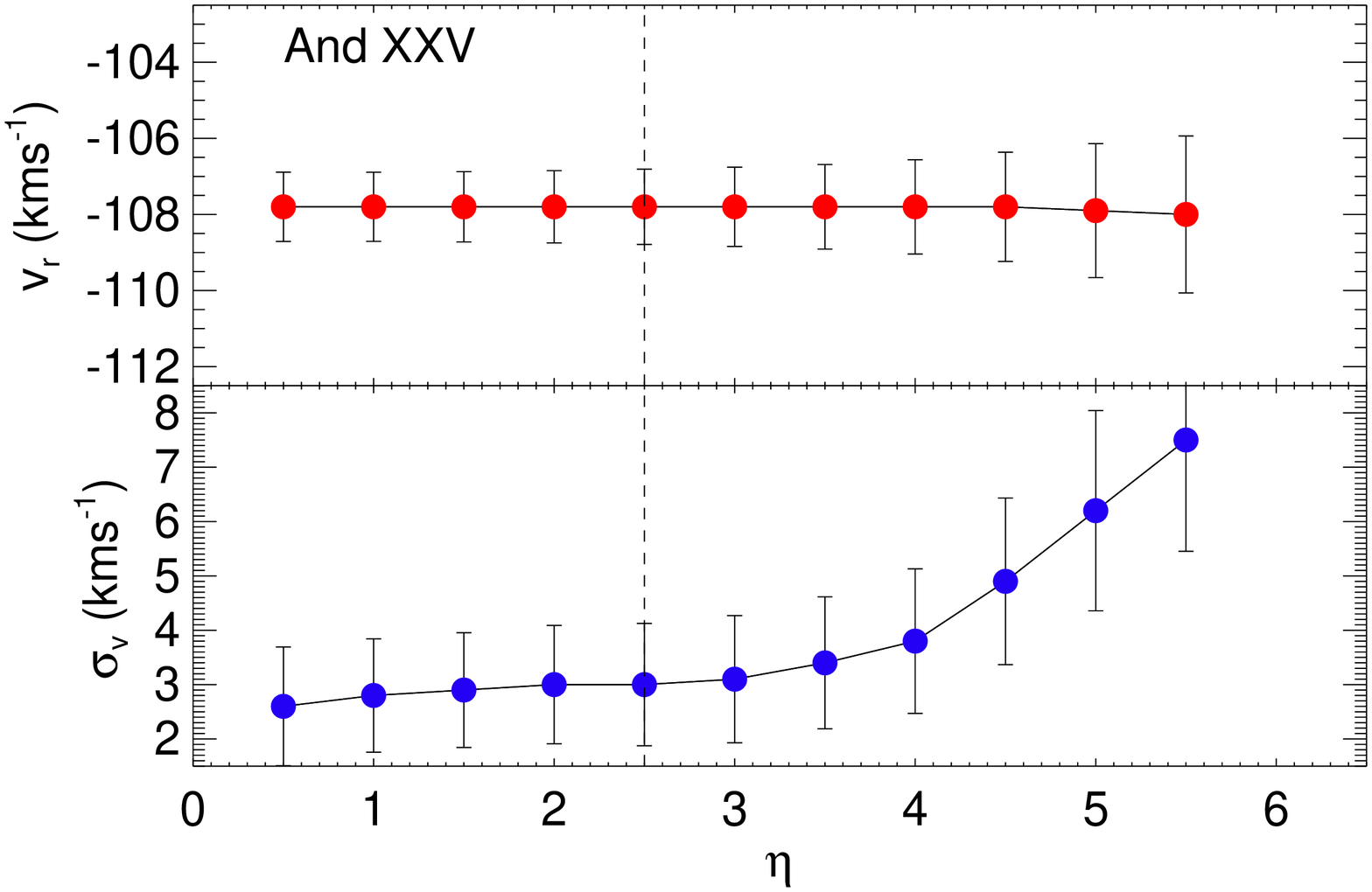}
    \caption{The effect of modifying $\eta$ -- a multiplicative weight applied
      to the dispersion of the identified substructure in our $P_{vel}$ term
      -- on the final derived systemic velocities ($v_r$) and dispersions
      ($\sigma_v$) for six of our dSphs.The effect of increasing $\eta$ is
      typically more pronounced in objects with systemic velocities close to
      that of Milky Way contamination (such as And XIX and XXV) than for other
      objects. Dashed lines represent the optimal value of $\eta$ for each
      system.}
\label{fig:etatest}
\end{center}
\end{figure*}

Here, to ensure we haven't biased our $P_{substr}$ strongly against stars that
lie within the wings of the Gaussian distribution of dwarf spheroidal
velocities, we have included a multiplicative free parameter, $\eta$, to our
derived value of $\sigma_v$. To determine the ideal value of $\eta$, we ran
our algorithm over all our datasets, changing the value of $\eta$ from
$0.5-10.5$ to see its effect on the final derived systemic velocities and
velocity dispersion. We find that in all cases, the solutions converge at
values of $\eta\sim2-4$. For dSphs whose kinematics are well separated from
contaminants, the derived kinematics can remain stable up to much larger
values of $\eta$, however for those with systemic velocities within the
velocity regime of the Milky Way, the solution quickly destabilizes as more
contaminants are included as probable members. We show this implicitly in
Fig.~\ref{fig:etatest}, where we present the effect of modifying $\eta$ on 6
dSphs, And XII, XIX, XXI, XXII, XXIII and XXV. These objects were selected as
they nicely probe our datasets with low numbers of probable member stars
($\sim8$), to those where we have 10s of probable members, as well as sampling
dSphs from highly contaminated to well isolated kinematic regimes. The value
of $\eta$ is therefore independently determined for each dataset separately,
and we report its final value in Table~\ref{tab:kprops}.

\begin{equation} 
\begin{aligned}
\label{eq:mlp}
P_{substr}=\frac{1}{\sqrt{2\pi(\sigma_{v,substr}^2+v_{err,i}^2)}}\times\\{\rm exp}\Big[-\frac{1}{2}\left(\frac{v_{r,substr}-v_{r,i}}{\sqrt{\sigma_{v,substr}^2+v_{err,i}^2}})]\right)^2\Big]
\end{aligned}
\end{equation}

\begin{figure}
  \begin{center}
    \includegraphics[angle=0,width=0.85\hsize]{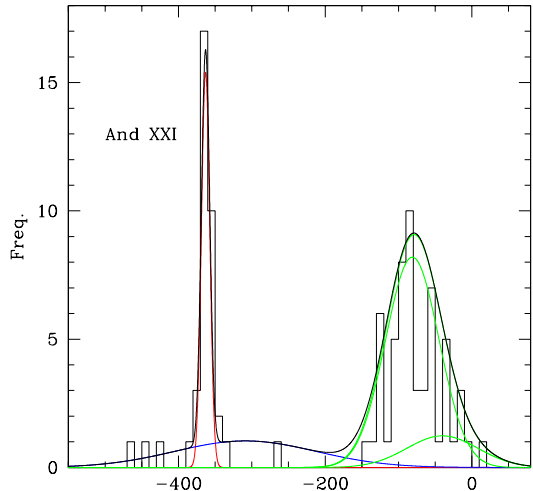}
    \caption{A velocity histogram for all observed stars in the field of And
      XXI. Our empirical Gaussian fits to the full Keck II data set for the
      M31 and MW halos are overlaid in blue and green respectively. A cold,
      kinematic peak at $-400\kms$ is also seen, and this is the likely
      signature of the dSph. Our coarse, initial ML procedure identifies this
      peak, and the values of $v_{sys}$ and $\sigma_v$ it measures are used to
      derive our kinematic probability, $P_{vel}$.}
\label{fig:vtest}
\end{center}
\end{figure}

The likelihood function can then be simply written as:

\begin{equation}
\rm{log}[\Lagr(v_r,\sigma_v)]=\sum_{i=1}^{N}\rm{log}\Big(
\alpha P_{i,{\rm halo}}+\beta P_{i,{\rm MW}}+\gamma P_{i,{\rm substr}}\Big)
\end{equation}

\noindent where $\alpha$, $\beta$ and $\gamma$ represent the Bayesian priors,
i.e., the expected fraction of stars to reside in each population. These are
determined by starting with arbitrary fractions (for example, 0.2, 0.5 and 0.3
respectively) and are then adjusted to the posterior distribution until priors
and posteriors match. This technique will therefore identify an additional
kinematic peak, independent of the MW and M31 halo populations, if it
exists. We stress that these are not the final systemic velocity and
dispersion of the dSph, but merely indicate a region in velocity space in
which an excess of stars above the two contaminant populations is seen. In
Fig.~\ref{fig:vtest}, we show the result of this process for the And XXI
dSph. Here, the substructure is clearly visible as a cold spike at
$\sim-400\kms$.

Now that we have a velocity profile for our three components (MW, M31 halo and
the dSph), we can assign probabilities for each star within our sample
belonging to each population using simple Bayesian techniques, i.e., the
probability that a given star belongs to the substructure, $P_{vel}$, is:

\begin{equation}
P_{vel}=\frac{\gamma P_{substr}}{\alpha P_{halo}+\beta P_{MW}+\gamma P_{substr}}
\end{equation}

\noindent and the probability of being a contaminant is:

\begin{equation}
P_{nvel}=\frac{\alpha P_{halo}+\beta P_{MW}}{\alpha P_{halo}+\beta P_{MW}+\gamma P_{substr}}
\end{equation}

\subsection{Measuring $v_r$ and $\sigma_v$}

\noindent Upon applying this to our data, we can identify the most probable members of
each dSph, without having to apply any additional constraints or cuts.

Having established the membership probability for each observed star (as
detailed above) we now calculate the kinematic properties of each dSph; namely
their systemic velocities ($v_r$) and velocity dispersions ($\sigma_v$). We
use the maximum likelihood technique of \citet{martin07}, modified to include
our probability weights for each star.  We sample a coarse grid in
$(v_r,\sigma_v)$ space and determining the parameter values that maximise the
likelihood function (ML), defined as:

\begin{equation} 
\label{eq:ml}
{\rm log}[\Lagr(v_r,\sigma_v)]=-\frac{1}{2}\sum_{i=1}^{N}\Big[P_i{\rm
  log}(\sigma_{\mathrm{tot}}^2)+P_i\frac{v_r-v_{r,i}}{\sigma_{\mathrm{tot}}}^2\\+P_i{\rm
 log}(2\pi)\Big]
\end{equation}

\noindent where $N$ is the number of stars in the sample, $v_{r,i}$ is the
radial velocity measured for the $i^\mathrm{th}$ star, $v_{err,i}$ is the
corresponding uncertainty and
$\sigma_{\mathrm{tot}}=\sqrt{\sigma_v^2+v_{err,i}^2}$. In this way, we aim to
separate the intrinsic dispersion of a system from the dispersion introduced
by the measurement uncertainties.

\subsection{Testing our probabilistic determination of membership and
  calculations of kinematic properties}
\label{sect:test}

Having developed the above technique, it is important for us to rigorously test
that it is robust enough to accurately determine the global kinematic
properties for each of our datasets. In Appendix A, we examine in detail a number of
potential issues that could cause our algorithm to return biased or incorrect
results. These are the inclusion of a velocity dependent term in our
probability calculation, the effect of including low S:N data (S:N$<5$\AA) in our
analysis and the effect of small sample sizes ($N_*<8$) on our measurements of
kinematic properties. We briefly summarize our findings here, and refer the
reader to Appendix A for a more detailed description.

This work has introduced the concept of assigning a probability of membership
for a given star to a dSph based on the prior knowledge of the velocity
profiles of our expected contaminant populations, $P_{vel}$, a technique that
has not previously been used in the study of M31 dSphs. To test that this is
not biasing our results, we can simply remove this term from
Eqn.~\ref{eqn:probtotal}, and follow the technique of T12, where they use only
$P_{CMD}$ and $P_{dist}$ terms and then cut all stars with $P_{i}<0.1$ and
velocities that lie greater than $3\sigma$ from the mean of this sample from
their final analysis. We find that both techniques produce very similar
results, however our algorithm is more robust in regimes where the systemic
velocity of the dSph is close to that of the MW, and in dSphs where our number
of probable member stars is low ($N_*<10$).

To test the effect of low S:N data on our calculations of $v_r$ and
$\sigma_v$, we use our datasets for And XXI, XXIII and XXV, all of which have
$\geq25$ associated members. For each dataset, we apply a series of cuts to
the sample based on S:N (at levels of S:N$>2,3,4$ and 5\AA) and rerun our
algorithm. In all cases we find that the derived probabilities do not
significantly differ when the low S:N data are included, justifying our
inclusion of all stars for which velocities are calculated by our pipeline.

We also test the ability of our algorithm to measure $v_r$ and $\sigma_v$
in the small $N_*$ regime. For some of our datasets, we are only able to
identify a handful of stars as probable members. In theory, one can calculate
velocity dispersions accurately from only 3 stars if one is confident of ones
measurement uncertainties, as is demonstrated by \citet{aaronson83}
measurement of the velocity dispersion for Draco from only 3 stars, which
remains consistent with modern day measurements from significantly larger
datasets \citep{walker09b}. We can test if our results are similarly robust
using our larger datasets (such as And XXI, XXIII and XXV) by randomly
selecting 4, 6, 8, 10, 15, 20 and 25 stars from these datasets and rerunning
our algorithm to determine $v_r$ and $\sigma_v$ from these subsets. We repeat
this exercise 1000 times for each sample size, and examine the mean and
standard deviations for the computed quantities. We find that on average, for all sample
sizes, our routine measures systemic velocities and velocity dispersions that
are entirely consistent with those measured from the full sample, with a
spread that is very comparable to typical errors produced by our ML routine in
these low $N_*$ regimes. As such, we conclude that our technique is able to
place sensible limits on these values, even when dealing with as few as four
member stars. 

Finally, as the individual positions, velocities and velocity uncertainties
for all the stars analyzed in T12 are publicly available (with the data from
the non-member stars having been kindly passed on to us by the SPLASH team),
we can check that our algorithm is able to reproduce the values they measure
for their M13 dSph sample. We find that, in all cases, we calculate systemic
velocities and velocity dispersions that agree with their measured values to
well within their $1\sigma$ uncertainties.

These tests demonstrate that our method is robust enough to accurately
determine the global kinematic properties of M31 dSphs across a wide range of
sample sizes and data quality. We therefore proceed to apply it to the
datasets of all of the dSphs for which our group has acquired Keck II DEIMOS
observations to date.

\begin{figure*}
\begin{center}
\includegraphics[angle=0,width=0.95\hsize]{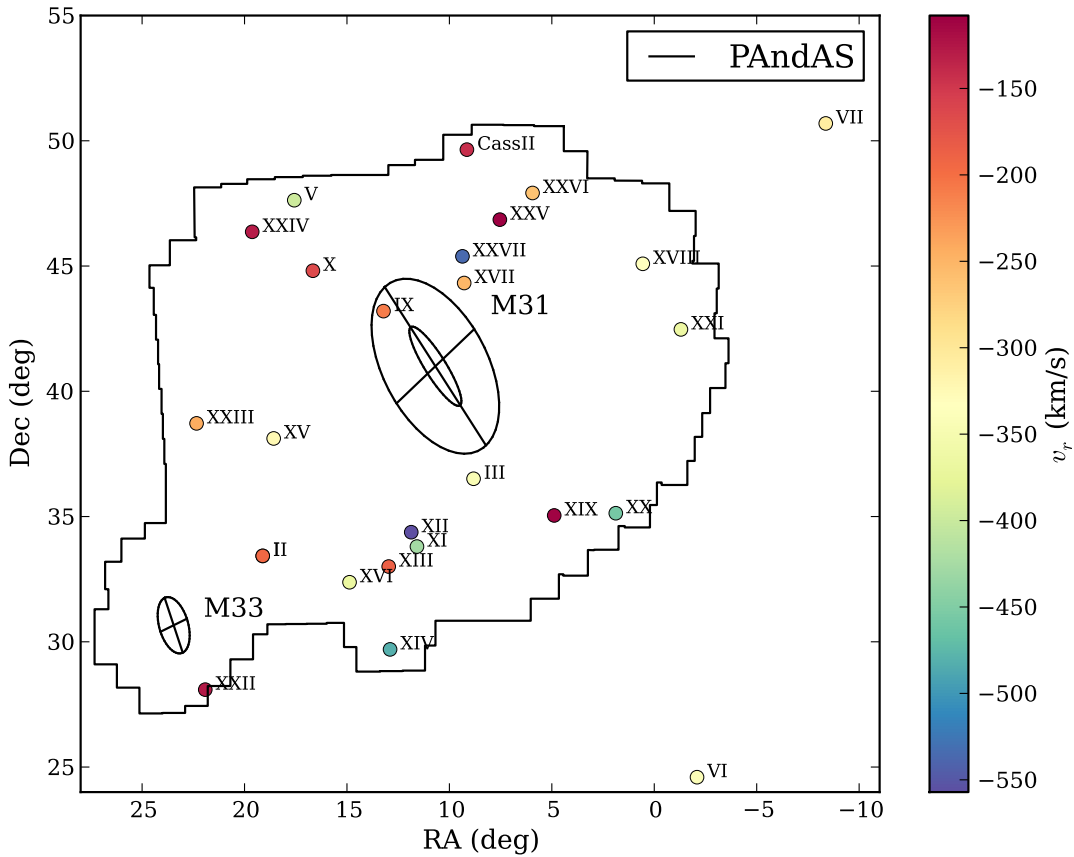}
\caption{Positions of 26 of the 28 Andromeda dSphs, color coded by their
  heliocentric velocity, $v_r$ (taken from this work, \citealt{kalirai10} and T12). The solid line represents the PAndAS survey
  foot print. M31 is shown by the double ellipse, where the outer ellipse
  shows a segment of a 55 kpc radius ellipse flattened to c/a = 0.6, and the
  major-axis and minor-axis are indicated with straight lines out to this
  ellipse. The inner ellipse corresponds to a disc of radius 2$\deg$ (27 kpc),
  with the same inclination as the main M31 disc. M33 is shown as a singular
  ellipse. }
\label{fig:vels}
\end{center}
\end{figure*}

\section{The kinematics of M31 dSphs}
\label{sect:results2}

With this vast dataset of dSph kinematics now in hand, we can begin to
statistically probe their structures more fully. In Fig.~\ref{fig:vels} we
display a summary of the velocities and positions of 26 of the 28 dSphs for
which kinematic data are available, where the values are taken from this work,
T12 and \citet{kalirai10}. In the following sections, we will discuss the
individual stellar kinematics, masses and chemistries of the dSphs analyzed within
this work.

\subsection{Andromeda XVII}
\label{sect:and17}

\begin{figure*}
\begin{center}
\includegraphics[angle=0,width=0.45\hsize]{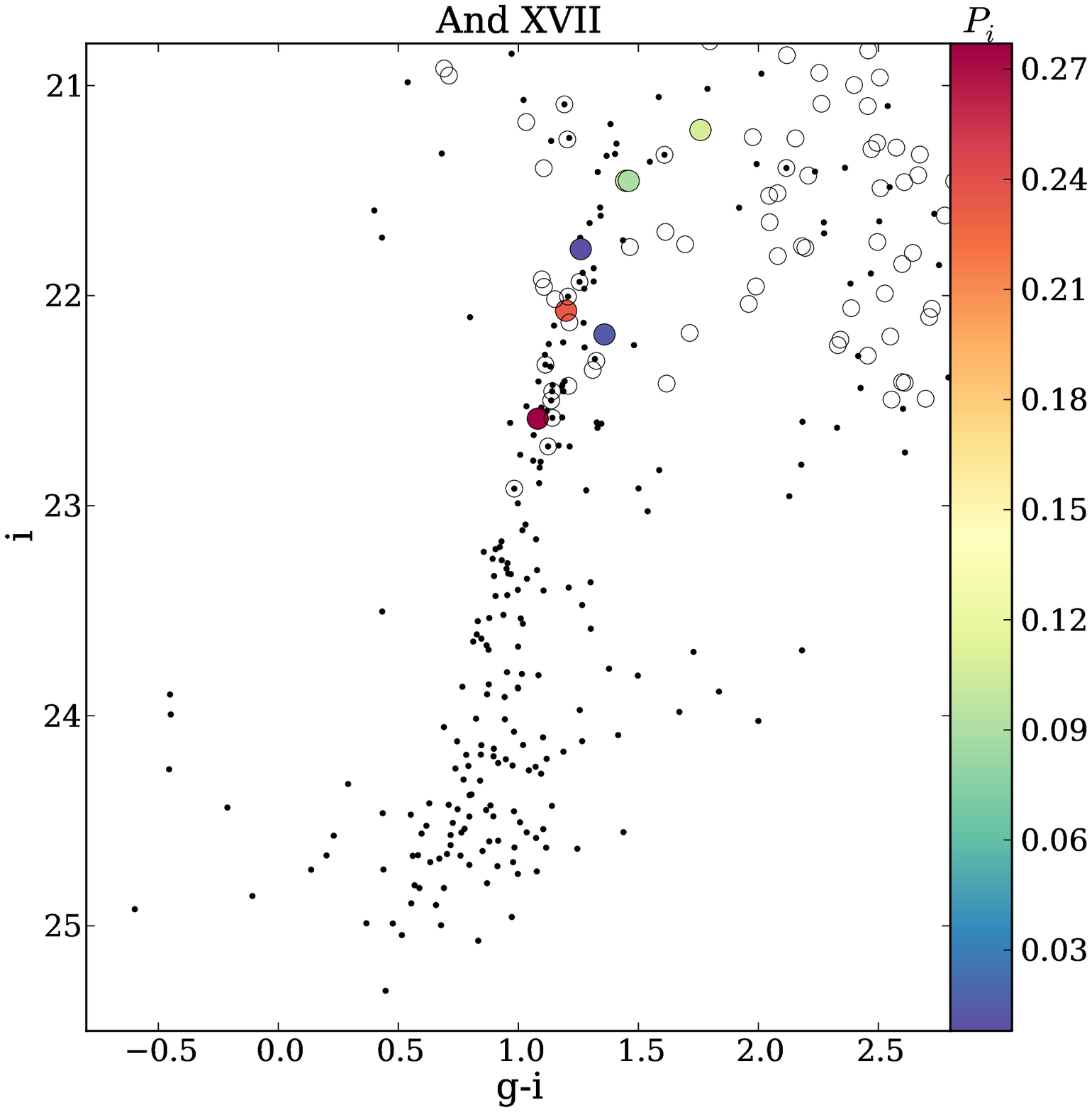}
\includegraphics[angle=0,width=0.45\hsize]{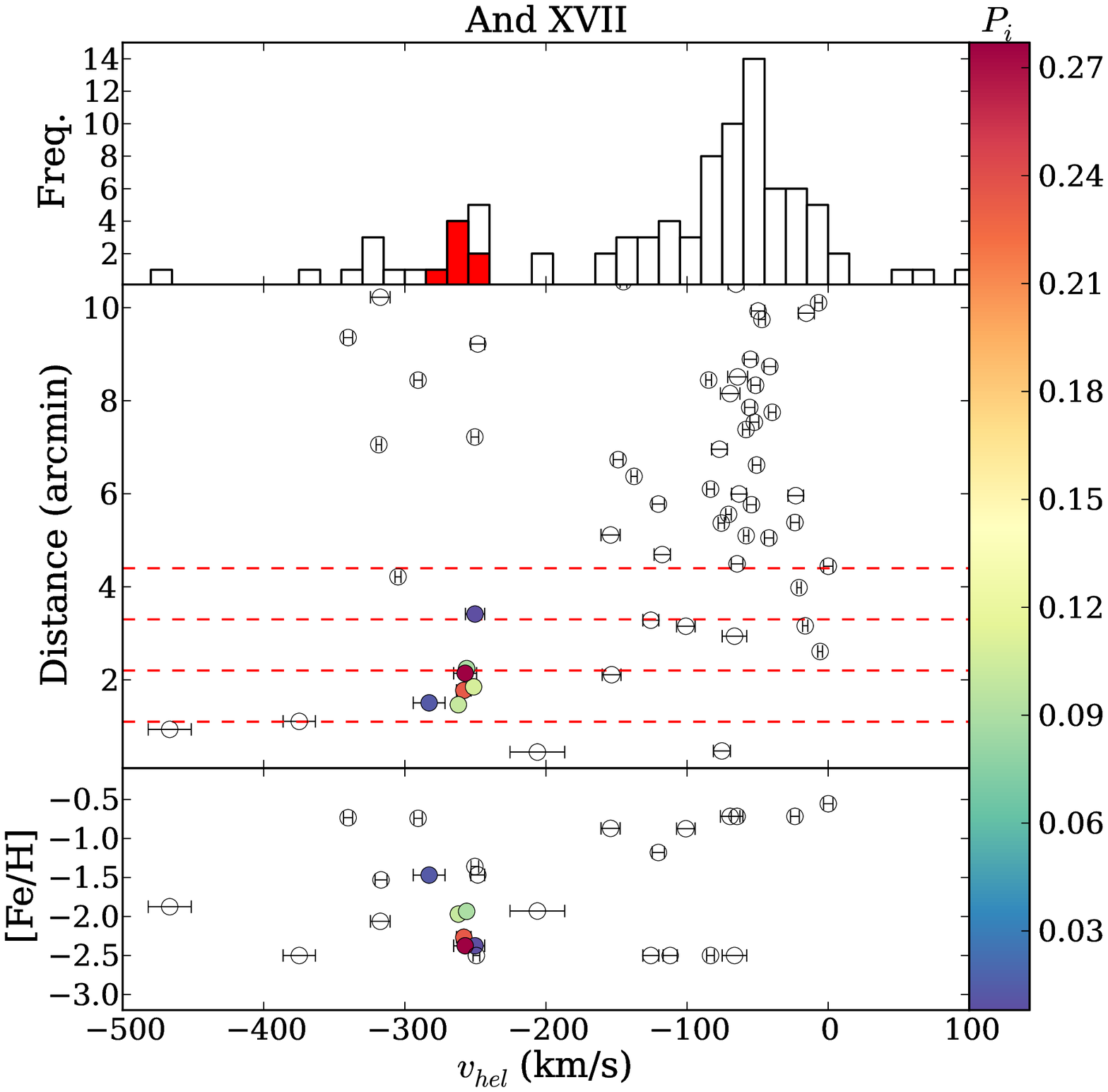}
\includegraphics[angle=0,width=0.9\hsize]{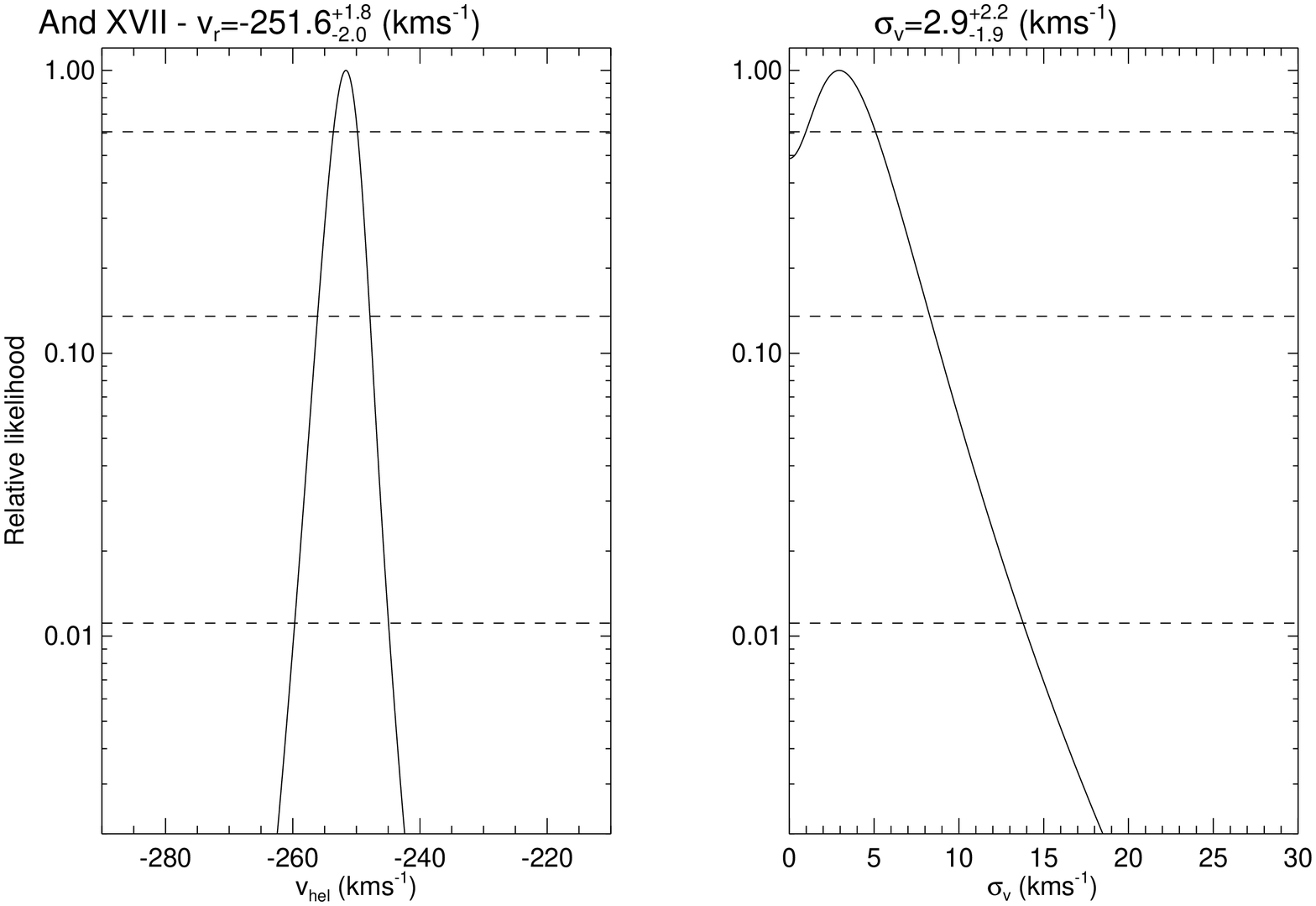}
\caption{CMD of And XVII, from PAndAS photometry (not extinction
   corrected). All stars observed with DEIMOS are color coded by their
   probability of membership. {\bf Top right: }Kinematic information for all
   stars observed with DEIMOS. The top subplot shows a velocity histogram for our
   sample, with all stars with $P_i>10^{-6}$ highlighted as a filled red
   histogram. The central subplot shows the distance of each star from the
   centre of And XVII as a function of velocity, and the lower subplot shows the
   photometric metallicity for each star, interpolated from \citet{dart08}
   isochrones in the CFHT-MegaCam $g-$ and $i-$bands (after correcting the
   observed colors of the stars for extinction using the \citet{schlegel98}
   dust maps), as a function of velocity. Stars that lie far from the RGB of the
   dSph are not well matched by the isochrones, and as a result there is no
   estimate of their [Fe/H].  {\bf Lower panels: }Resulting probability distributions
   from ML analysis of the kinematics of And XVII. The left hand panel shows
   the systemic velocity ($v_r$) likelihood distribution and the right shows
   the intrinsic velocity dispersion ($\sigma_v$) likelihood distribution. The
   dashed lines represent canonical $1, 2$ and $3\sigma$ confidence intervals,
   derived assuming Gaussian uncertainties.}
\label{fig:And17}
\end{center}
\end{figure*}

And XVII was discovered by \citet{irwin08}, and it is located at a projected
distance of $\sim40$~kpc to the North West of Andromeda. A detailed study of
deep imaging obtained with the Large Binocular Camera on the LBT was also
performed by \citet{brasseur11a}, and throughout, we use the structural
properties as determined from this work. It is a faint, compact galaxy
($M_V=-8.61, r_{\rm half}=1.24'$ or 262$^{+53}_{-46}$ pc). In the left panel
of Fig.~\ref{fig:And17} we display the PAndAS color magnitude diagram for And
XVII. Over-plotted we show the observed DEIMOS stars, color-coded by their
probabilities of membership. The open symbols represent stars for which
$P_i<10^{-6}$. We employ this cut solely to make clarify which stars have the
highest probability of belonging to And XVII. In the right hand panel, we
display the basic kinematic information for And XVII. In the top panel of this
subplot, we show a velocity histogram for all stars observed within the LRIS
mask, and stars with $P_{i}>10^{-6}$ are highlighted with a filled red
histogram. The centre panel shows the velocities as a function of distance
from the centre of And XVII (and the red dashed lines indicate $1,2,3$ and
$4\times r_{\rm half}$), and the lower panel shows the photometric
metallicities for all stars, as determined using \citet{dart08} CFHT
isochrones. Again, all points are color-coded by their probability of
membership. Finally, the two lower panels show the resulting, one dimensional,
probability weighted, marginalized maximum likelihood distributions for $v_r$
and $\sigma_v$ for this data set. From the kinematics presented in
Fig.~\ref{fig:And17}, which represent the first spectroscopic observations of
this object, we see the signature of the dwarf galaxy as a cold spike at
$v_r\sim-250\kms$. From the lower panels of this figure and the accompanying
CMD we see that there is a cluster of 7 stars sitting within this spike that
are centrally concentrated and are consistent with the RGB of the dwarf
itself, leading us to believe that our algorithm has cleanly detected the
signature of the galaxy. Interestingly, we also see 3 stars that,
kinematically, are indistinguishable from the stars that have been dubbed as
probable members in our analysis. However, they all sit at large distances
from the centre of And XVII, equivalent to greater than 6 times the half-light
radius of the dwarf, and hence the routine has classified them as likely
members of the M31 halo rather than And XVII members. But, given their tight
correlation in velocity to the systemic velocity of And XVII the possibility
exists that these are extra-tidal stars of And XVII. No sign of extra-tidal
features were cited in either the discovery paper of And XVII or the LBT
followup, but given its position in the north M31 halo where contamination
from the MW becomes increasingly problematic, and its relatively low
luminosity ($M_V=-8.5$), such features may be difficult to see within the
imaging. However, at present they are considered unlikely members by our
routine, and do not factor into our calculation of global kinematic properties
for this object. We find $v_r=-251.6^{+1.8}_{-2.0}\kms$ and
$\sigma_v=2.9^{+2.2}_{-1.9}\kms$.

\subsection{Andromeda XVIII}
\label{sect:and18}

\begin{figure*}
\begin{center}
\includegraphics[angle=0,width=0.45\hsize]{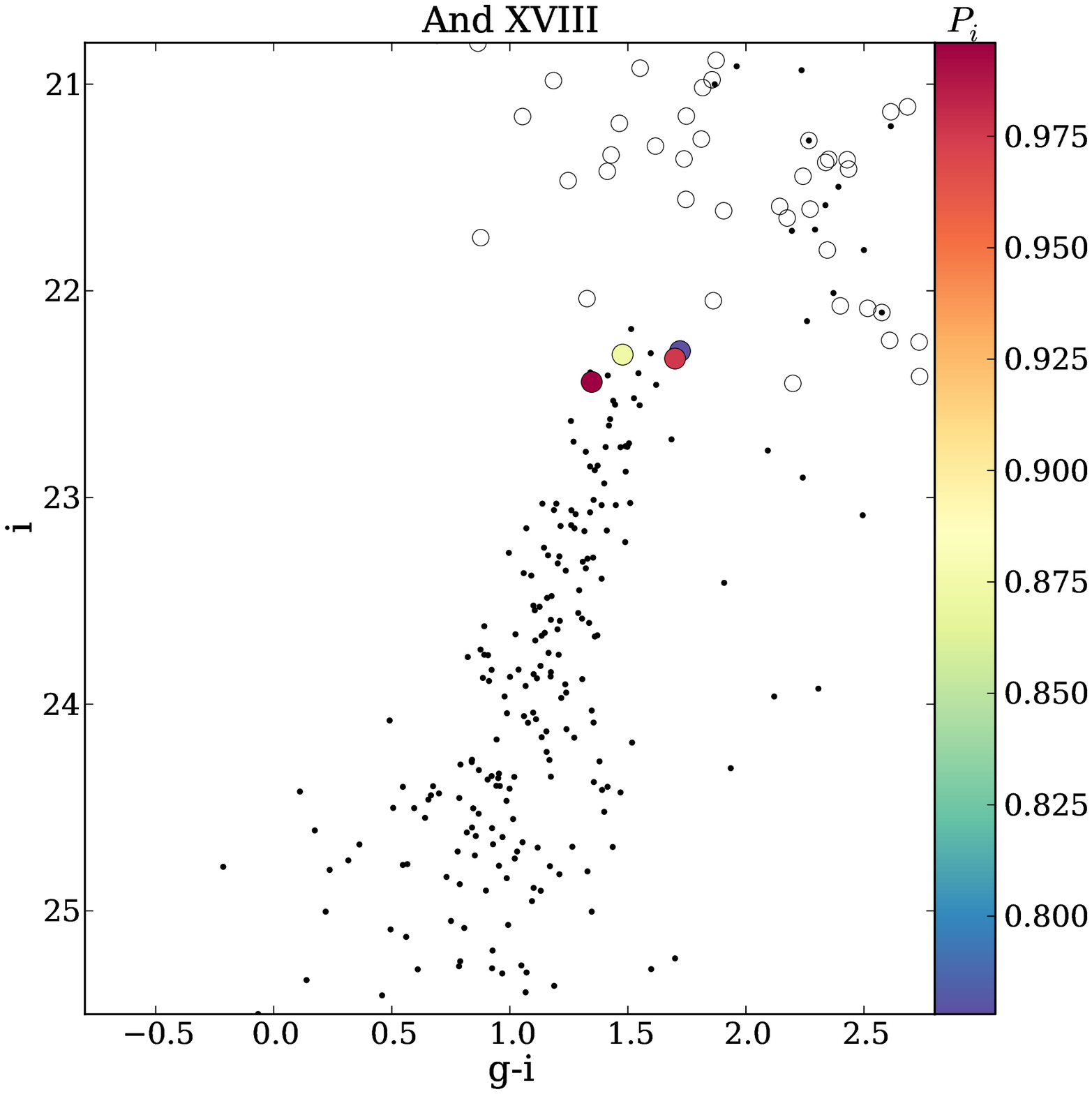}
\includegraphics[angle=0,width=0.45\hsize]{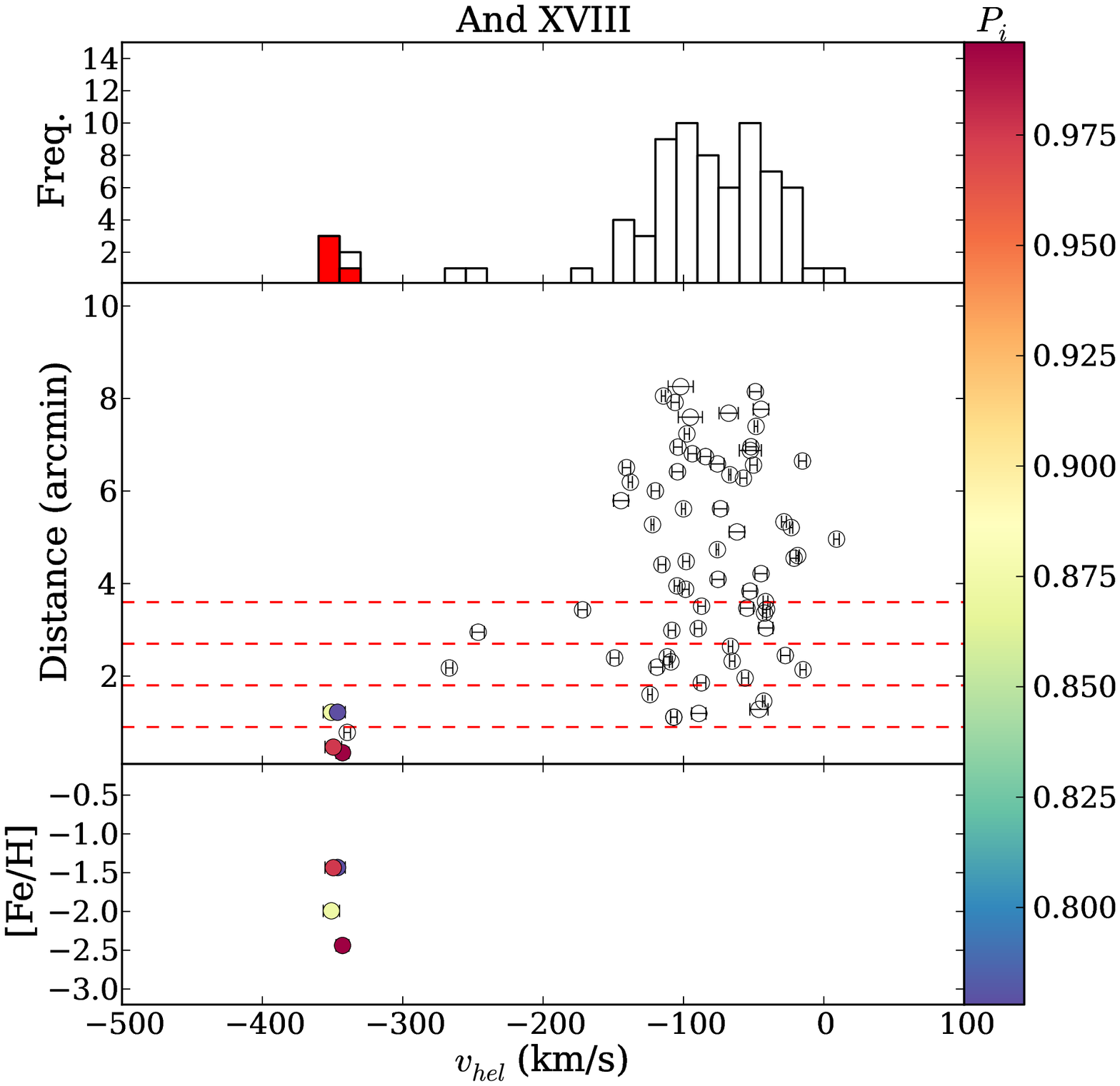}
\includegraphics[angle=0,width=0.9\hsize]{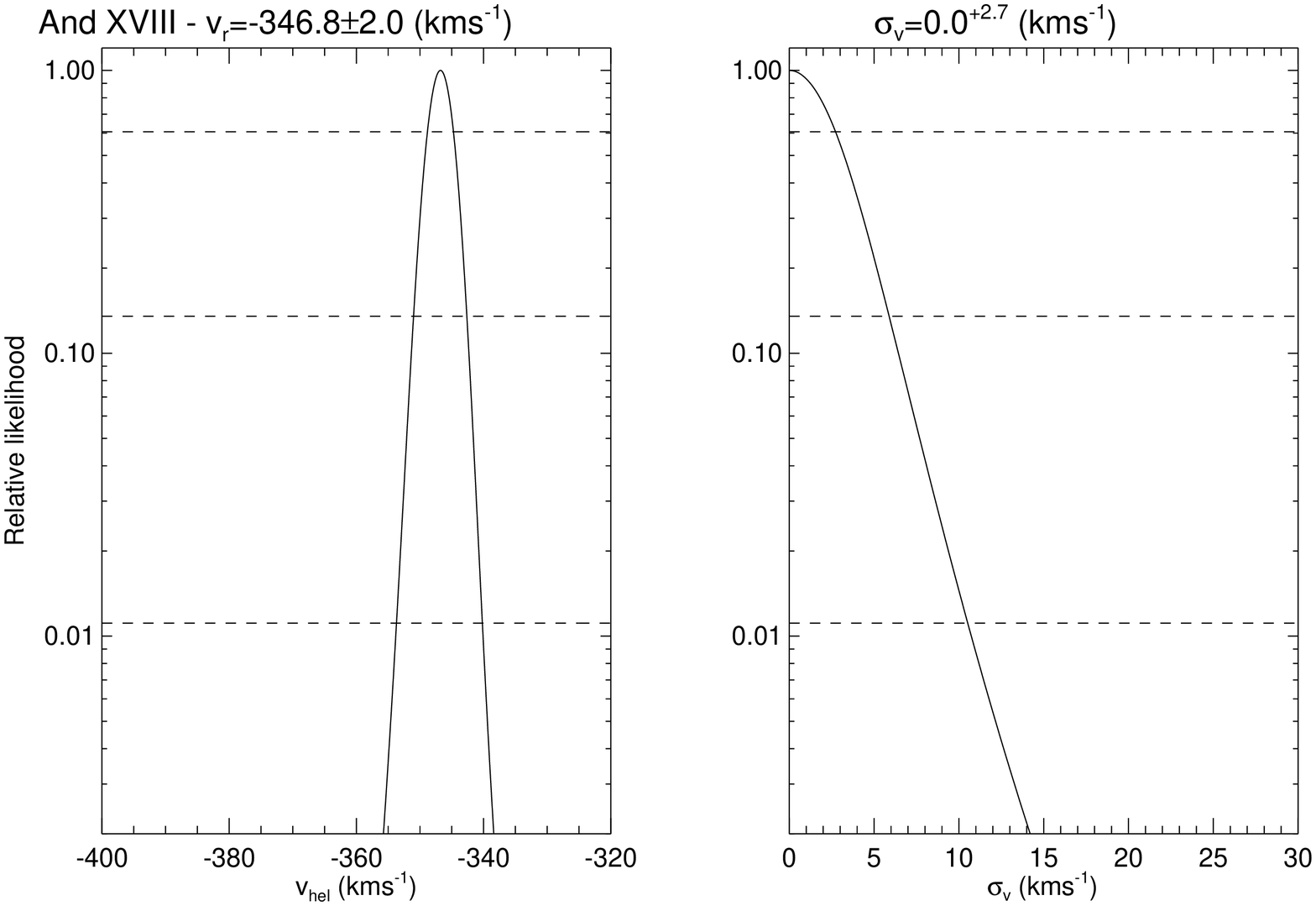}
\caption{As Fig.~\ref{fig:And17}, but for And XVIII.}
\label{fig:And18}
\end{center}
\end{figure*}

Andromeda XVIII (And XVIII) was detected by \citet{mcconnachie08} in the
PAndAS CFHT maps. Located at a projected distance of $\sim110$~kpc to the
North-West of M31, it is one of the most distant of its satellites, sitting
$\sim600$~kpc behind the galaxy, making spectroscopic observations of its
individual RGB stars taxing, as they are all relatively faint
($i\gta22.2$). Thus, from our 1 hour DEIMOS observation, we were only able to
confirm 4 stars as probable members (see Fig.~\ref{fig:And18}). We determine
the global systemic velocity to be $v_r=-346.8\pm2.0\kms$, and we are unable
to resolve a velocity dispersion, finding $\sigma_v=0.0^{+2.7}\kms$ where the
upper bound is determined from the formal $1\sigma$ confidence interval
produced by our maximum likelihood analysis. This suggests that the 4 stars we
are able to confirm as members do not adequately sample the underlying
velocity profile. The systemic velocity we measure is different to that
presented in T12 of $v_r=-332.1\pm2.7$ at a level of 3.4$\sigma$. Our
1$\sigma$ limit of 2.7$\kms$ is also at odds with the dispersion determined by
T12 ($\sigma_v=9.7\pm2.3$). They were able to measure velocities for
significantly more probable member stars (22 vs. 4) owing to their longer
integration of 3 hours.  The faintness of our targets and shorter exposure
time could mean that strong sky absorption lines have systematically skewed
our velocity measurements for these stars, and this could explain the
discrepancy of our measurements with respect to those of T12. To check against
this, we again perform weighted cross-correlations to each of the 3 Ca II
lines individually. We find that the values we obtain, and their average, are
fully consistent with that derived from the technique described in
\S~\ref{sect:specobs}, differing by less than $3\kms$ from those values. The
true systemic velocity of And XVIII therefore remains unclear. However, given
their larger sample size, the T12 systemic properties are more statistically
robust than those we present here.

\subsection{Andromeda XIX}
\label{sect:and19}

\begin{figure*}
\begin{center}
\includegraphics[angle=0,width=0.45\hsize]{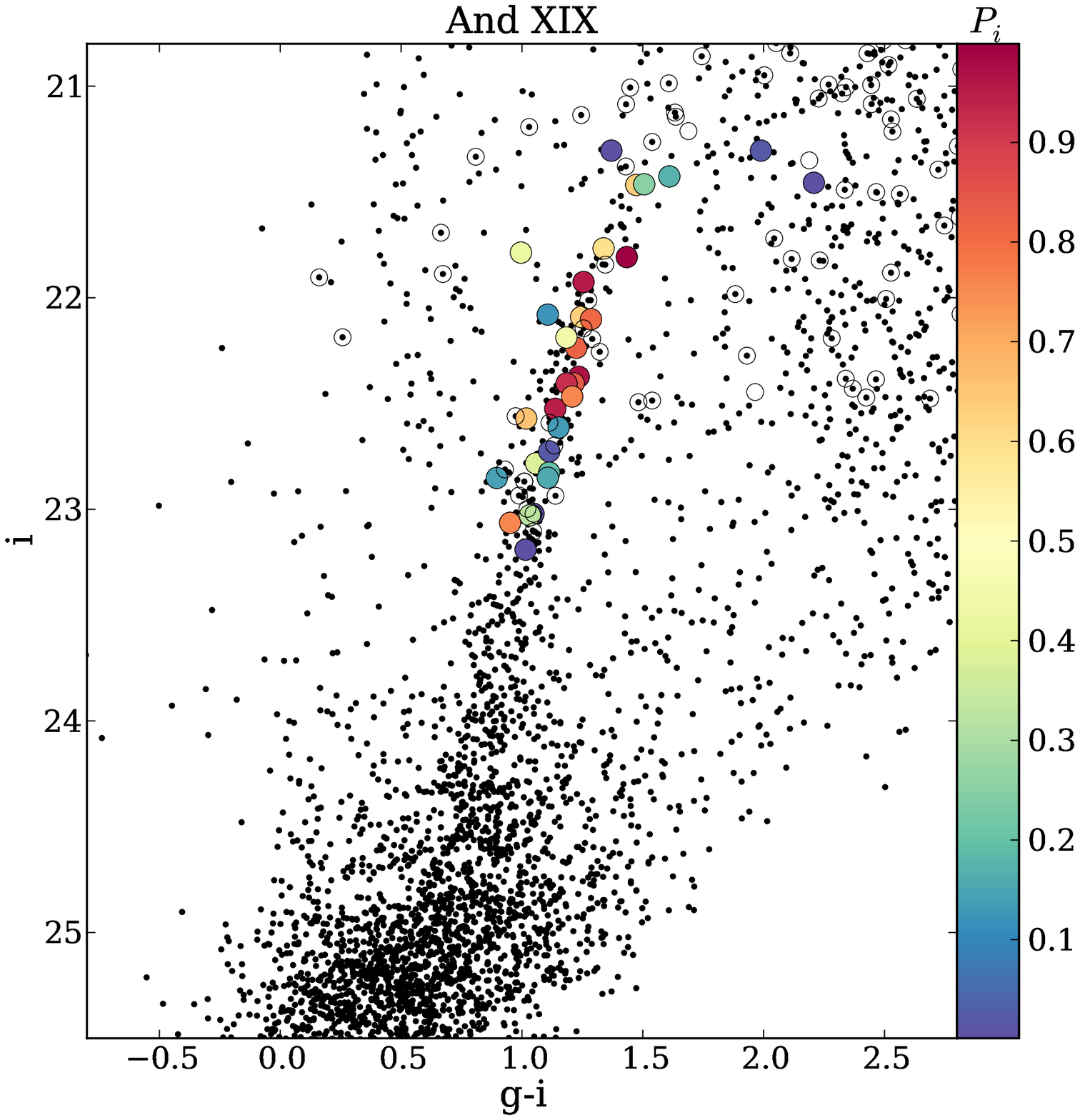}
\includegraphics[angle=0,width=0.45\hsize]{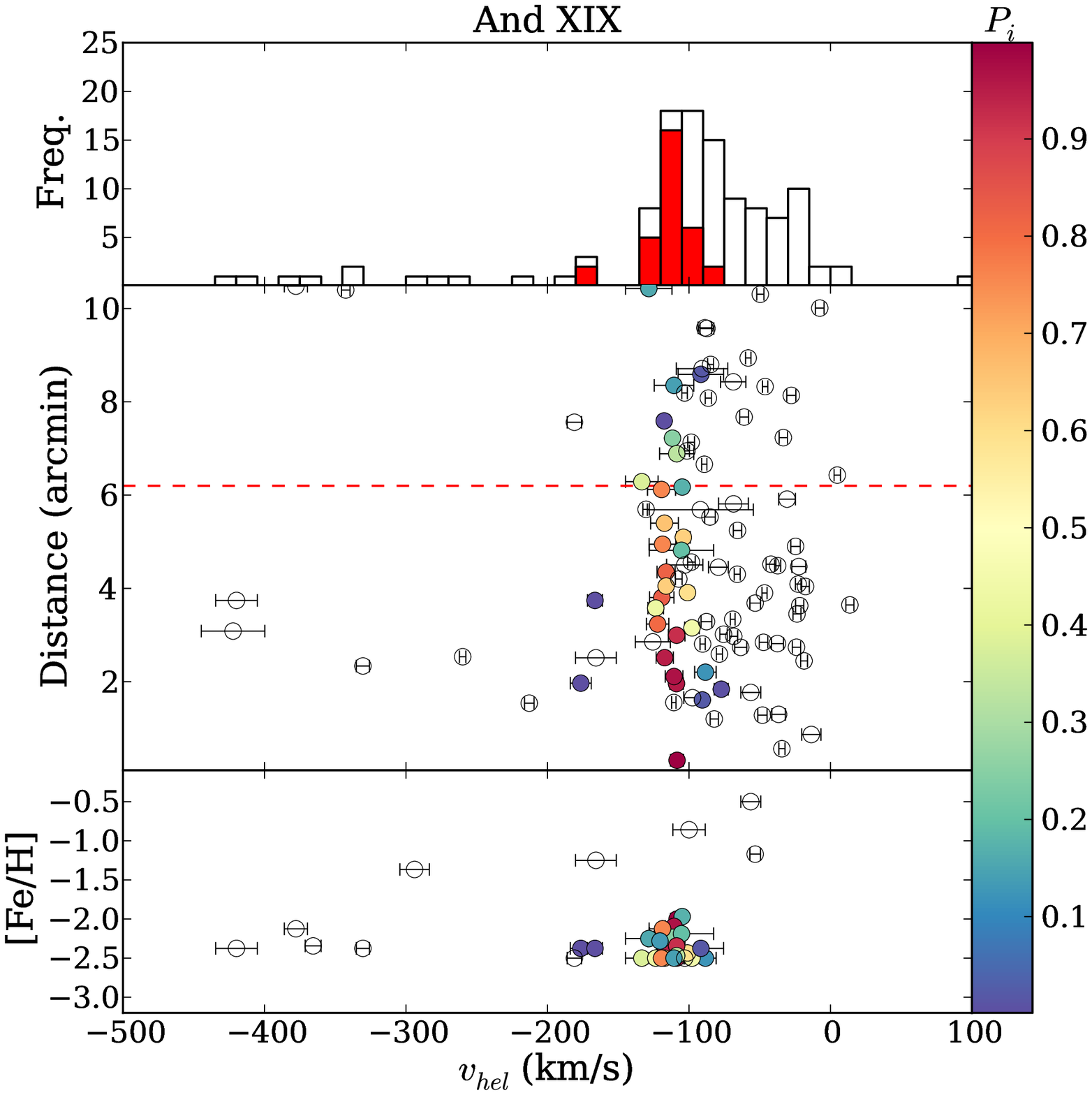}
\includegraphics[angle=0,width=0.9\hsize]{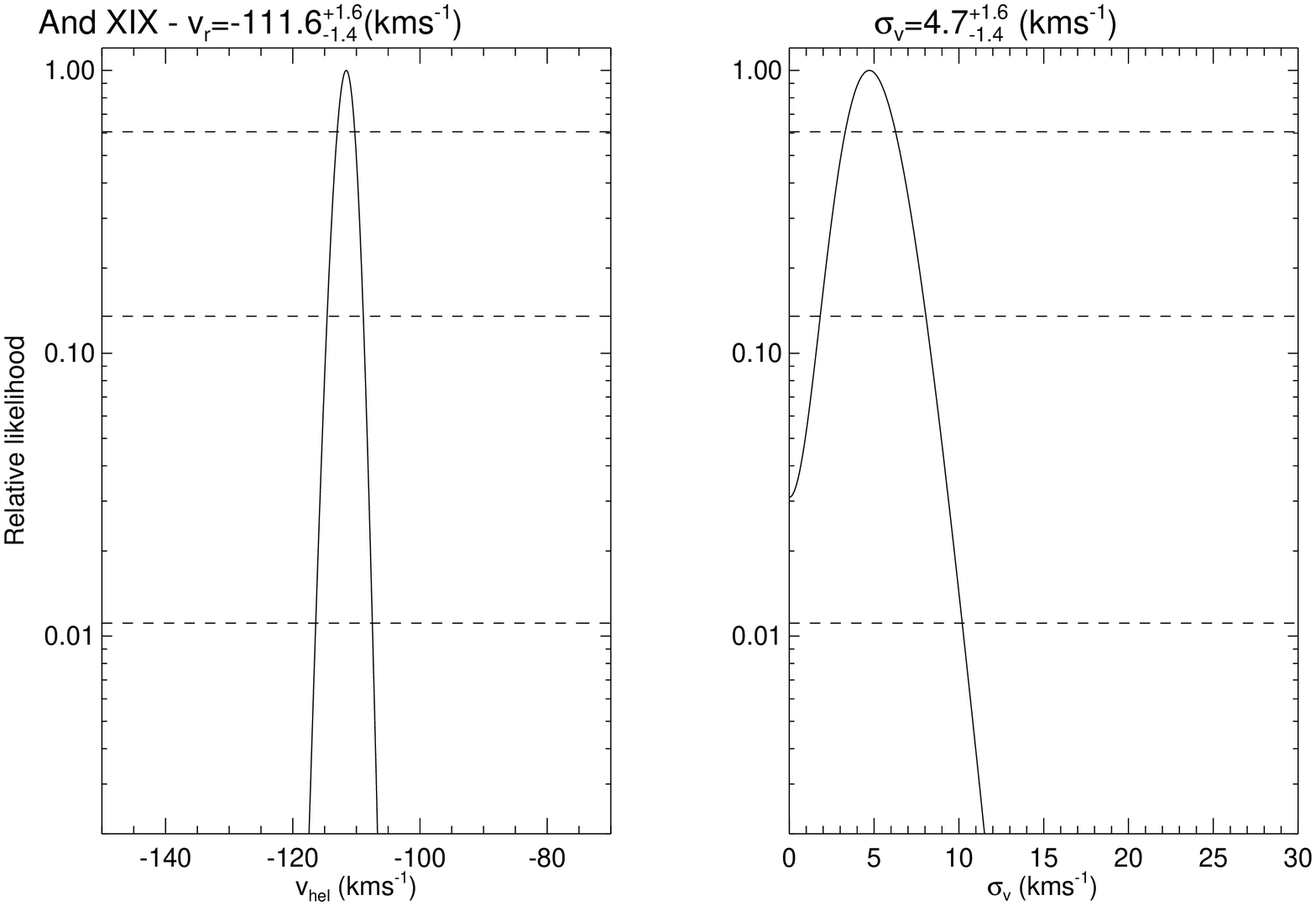}
\caption{As Fig.~\ref{fig:And17}, but for And XIX.}
\label{fig:And19}
\end{center}
\end{figure*}

Andromeda XIX (And XIX) was first reported in \citet{mcconnachie08}, and is a
relatively bright, very extended ($M_V=-9.3$, $r_{\rm half}=1.5$~kpc) dSph,
located at a projected distance of $\sim180$~kpc to the south west of M31. Its
unusual morphology, very low surface brightness $\Sigma_v=30.2$mag/arcsec$^2$,
and evidence in the photometry for a possible link to the major axis
substructure reported in \citet{ibata07} caused \citet{mcconnachie08} to
question whether And XIX was truly a dynamically relaxed system, or whether it
had experienced a significant tidal interaction. Here, we present the first
spectroscopic observations of the And XIX satellite in Fig.~\ref{fig:And19}
from two DEIMOS masks placed at different position angles. These data allow us
to comment on its dark matter content, and on the likelihood of a tidal origin
for its unique structure. We identify 27 stars where $P_i>10^{-6}$ within the
system. These measurements were made increasingly challenging as the systemic
velocity we measure is $v_r=-111.6^{+1.6}_{-1.4}\kms$, placing it within the
regime of Galactic contamination. However, we are confident that our algorithm
is robust to this unfortunate location of And XIX in velocity space (see
discussion in \S~\ref{sect:test} and Appendix A). As a further check that none
of the stars we define as probable members are actually foreground
contaminants, we measure the equivalent widths of the Na I doublet lines
(located at $\sim8100$\AA). These gravity-sensitive absorption lines are
typically significantly stronger in foreground dwarf stars than M31 RGB stars,
although there is some overlap between the two populations. For the stars
tagged as probable members by our algorithm, we find no evidence of strong
absorption in the region of the Na I doublet, indicating that we are not
selecting foreground stars as members. We measure a relatively cold velocity
dispersion for this object of $\sigma_v=4.7^{+1.6}_{-1.4}\kms$, which is
surprising given the radial extent of this galaxy. This result will be
discussed further in \S~\ref{sect:mass}, and in a follow-up paper (Collins et
al. in prep).

\subsection{Andromeda XX}
\label{sect:and20}

\begin{figure*}
\begin{center}
\includegraphics[angle=0,width=0.45\hsize]{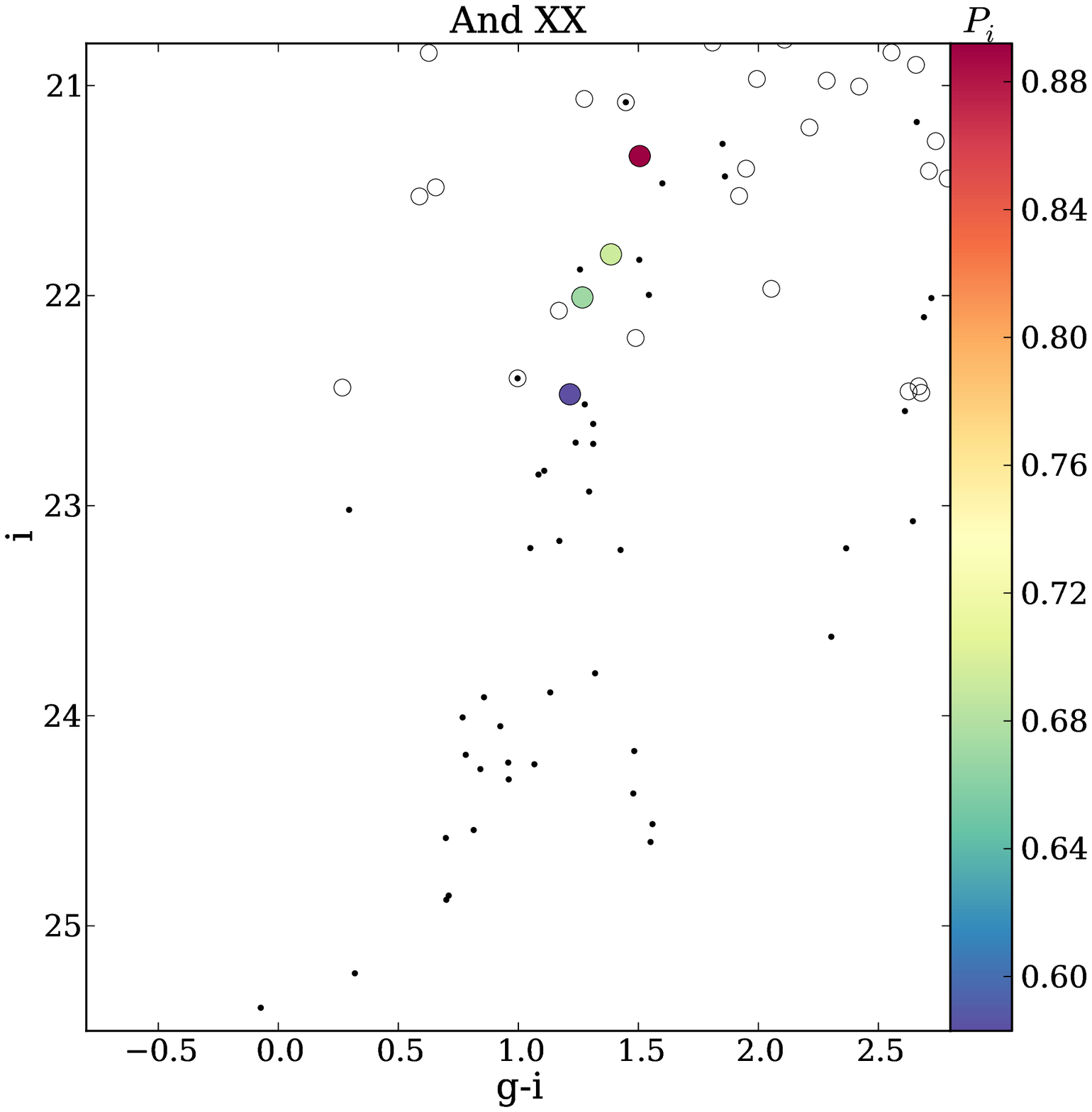}
\includegraphics[angle=0,width=0.45\hsize]{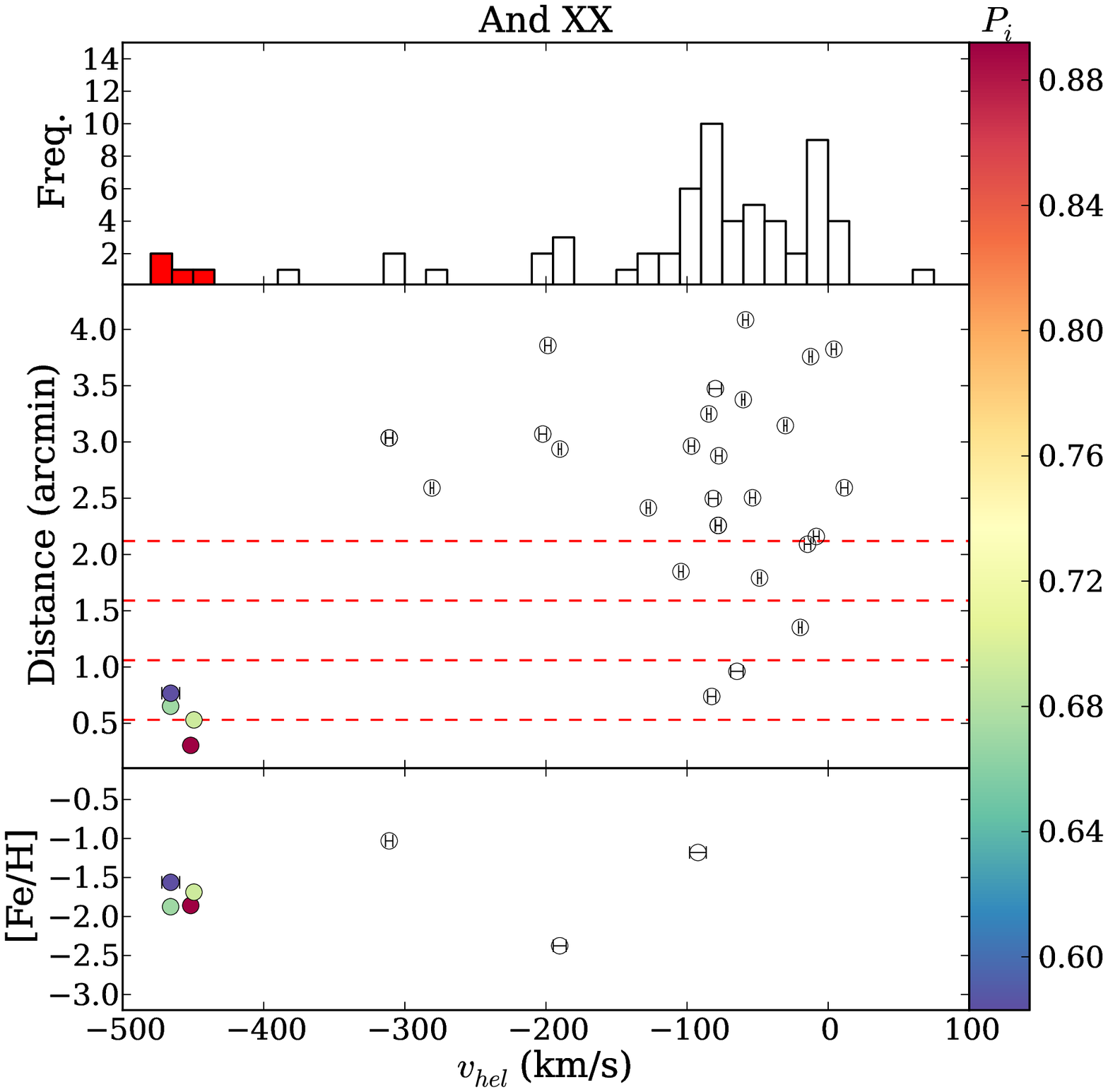}
\includegraphics[angle=0,width=0.9\hsize]{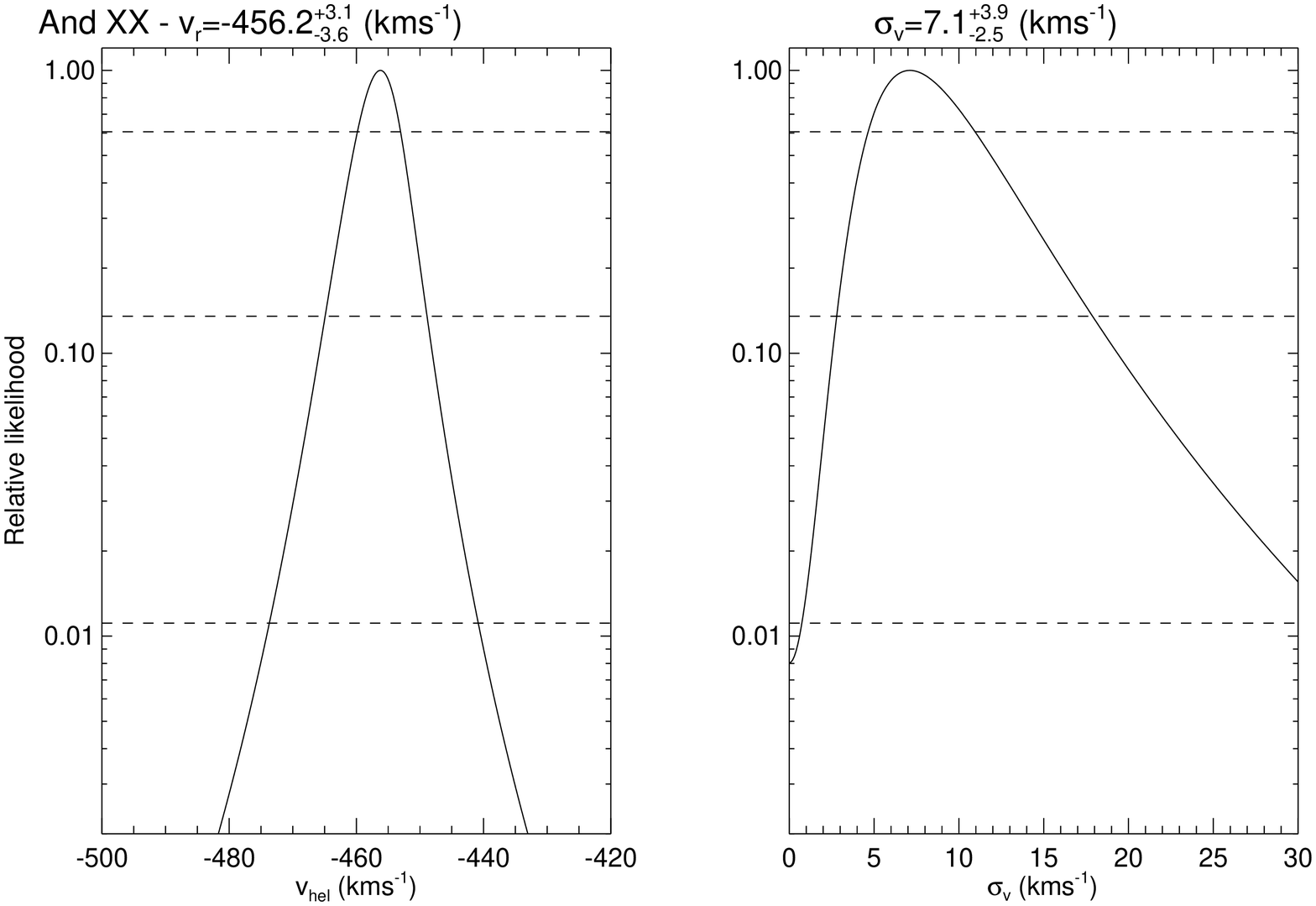}
\caption{As Fig.~\ref{fig:And17}, but for And XX.}
\label{fig:And20}
\end{center}
\end{figure*}

And XX was the third of three dSphs discovered by \citet{mcconnachie08}, and
is notable for being one of the faintest dSph companions detected surrounding
Andromeda thus far. With $M_V=-6.3$ and $r_{\rm half}=114^{+31}_{-12}$~pc, it
is a challenging object to study spectroscopically as there are very few stars
available to target on its RGB, as shown in the top left subplot in
Fig.~\ref{fig:And20}. As a result, our algorithm is only able to find 4 stars
for which $P_i>10^{-6}$. These are found to cluster around
$v_r=-456.2^{+3.1}_{-3.6}\kms$, with a dispersion of
$\sigma_v=7.1^{+3.9}_{-2.5}\kms$. Despite the low number of stars, we are
confident in this detection, as the systemic velocity places it in the outer
wings of the velocity profile of the M31 halo. And XX is also located at a
large projected distance from M31 of $\sim130$~kpc, where we expect the
density of the M31 halo to be very low. As such, seeing 4 halo stars so
tightly correlated in velocity in the wings of the halo velocity profile
within such a small area of the sky (all stars are within 1 arcmin of the
centre of And XX) is highly unlikely. We caution the reader that, while we are
confident that our algorithm is able to measure velocity dispersions for
sample sizes as small as 4 stars, as we are not probing the full velocity
profile of this object this measurement ideally needs to be confirmed with
larger numbers of member stars.

\subsection{Andromeda XXI}
\label{sect:and21}

\begin{figure*}
\begin{center}
\includegraphics[angle=0,width=0.45\hsize]{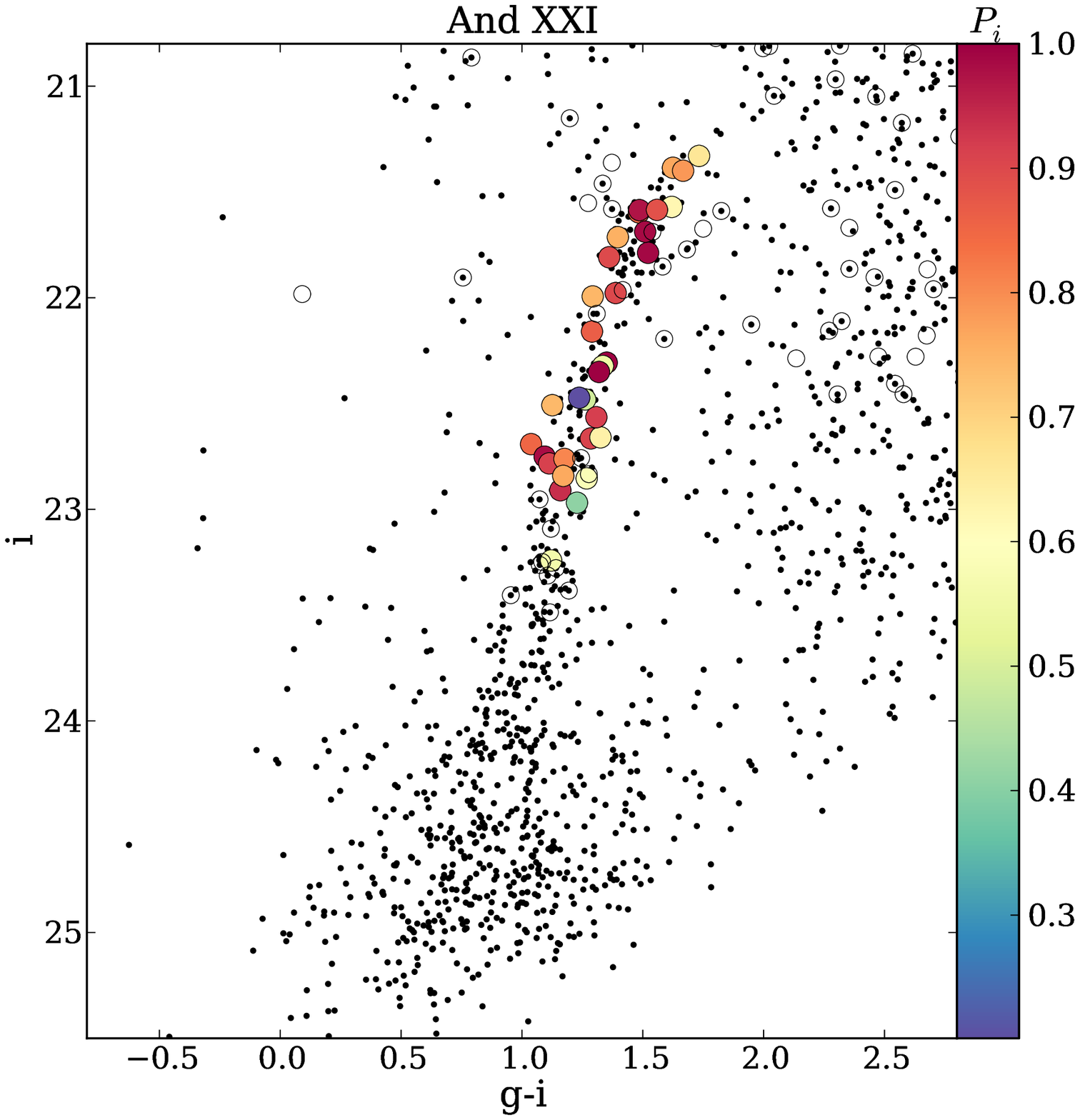}
\includegraphics[angle=0,width=0.45\hsize]{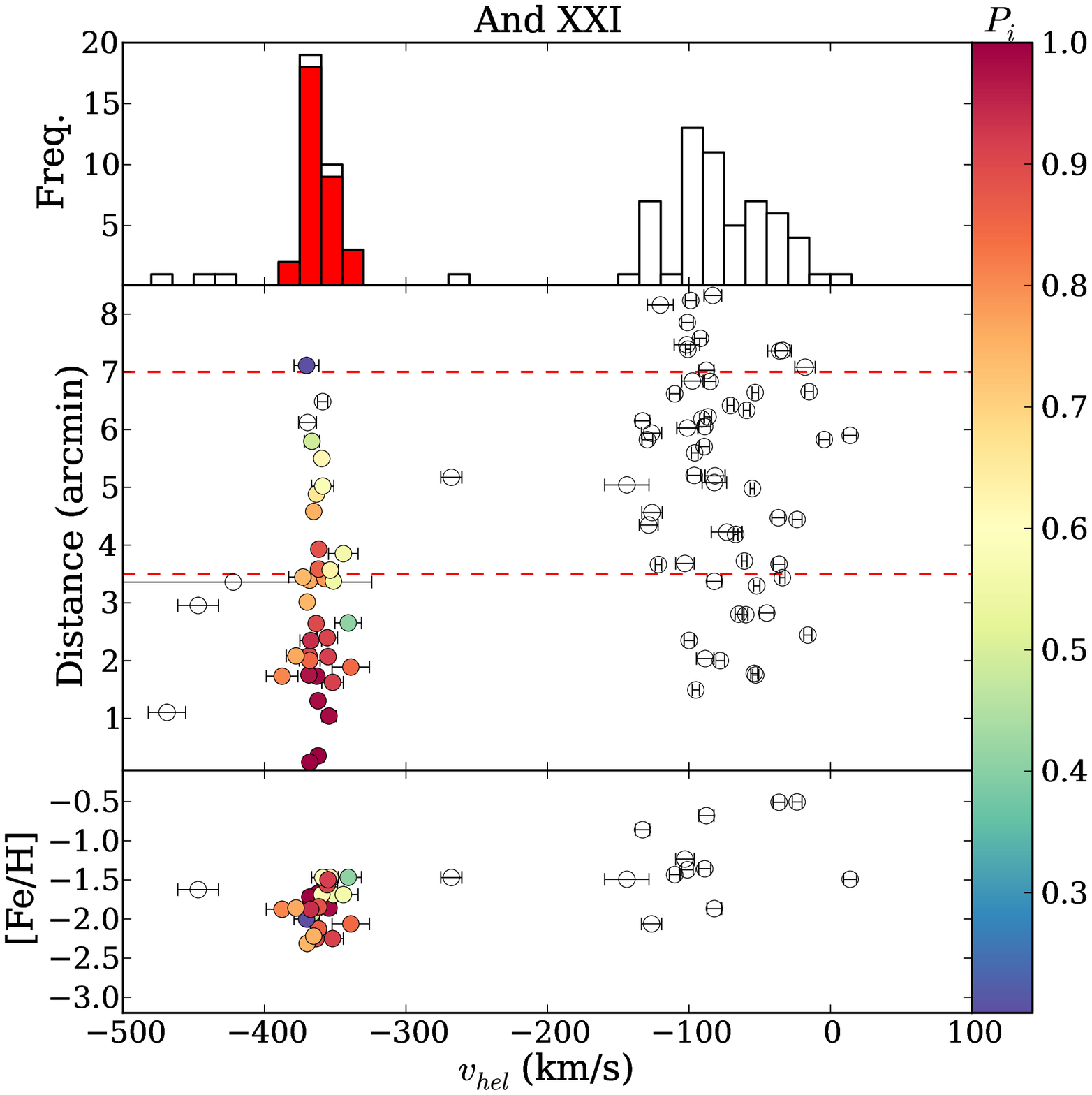}
\includegraphics[angle=0,width=0.9\hsize]{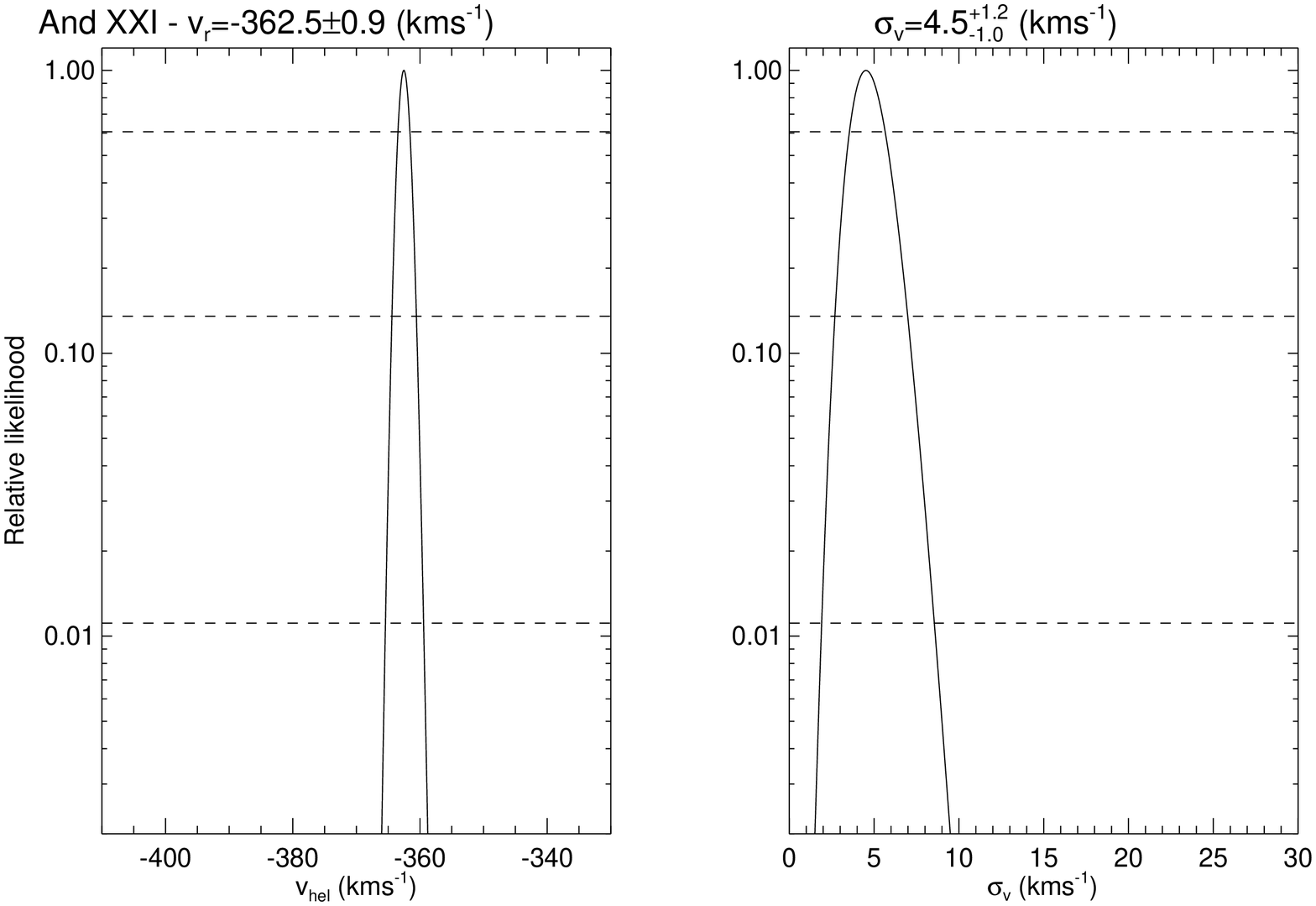}
\caption{As Fig.~\ref{fig:And17}, but for And XXI}
\label{fig:And21}
\end{center}
\end{figure*}

Andromeda XXI (And XXI) was identified within the PAndAS imaging maps by
\citet{martin09}. It is a relatively bright dSph ($M_V=-9.9$), located at a
projected distance of $\sim150$~kpc from M31, and it has a half-light radius
of $r_{\rm half}=842\pm77$~pc. We present our spectroscopic observations for
this object in Fig.~\ref{fig:And21}, and in the top right subplot, we can
clearly see the signature of And XXI as a cold spike in velocity with 32
probable member stars, located at $v_r=-362.5\pm0.9\kms$, with a curiously low
velocity dispersion of only $\sigma_v=4.5^{+1.2}_{-1.0}\kms$. These results are
completely consistent with those of T12, where they measured
$v_r=-361.4\pm5.8\kms$ and $\sigma_v=7.2\pm5.5\kms$. As their sample contained
only 6 likely members compared with the 29 we identify here, ours constitute a more
statistically robust measurement of the global kinematics for this object than
those presented in T12.

\subsection{Andromeda XXII}
\label{sect:and22}

Andromeda XXII (And XXII) was identified within the PAndAS imaging maps by
\citet{martin09}, and is a relatively faint dSph, with $M_V=-6.5$. Its
physical position in the halo, located at a distance of 224 kpc in projection
from M31, but only 42 kpc in projection from M33, led the authors to postulate
that it could be the first known dSph satellite of M33. Subsequent work
analysing the kinematics of And XXII by T12 measured a systemic velocity for
And XXII of $-126.8\pm3.1\kms$ from 7 stars, more compatible with the systemic velocity of
M33 ($-178\kms$, \citealt{mateo98}) than that of M31. Another study by
\citet{chapman12} using the same data and the same method we present here
concluded the same, measuring a systemic velocity for the satellite of
$-129.8\pm2.0$ from 12 probable member stars, consistent with the T12
value. \citet{chapman12} also compare the position and kinematics of And XXII
with a suite of $N-$body simulations of the M31-M33 system, concluding that
And XXII was a probable M33 satellite.

\begin{figure*}
\begin{center}
\includegraphics[angle=0,width=0.45\hsize]{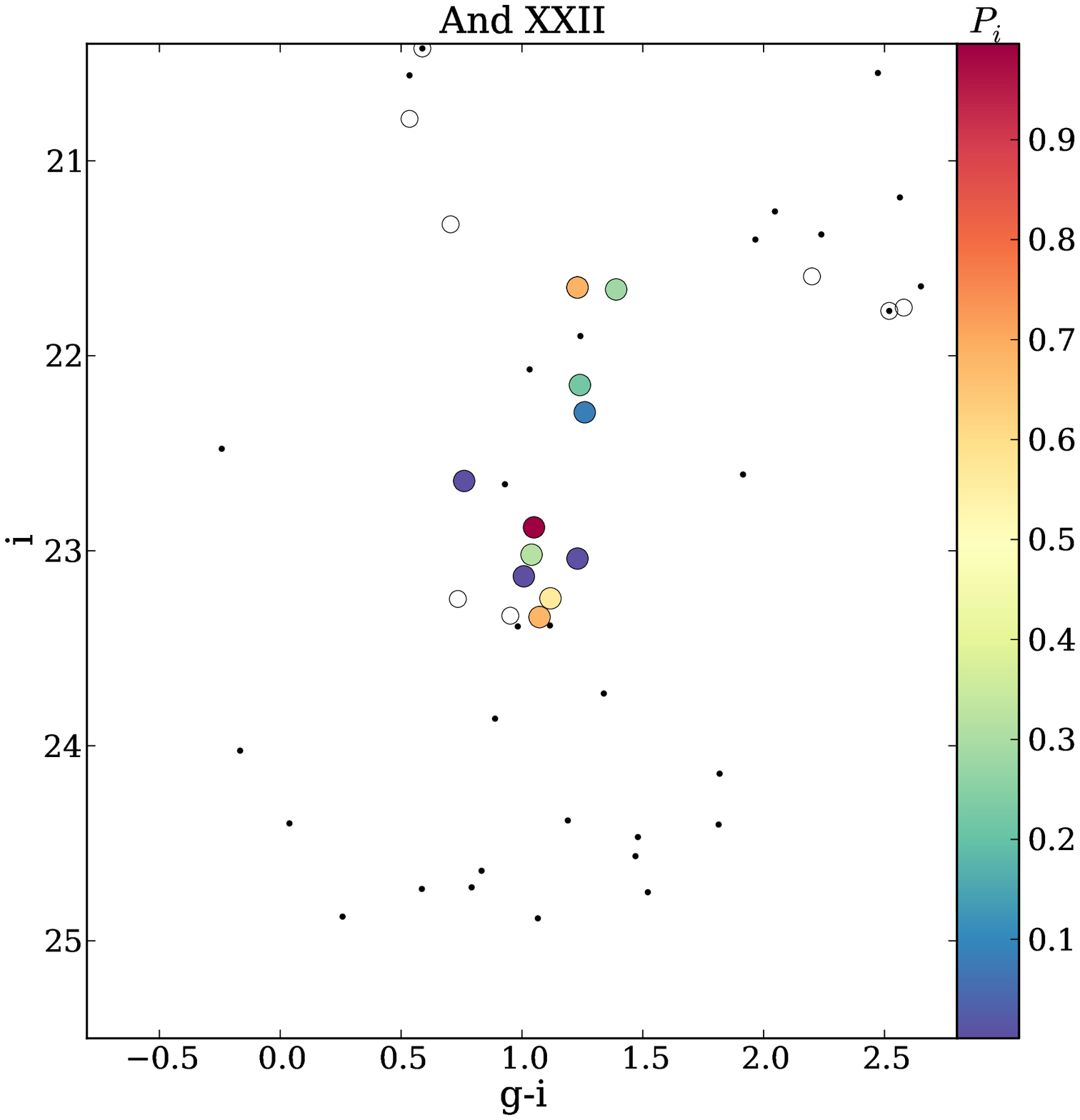}
\includegraphics[angle=0,width=0.45\hsize]{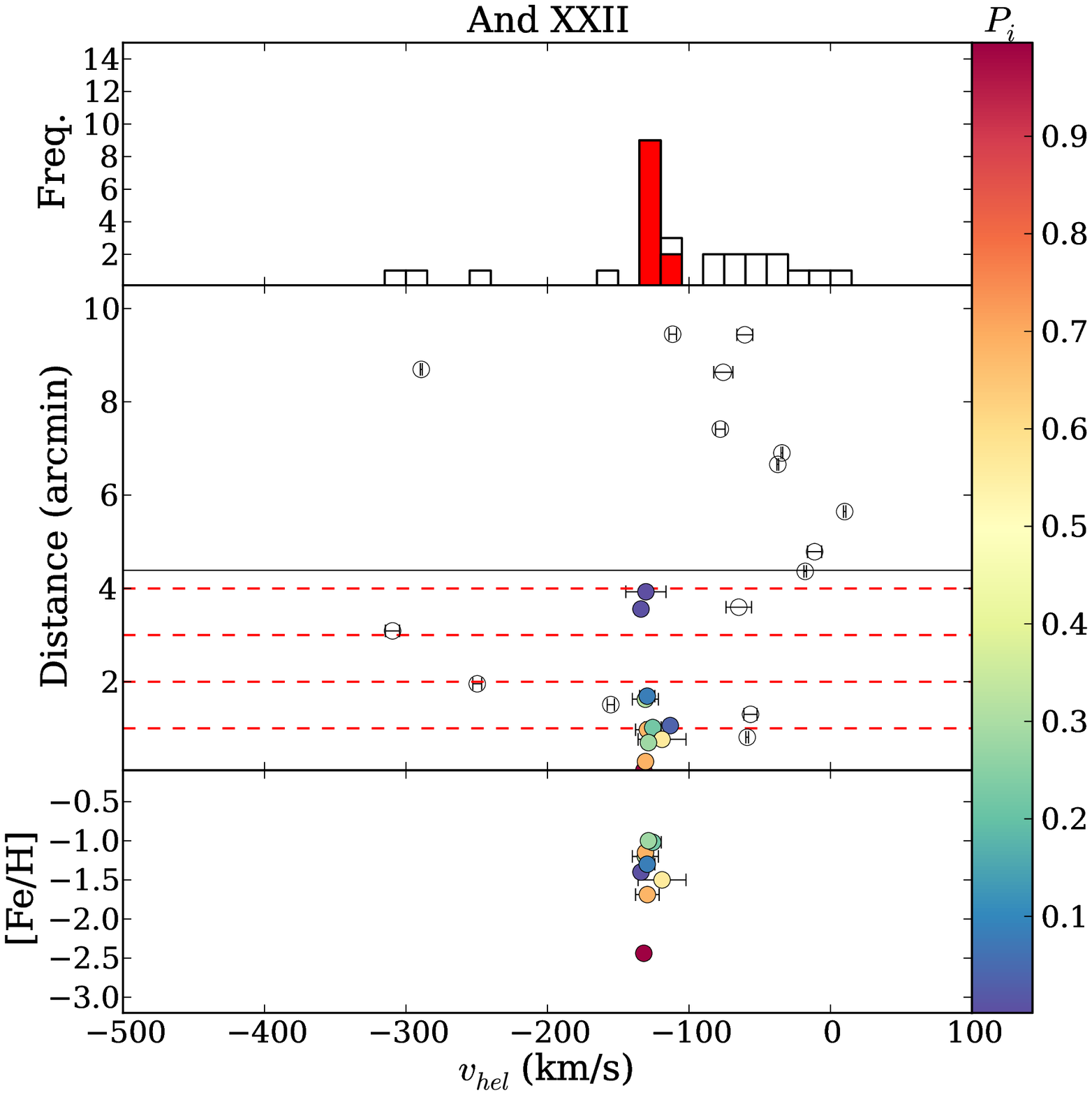}
\includegraphics[angle=0,width=0.9\hsize]{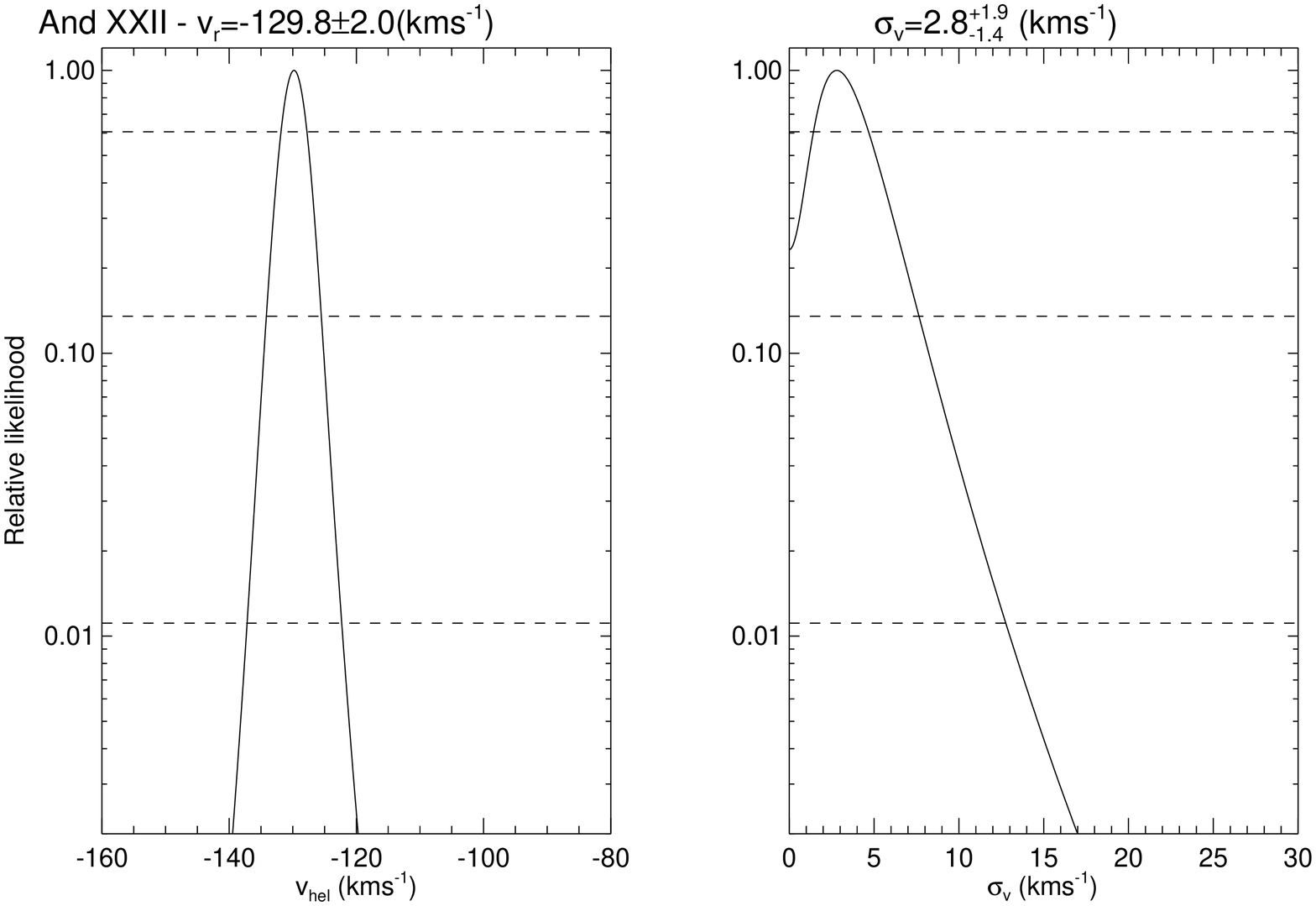}
\caption{As Fig.~\ref{fig:And17}, but for And XXII}
\label{fig:And22}
\end{center}
\end{figure*}

In Fig.~\ref{fig:And22}, we present the same data as analyzed by
\citet{chapman12} for completeness. The velocity dispersion of And XXII is
just resolved at $\sigma_v=2.8^{+1.9}_{-1.4}\kms$, completely consistent with
the value of $\sigma_v=3.5^{+4.2}_{-2.5}\kms$ from T12. As our values are
calculated from a 50\% greater sample size, we posit that they are the more
statistically robust.

\subsection{Andromeda XXIII}
\label{sect:and23}

\begin{figure*}
\begin{center}
\includegraphics[angle=0,width=0.45\hsize]{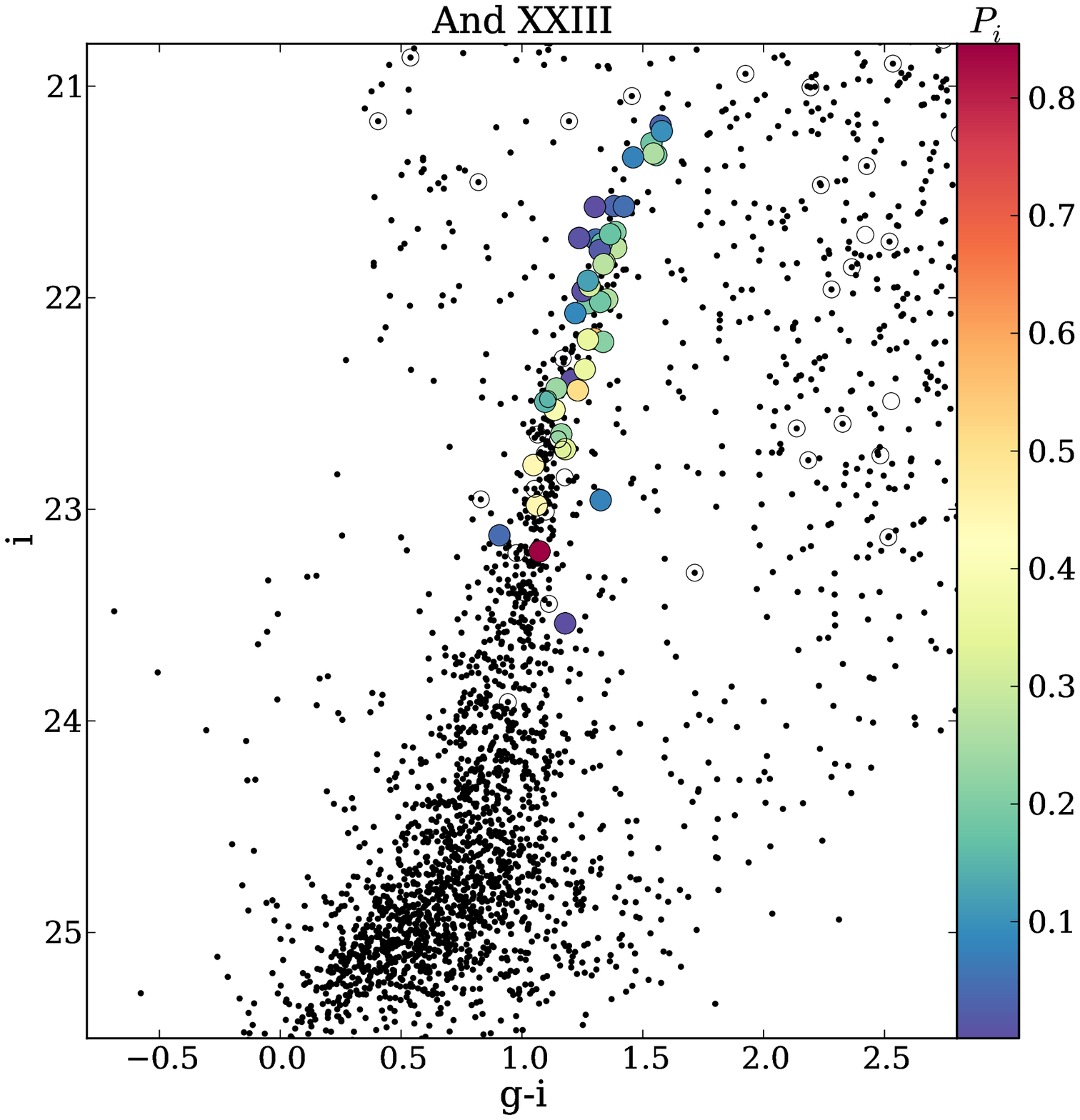}
\includegraphics[angle=0,width=0.45\hsize]{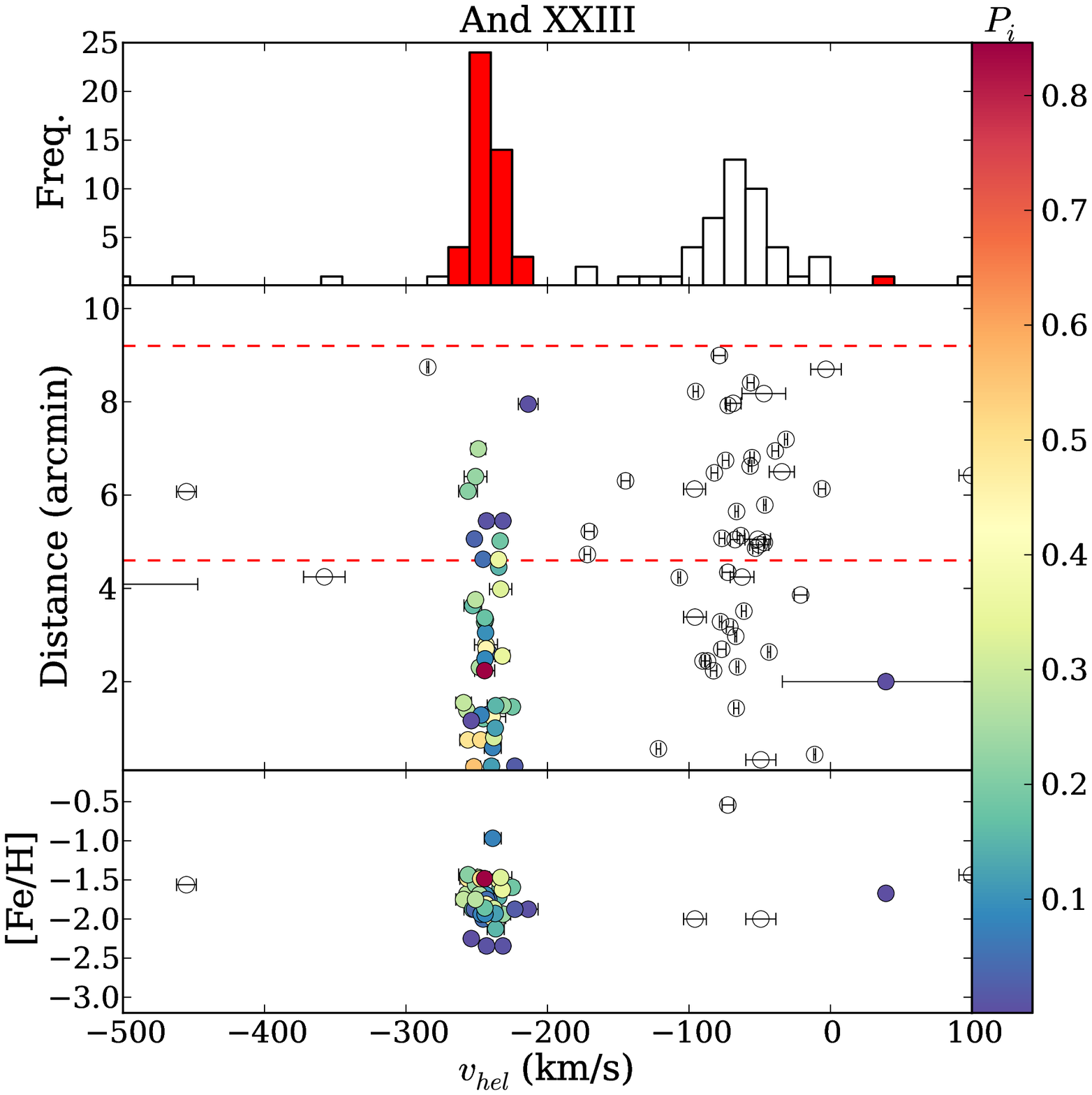}
\includegraphics[angle=0,width=0.9\hsize]{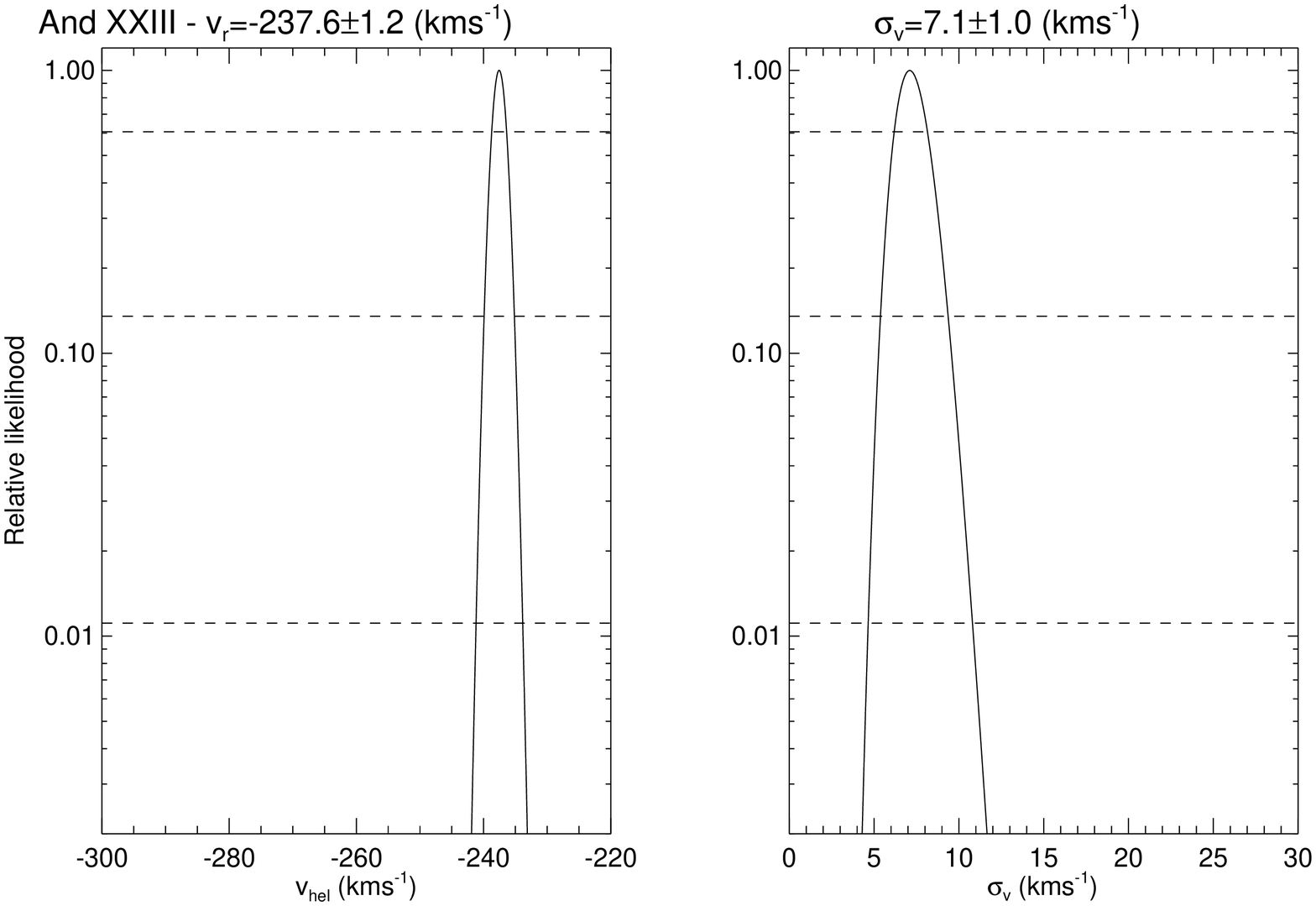}
\caption{As Fig.~\ref{fig:And17}, but for And XXIII.}
\label{fig:And23}
\end{center}
\end{figure*}

Andromeda XXIII (And XXIII) was the first of five M31 dSphs identified by
\citet{richardson11}. Located at a projected distance of $\sim130$~kpc to the
east of Andromeda, it is relatively bright, with $M_V=-10.2$, and extended,
with $r_{\rm half}=1001^{+53}_{-52}$~pc. Our routine clearly
detects a strong cold kinematic peak for And XXIII located around
$-230\kms$ and calculates a systemic velocity
of $v_r=-237.7\pm1.2\kms$, and a velocity dispersion of
$\sigma_v=7.1\pm1.0\kms$ from 40 probable member stars, as show in
Fig.~\ref{fig:And23}. This small, positive velocity relative to M31, combined
with its large projected distance from the host suggests that And XXIII is not
far past the apocentre of its orbit, heading back towards M31. 

\subsection{Andromeda XXIV}
\label{sect:and24}

\begin{figure*}
\begin{center}
\includegraphics[angle=0,width=0.45\hsize]{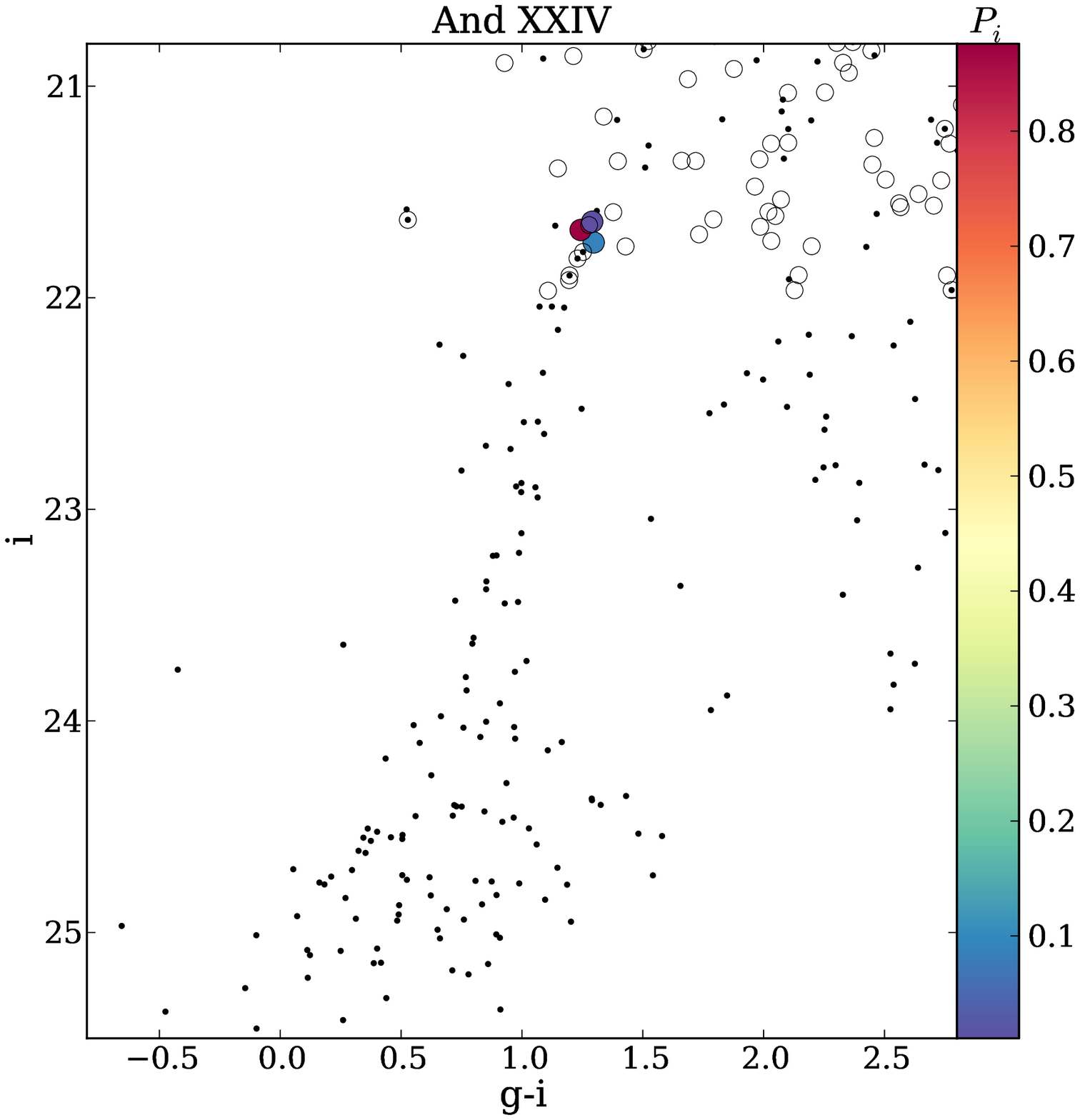}
\includegraphics[angle=0,width=0.45\hsize]{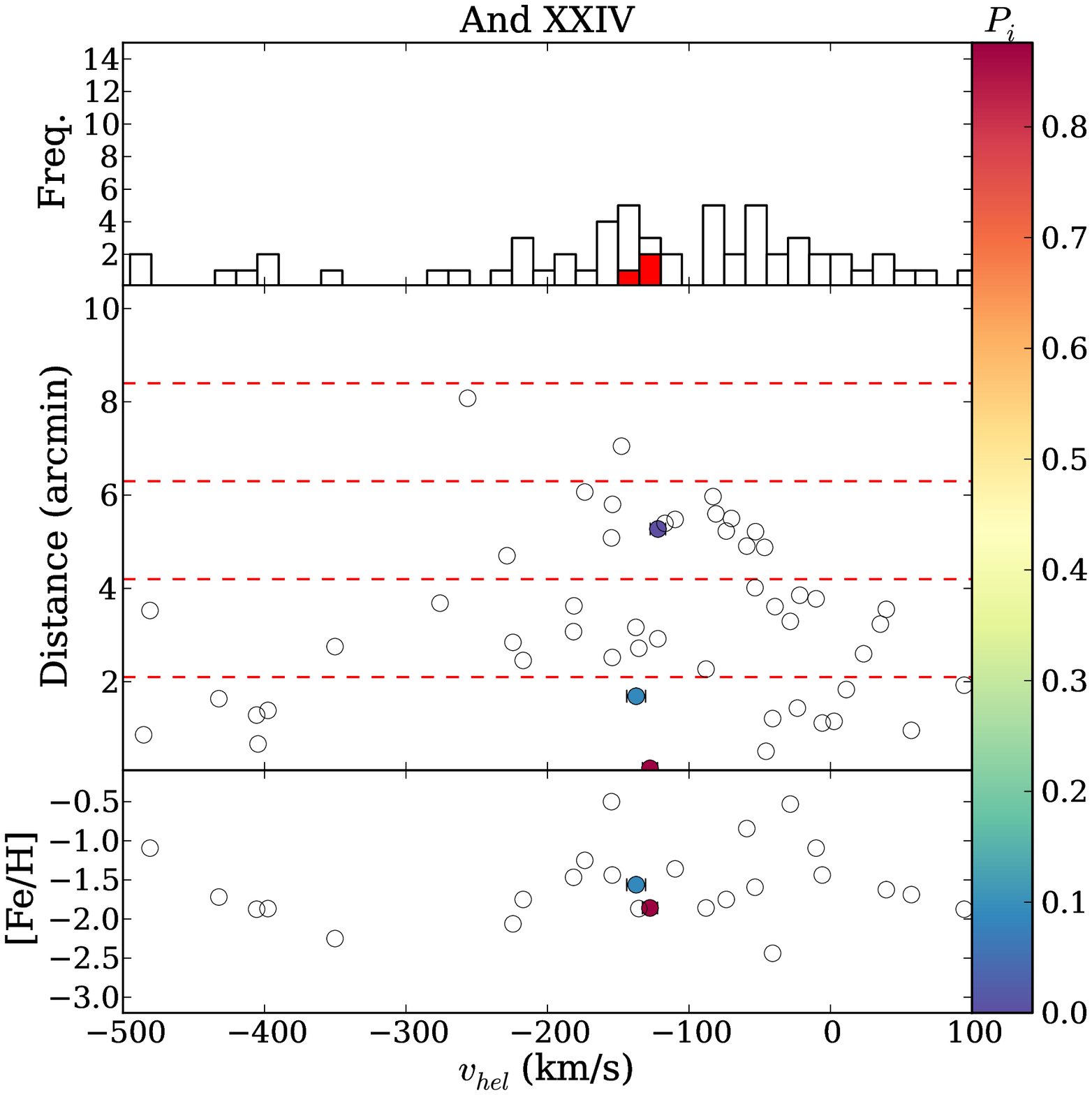}
\includegraphics[angle=0,width=0.9\hsize]{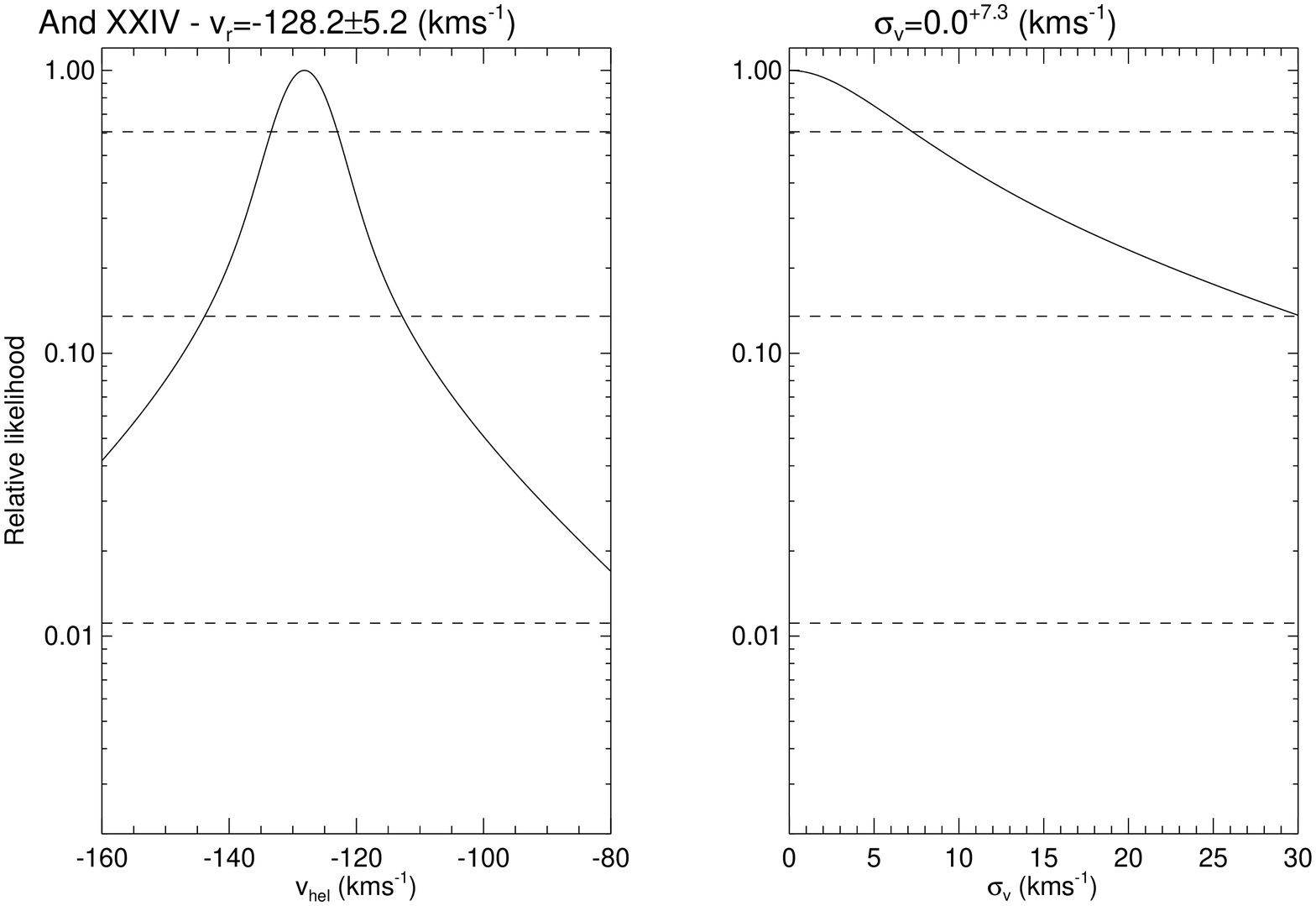}
\caption{As Fig.~\ref{fig:And17}, but for And XXIV.}
\label{fig:and24}
\end{center}
\end{figure*}

And XXIV was also first reported in \citet{richardson11}. Relatively faint and
compact ($M_V=-7.6$, $r_{\rm half}=548^{+31}_{-37}$~pc), spatially it is
located $\sim200\kpc$ from M31, along its northern major axis. And XXIV was
observed on two separate occasions as detailed in Table~\ref{tab:specobs}. For
the first mask, there was an error in target selection, and as a result, only
one star that lay on the RGB of And XXIV was observed. The second mask was
observed in May 2011, however owing to target visibility, only a short
integration of 45 minutes was obtained, which resulted in higher velocity
uncertainties than typically expected ($\sim8\kms$ vs. $\sim5\kms$). For this
reason, we have only included stars from this mask with $i<22.0$, as the
spectra for fainter stars were too noisy to determine reliable velocities
from. The systemic velocity of And XXIV also unsatisfactorily coincides with
that of the MW halo contamination, as can be seen in
Fig.~\ref{fig:and24}. As
for And XIX, we check the strength of the Na I doublet of all the stars
classified as potential members for And XXIV, and find no significant
absorption, making them unlikely foreground contaminants. But, given the
lower quality of this dataset, this check is far from perfect, and it is
possible that we have included contaminants from the MW within our
sample. Owing to the larger velocity uncertainties of the And XXIV dataset,
and the overlap of And XXIV with the MW, the determination of probability of
membership for stars within this dataset is based largely on their position in
the color magnitude diagram (e.g., location on the RGB) and their distance
from the centre of And XXIV.

When we run our machinery over the data acquired from both masks, we identify
only 3 probable members and determine a systemic velocity of
$v_r=-128.2\pm5.2\kms$ and we resolve a velocity dispersion of
$\sigma_v=0.0^{+7.3}\kms$. Given the lower quality of this dataset in
comparison to the remainder of those we present in this work, and the overlap
of And XXIV in velocity space with contamination from the MW, a robust
kinematic detection and characterisation of this galaxy is made incredibly
challenging. As such, we present these results as a tentative identification
of And XXIV, and do not include its measured properties in the remainder of
our analysis. Further kinematic follow up of And XXIV is required to
understand this system. We present the velocities of all bright stars for
which velocity measurements were possible in Table~\ref{tab:members} so that
they may be helpful for any future kinematic analysis of this system.

\begin{deluxetable*}{lccccccccc}
\tabletypesize{\footnotesize}
\tablecolumns{10} 
\tablewidth{0pt}
\tablecaption{Details of probable dSph members (This table is available in its entirety in machine-readable form in the online journal. A portion is shown here for guidance regarding its form
and content)\label{tab:members}}
\tablehead{
\colhead{Field} &	   \colhead{Star ID} &	 \colhead{$\alpha_{0,J2000}$}& \colhead{$\delta_{0,J2000}$}&
\colhead{$g$}	&  \colhead{$i$} &   \colhead{$v_{hel} (\kms)$}	&\colhead{$v_{err} (\kms)$}&	   \colhead{S:N (\AA$^{-1}$)}	&     \colhead{$P_i$}}
\startdata
And V    &    9  &      1:10:2.38       &47:37:48.5 &   23.640 &    22.402 &
-371.270 &    5.080  &   1.700  &   0.014 \\
And V   &    12  &      1:10:5.540 &      47:36:41.6  &  23.100 &   21.530 &
-406.500 &     3.600&     2.000 &    0.111\\
...&...&...&...&...&...&...&...&...&...\\
\enddata
\end{deluxetable*}

\subsection{Andromeda XXV}
\label{sect:and25}

\begin{figure*}
\begin{center}
\includegraphics[angle=0,width=0.45\hsize]{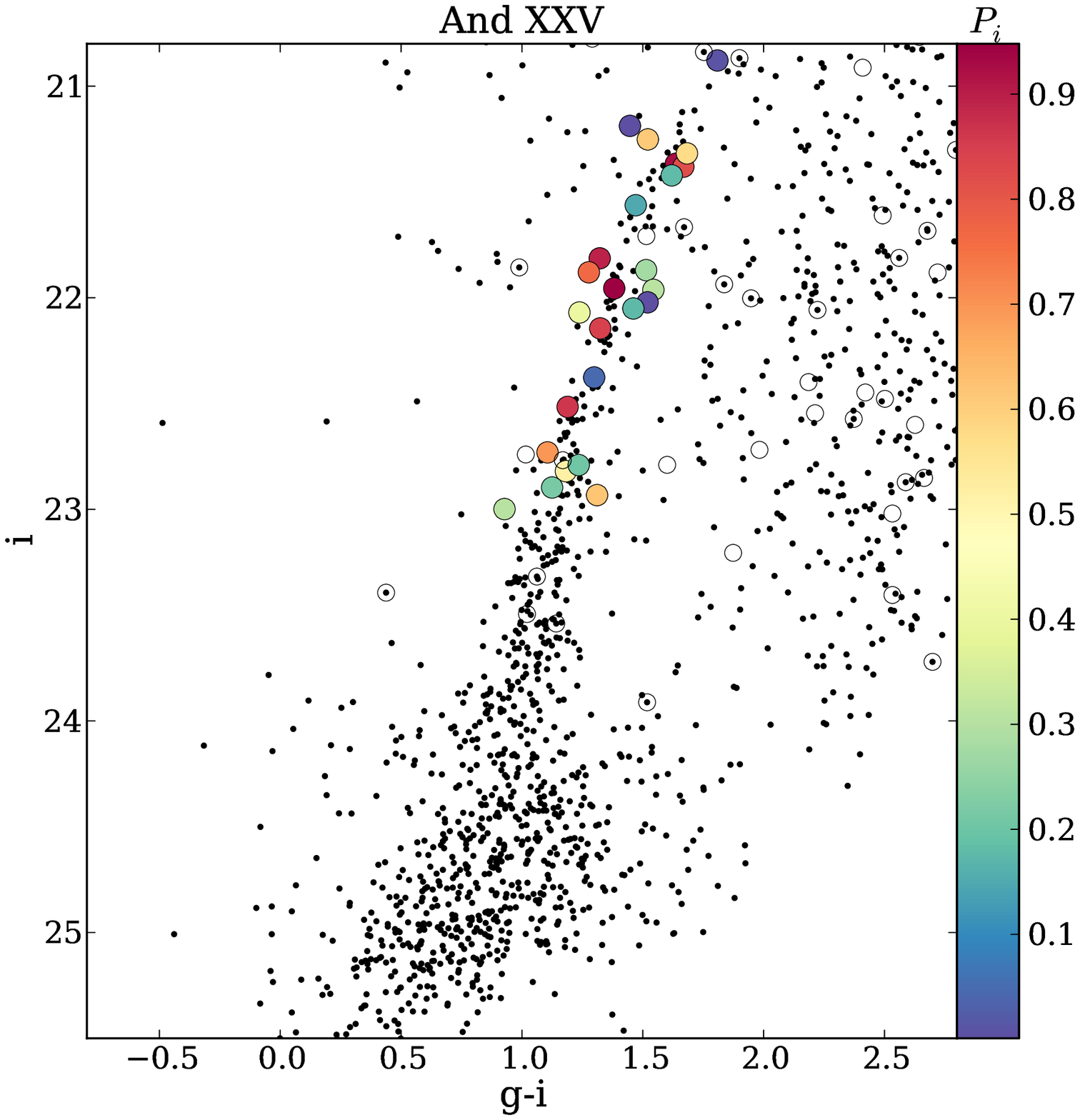}
\includegraphics[angle=0,width=0.45\hsize]{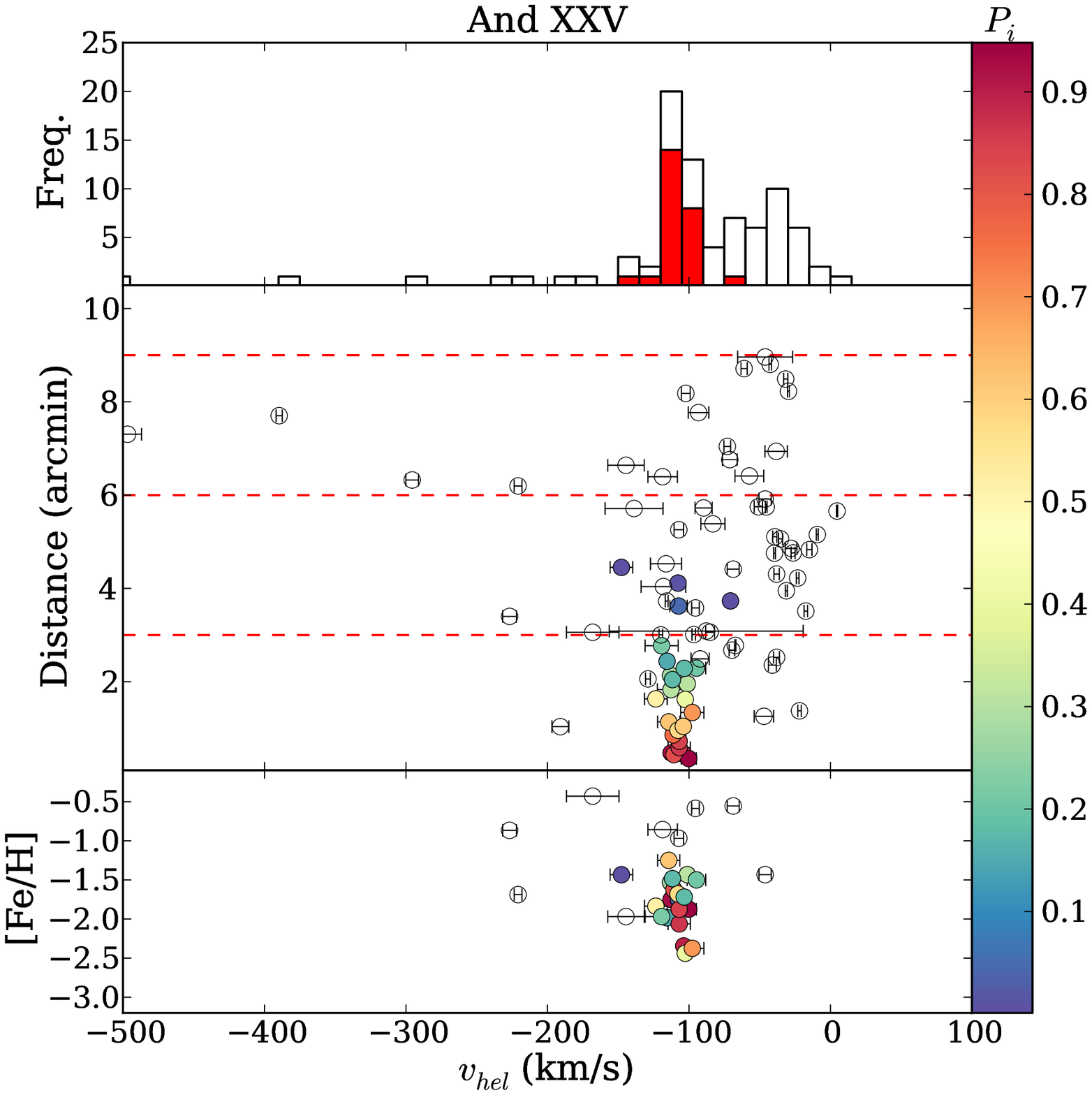}
\includegraphics[angle=0,width=0.9\hsize]{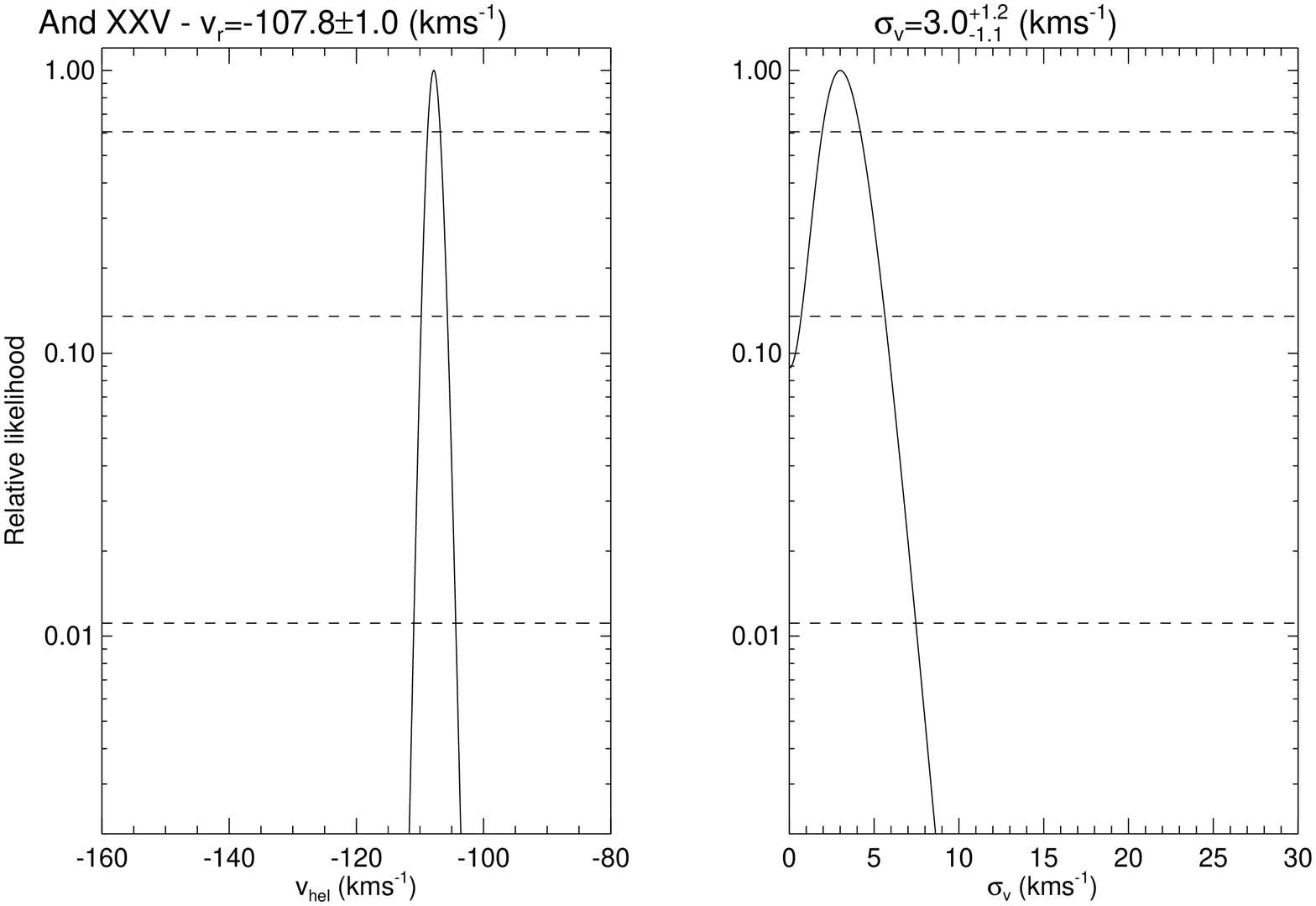}
\caption{As Fig.~\ref{fig:And17}, but for And XXV. }
\label{fig:And25}
\end{center}
\end{figure*}

And XXV was identified in \citet{richardson11} as a relatively bright
($M_V=-9.7$), extended ($r_{\rm half}=642^{+47}_{-74}$~pc) dwarf spheroidal,
located at a projected distance of $\sim90$~kpc to the north west of M31. As
with And XXIII, we present here a kinematic analysis of And XXV. The
results are displayed in Fig.~\ref{fig:And25}. We see
that the systemic velocity of And XXV ($v_r=-107.8\pm1.0\kms$), places
it in the regime of the Galactic foreground. However, given the strong
over-density of stars with this velocity relative to the expected contribution
of MW stars, we are confident that our routine has detected 25 likely members
for this object. We check the strength of the Na I doublet in
these likely members, and find no significant absorption, making them unlikely
foreground contaminants.  As for And XIX and XXI, we find that And XXV has a
curiously low velocity dispersion for its size, with
$\sigma_v=3.0^{+1.2}_{-1.1}\kms$. 
We
discuss the significance of this further in \S~\ref{sect:mass} and Collins et
al (2013, in prep).

\subsection{Andromeda XXVI}
\label{sect:and26}

\begin{figure*}
\begin{center}
\includegraphics[angle=0,width=0.45\hsize]{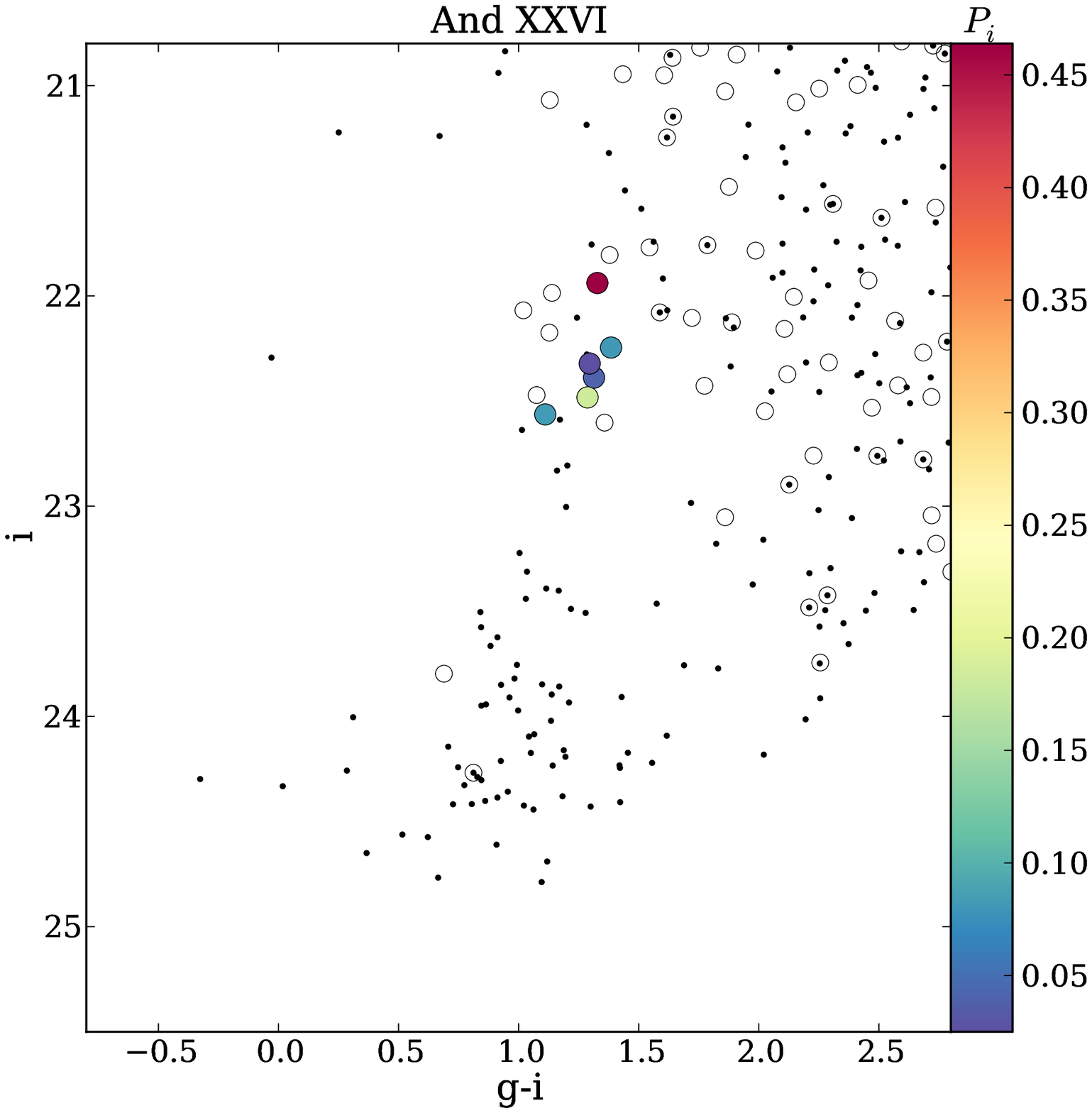}
\includegraphics[angle=0,width=0.45\hsize]{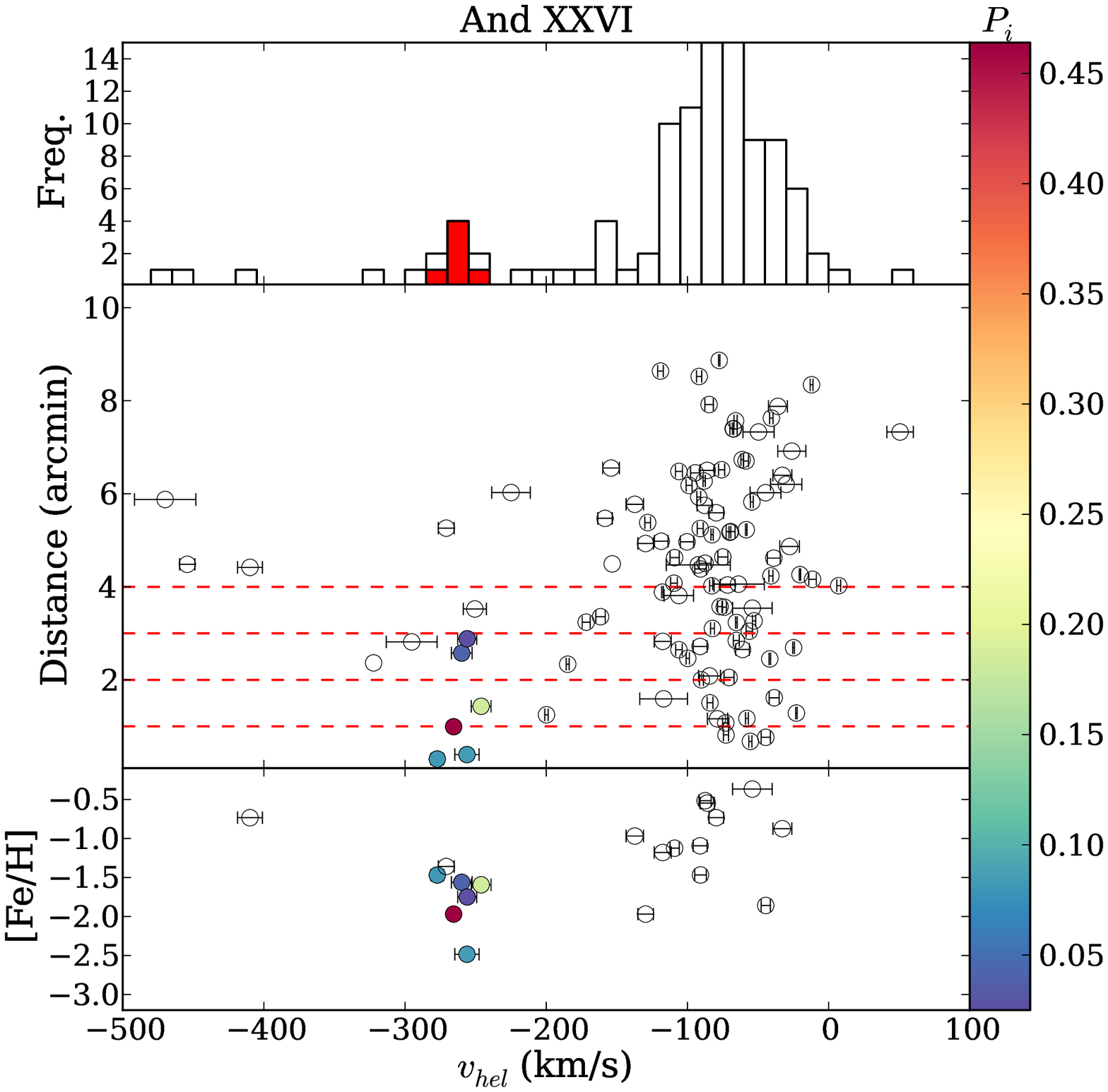}
\includegraphics[angle=0,width=0.9\hsize]{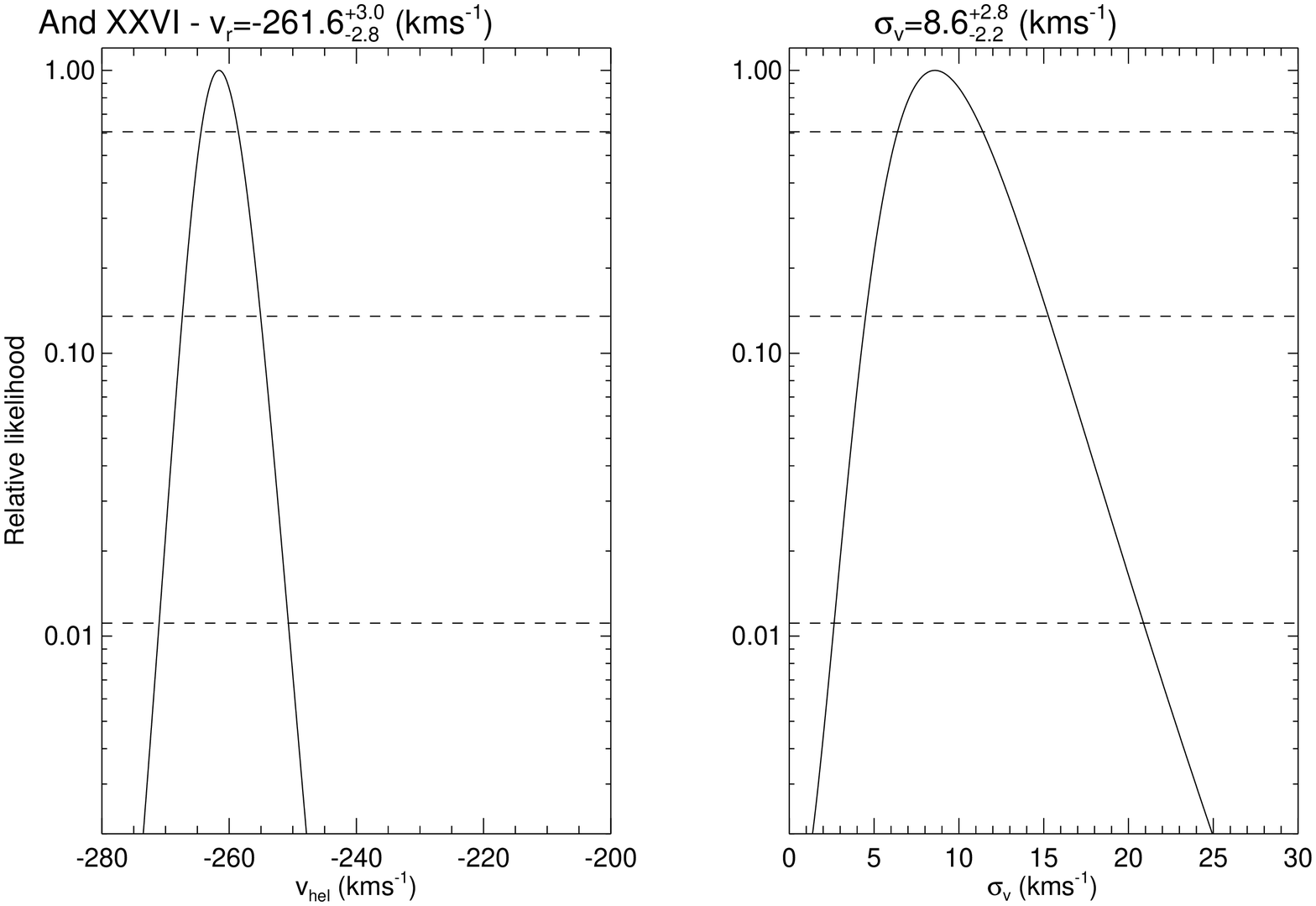}
\caption{As Fig.~\ref{fig:And17} but for And XXVI. }
\label{fig:And26}
\end{center}
\end{figure*}

And XXVI is a relatively faint ($M_V=-7.1$) dSph with $r_{\rm half}=219^{+67}_{-52}$~pc, also
first reported in \citet{richardson11}. Its low luminosity makes observing large numbers of
member stars difficult, owing to the paucity of viable targets on the RGB that
can be observed with DEIMOS. As a result, our routine has identified only 6
stars as potential members, highlighted in Fig.~\ref{fig:And26}. The dwarf has
a systemic velocity of $v_r=-261.6^{+3.0}_{-2.8}\kms$, and a fairly typical
velocity dispersion of $\sigma_v=8.6^{+2.8}_{-2.2}\kms$. As with And XX, while
we believe our routine can robustly measure the velocity dispersions of
systems with only 6 confirmed members, to be truly confident of this value,
follow up of And XXVI to increase the number of likely members is required.

In \citet{conn12}, from an analysis of the photometry of And
XXVI, they determined a distance modulus to the object of
$(m-M)_0=24.39^{+0.55}_{-0.53}$ from a Markov-Chain-Monte-Carlo analysis of
the PAndAS photometry of And XXVI. This value corresponds to an $i-$band
magnitude for the TRGB of And XXVI of $m_{i,0}=21.1^{+0.55}_{-0.53}$. Our CMDs
for the dwarfs are not extinction corrected, but using the extinction values
from \citet{richardson11} of $E(B-V)=0.110$ \citep{schlegel98}, this would
correspond to an $i-$band magnitude of
$m_{i,TRGB}=21.3^{+0.55}_{-0.53}$. Three targets were observed with magnitudes
and colors that should be consistent with their belonging to And
XXVI. However. we find that all these objects have velocities that are
consistent with being Galactic foreground contaminants. Given the position of
And XXVI in the northern M31 halo, where contamination from the MW increases,
this is not unexpected. The brightest star we observe that is likely
associated with And XXVI has $m_i=21.9$ ($m_{i,0}=21.7$). Assuming that this
brightest confirmed member star of And XXVI sits at the RGB tip, the distance
estimate becomes 1.1 Mpc. While this value is higher than that of
\citet{conn11}, it is still within their upper distance
estimate. Additionally, as not every star brighter than the member with
$m_{i,0}=21.7$, with colors consistent with the And XXVI RGB was observed,
this value merely represents an upper limit, on the distance to And XXVI and
highlights the difficulty of calculating distances to these faint galaxies
where RGB stars are sparse.

\subsection{Andromeda XXVII}
\label{sect:and27}

Andromeda XXVII (And XXVII) is a somewhat unusual object as it is currently
undergoing tidal disruption, spreading its constituent stars into a large
stellar stream, named the northwestern arc, discovered in the PAndAS survey by
\citet{richardson11}. As such, it is unlikely to be in virial equilibrium, if
it remains bound at all.

\begin{figure*}
\begin{center}
\includegraphics[angle=0,width=0.45\hsize]{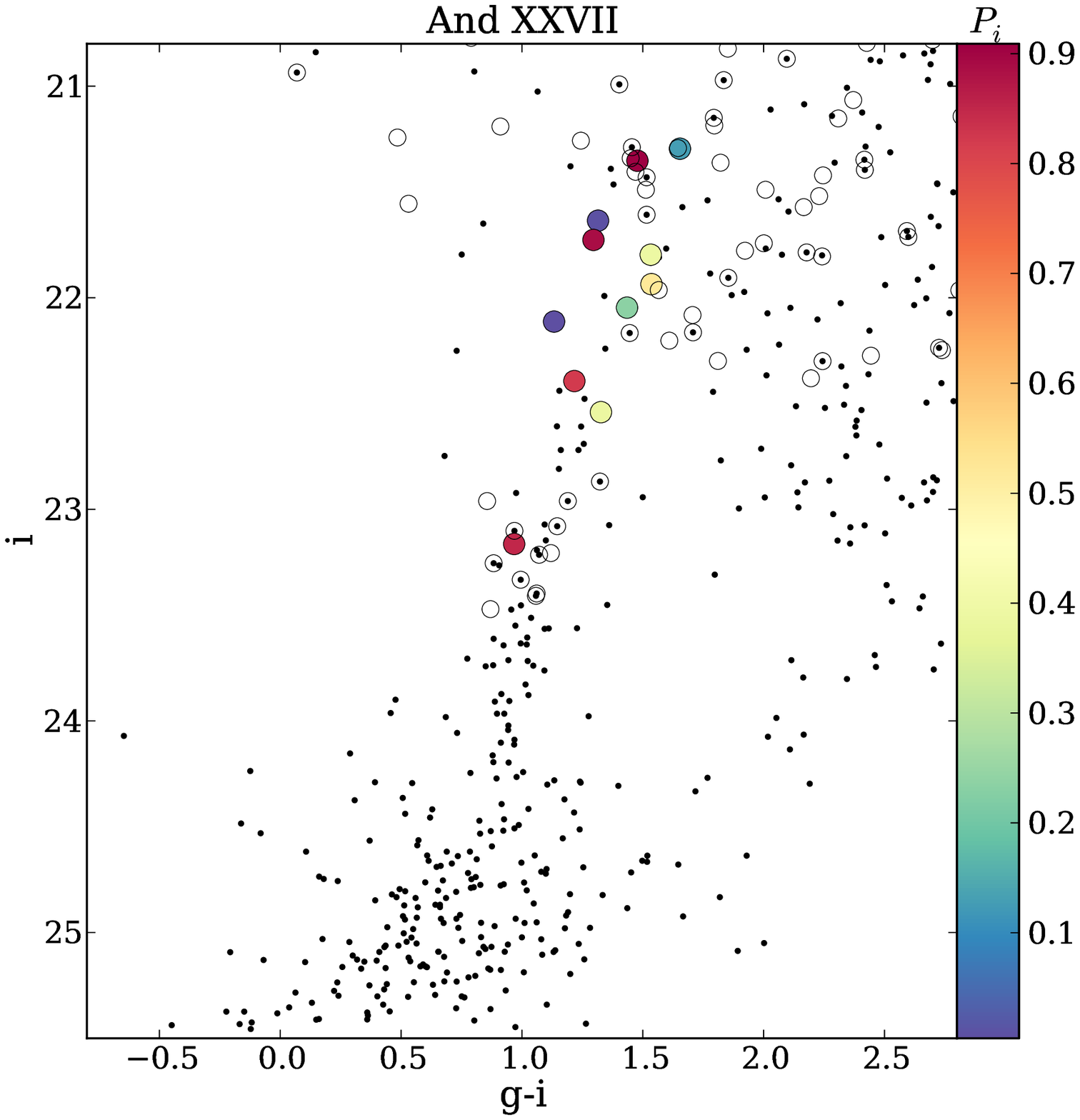}
\includegraphics[angle=0,width=0.45\hsize]{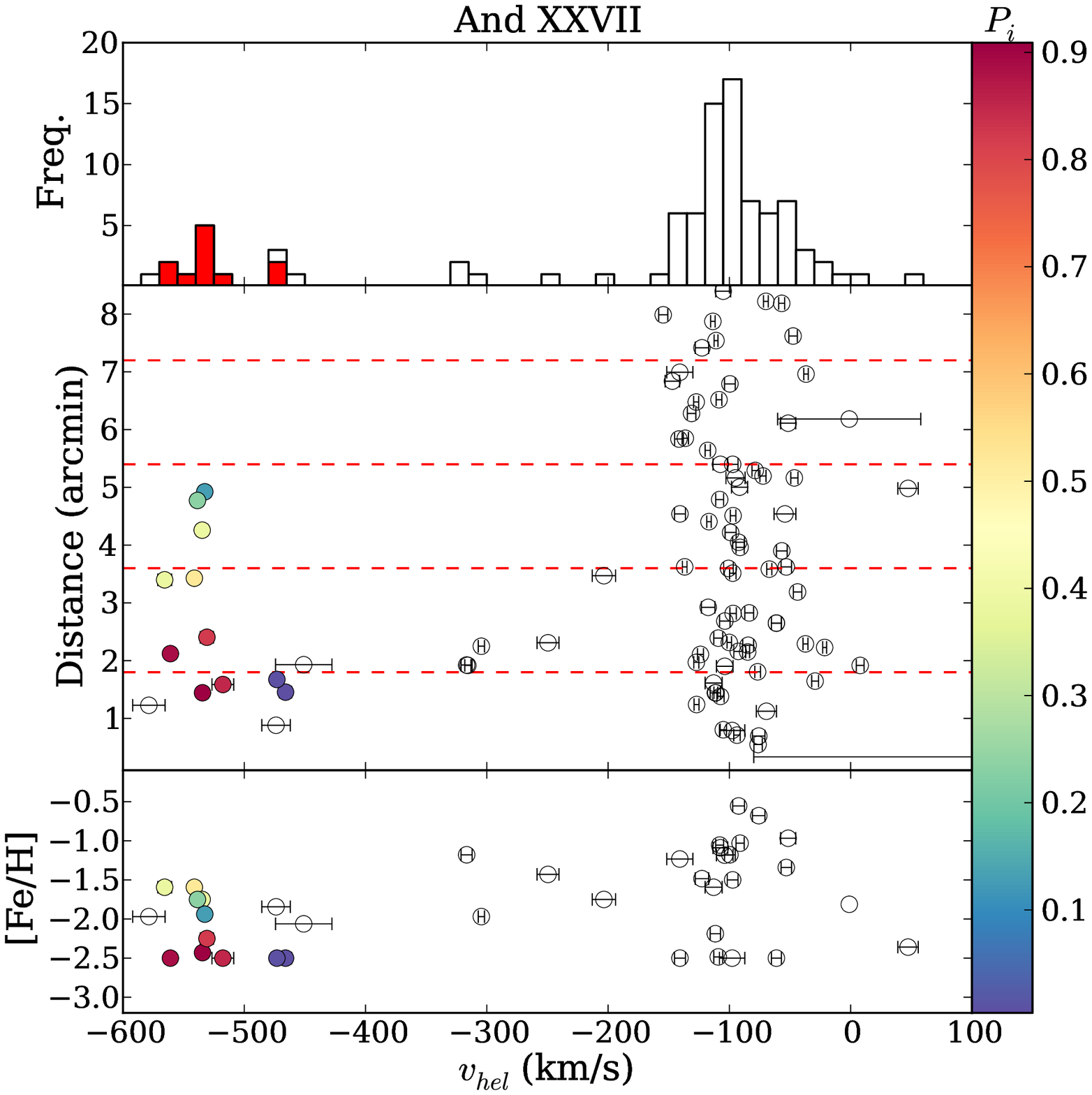}
\includegraphics[angle=0,width=0.9\hsize]{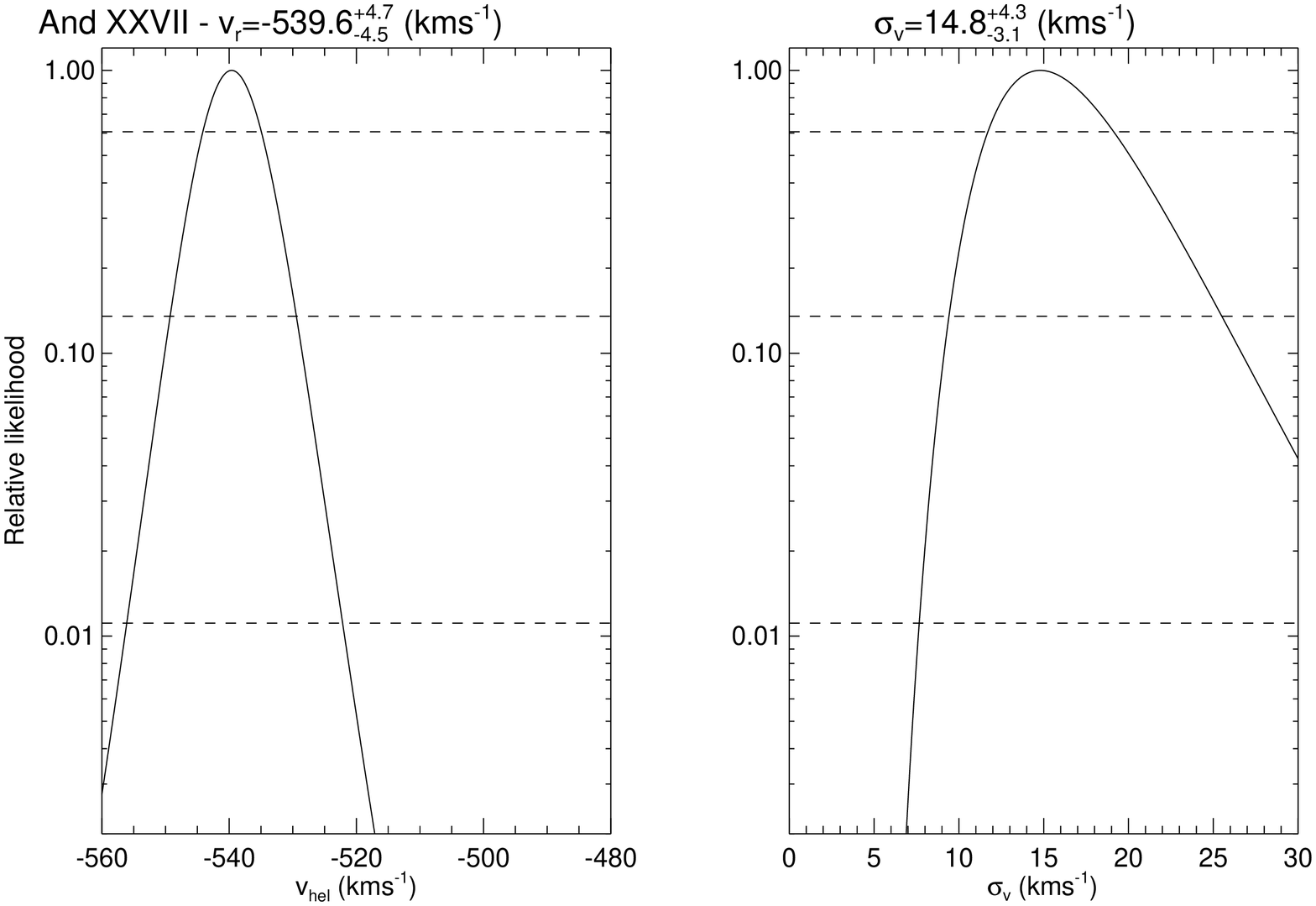}
\caption{As Fig.~\ref{fig:And17}, but for And XXVII.}
\label{fig:And27}
\end{center}
\end{figure*}

When determining the kinematics of And XXVII, we find the results somewhat
unsatisfactory. Our routine determines $v_r=-539.6^{+4.7}_{-4.5}\kms$ and
$\sigma_v=14.8^{+4.3}_{-3.1}\kms$ from 11 stars. However, from an inspection of
Fig.~\ref{fig:And27}, we see that there is significant substructure around
$v_{hel}\sim-500\kms$, much of which is considered to be unassociated with And
XXVII in this analysis as it does not fall within a cold, well-defined
Gaussian velocity peak. Given the disrupting nature of And XXVII, it is likely
that a different analysis is required for this object, and we shall discuss
this further in a future analysis, where the kinematics of the northwestern
arc itself are also addressed. From this first pass however, it would appear
that And XXVII may no longer be a gravitationally bound system.

\subsection{Andromeda XXVIII}
\label{sect:and28}
\begin{figure*}
\begin{center}
\includegraphics[angle=0,width=0.45\hsize]{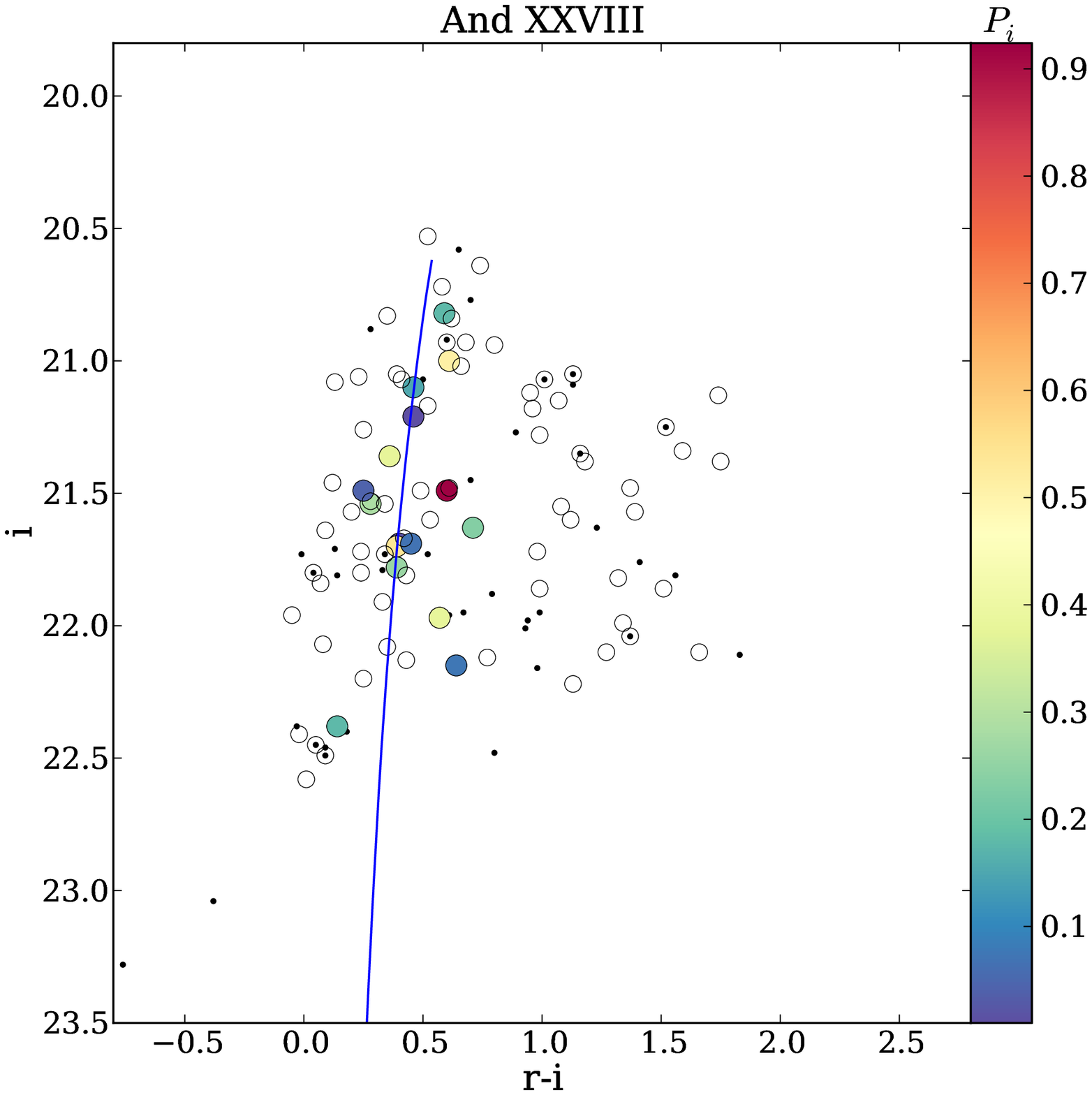}
\includegraphics[angle=0,width=0.45\hsize]{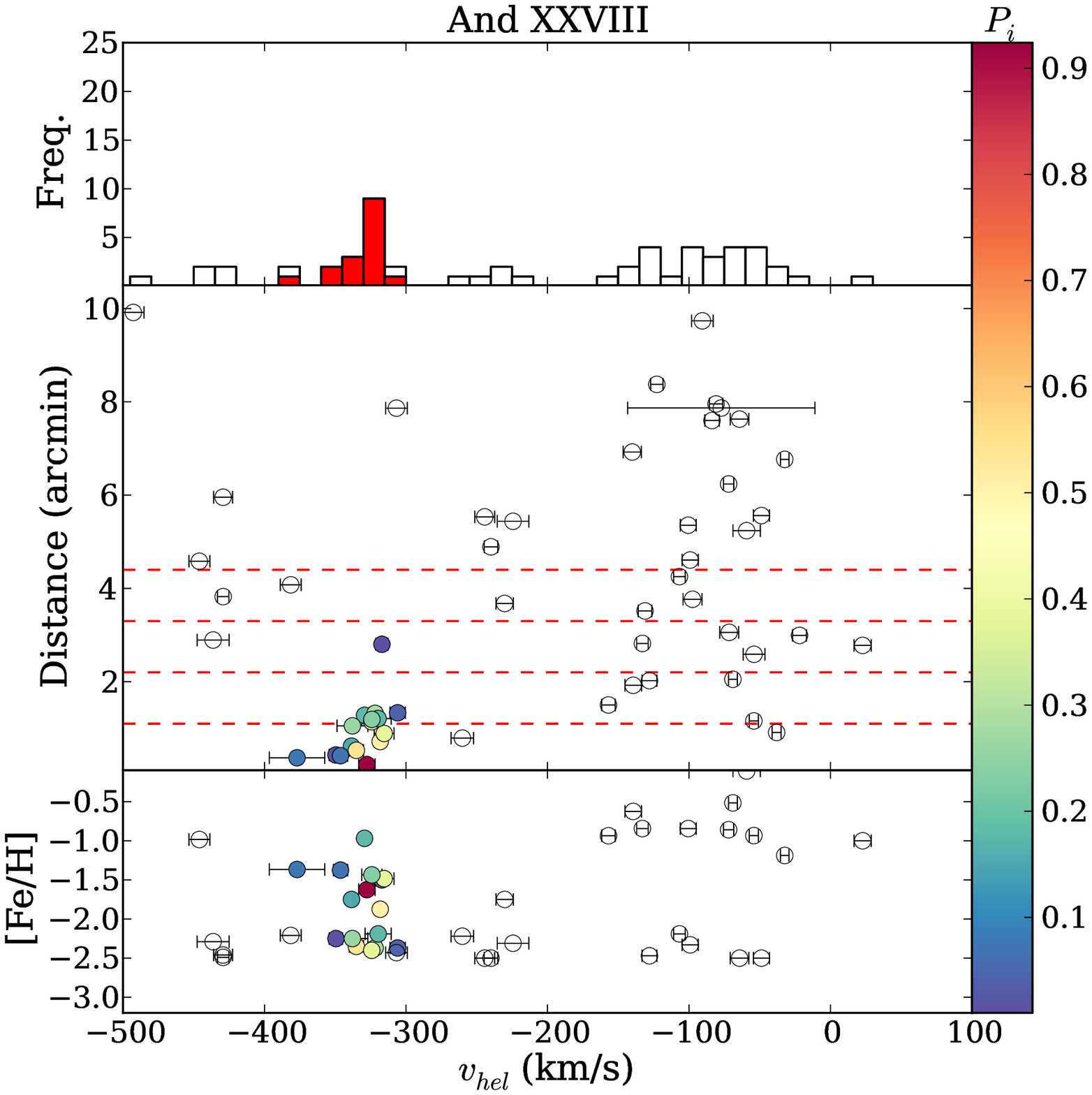}
\includegraphics[angle=0,width=0.9\hsize]{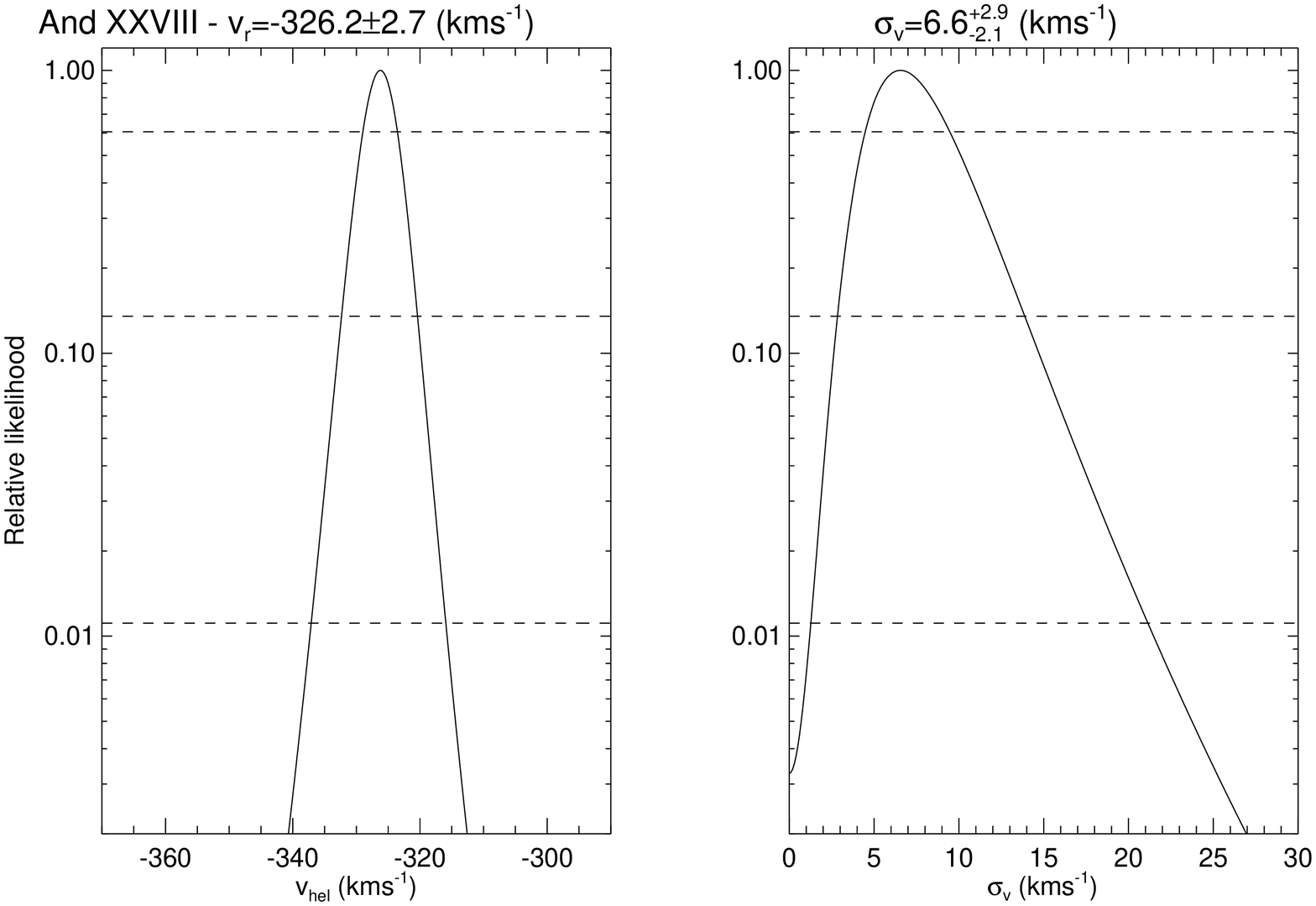}
\caption{As Fig.~\ref{fig:And17}, but for And XXVIII.}
\label{fig:And28}
\end{center}
\end{figure*}

And XXVIII was recently discovered in the 8th data release of the SDSS survey
\citep{slater11}. It has $M_v=-8.5$ and $r_{\rm half}=210^{+60}_{-50}$~pc. It
is also potentially one of Andromeda's most distant satellites, with a
host-satellite projected separation of $365^{+17}_{-1}$ kpc. The And XXVIII
satellite is not covered by the PAndAS footprint, so we must instead use the
original SDSS photometry for our analysis. A CMD with the SDSS $i-$band and
$r-i$ colors for And XXVIII is shown in Fig.~\ref{fig:And28}, where all
targets brighter that $i\sim23.5$ within $3r_{\rm half}$ are shown. The
photometry here do not show an RGB that is as convincing as those from the
PAndAS survey, so to guide the eye, we also overplot an isochrone from
\citet{dart08} with a metallicity of $\feh=-2.0$, corrected for the distance
of And XXVIII as reported in \citet{slater11}. 

In a recent paper, \citet{tollerud13} discussed the kinematics of this object
as derived from 18 members stars.  They find $v_r=-328.0\pm2.3\kms$ and
$\sigma_v=8.1\pm1.8\kms$ from their full sample. They then remove two stars
that they categorize as outliers based on their distance from the centre of
And XXVIII, which alters their measurements to $v_r=-331.1\pm1.8\kms$ and
$\sigma_v=4.9\pm1.6\kms$. Analyzing our own DEIMOS dataset for this object, we
find $v_r=-326.1\pm2.7\kms$ and $\sigma_v=6.6^{+2.9}_{-2.2}\kms$ based on 17
probable members. This is fully consistent with the results from the
full sample in \citet{tollerud13}. However, the systemic velocity we measure
is offset at a level of $\sim1\sigma$ from their final value (calculated after
excluding 2 outliers). This offset is small, and is probably
attributable to our differing methodologies for classifying stars as
members. As we believe our method is more robust (as discussed in
\S~\ref{sect:test} and Appendix A), we will use our derived parameters
for this object in the remainder of our analysis.

\subsection{And XXX/Cassiopeia II}
\label{sect:and30}

And XXX -- also known as Cass II owing to its spatial location, overlapping
the Cassiopeia constellation -- is a recently discovered dSph from the PAndAS
survey (Irwin et al. in prep). It has $M_v=-8.0$ and $r_{\rm
  half}=267^{+23}_{-36}$~pc. Located to the north west of Andromeda, it sits
within 60 kpc of the two close dwarf elliptical M31 companions, NGC 147 and
NGC 185. With these 3 objects found so close together in physical space, it is
tempting to suppose them a bound system within their own right, but this
can only be borne out by comparing their kinematics.

\begin{figure*}
\begin{center}
\includegraphics[angle=0,width=0.45\hsize]{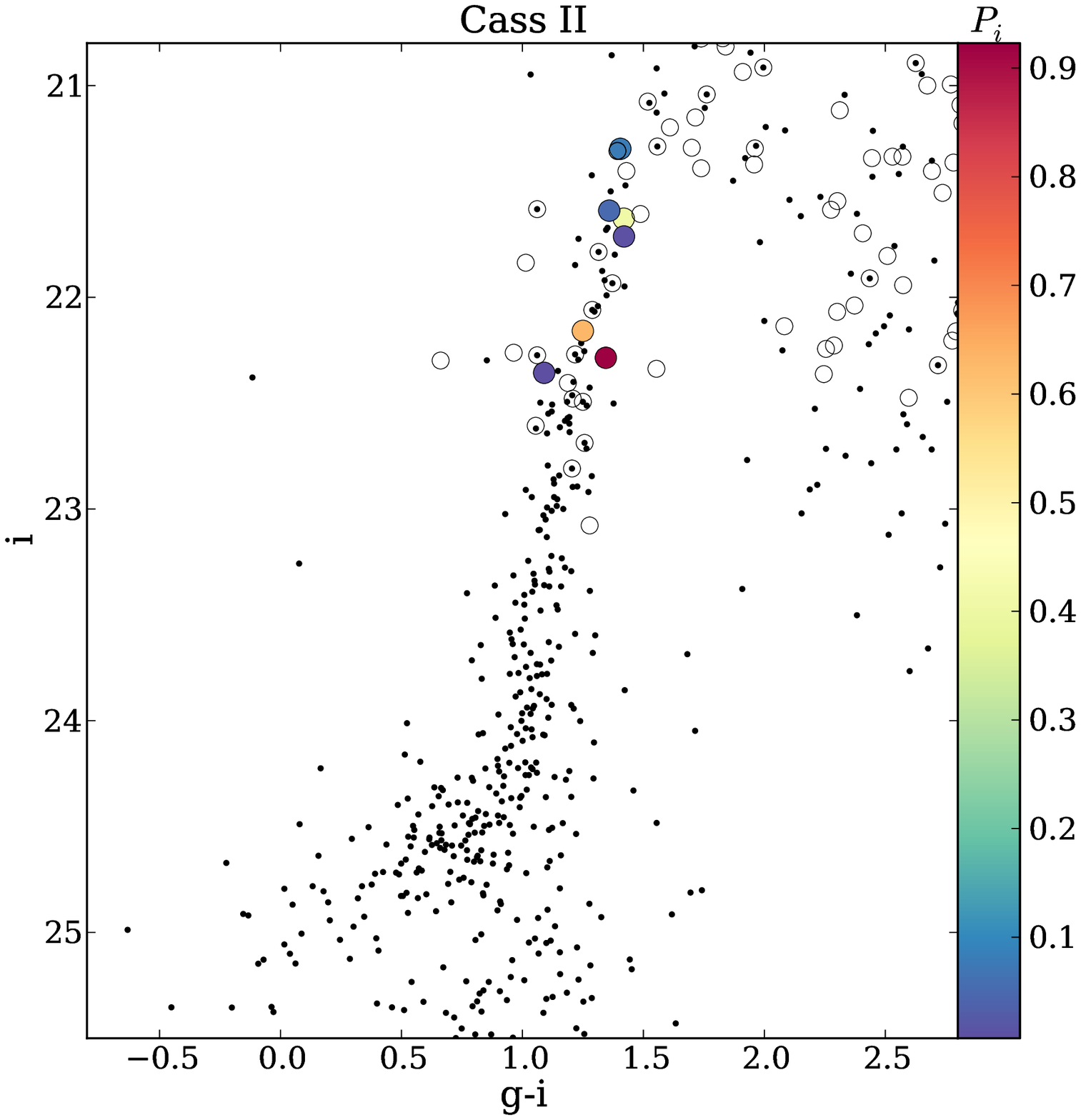}
\includegraphics[angle=0,width=0.45\hsize]{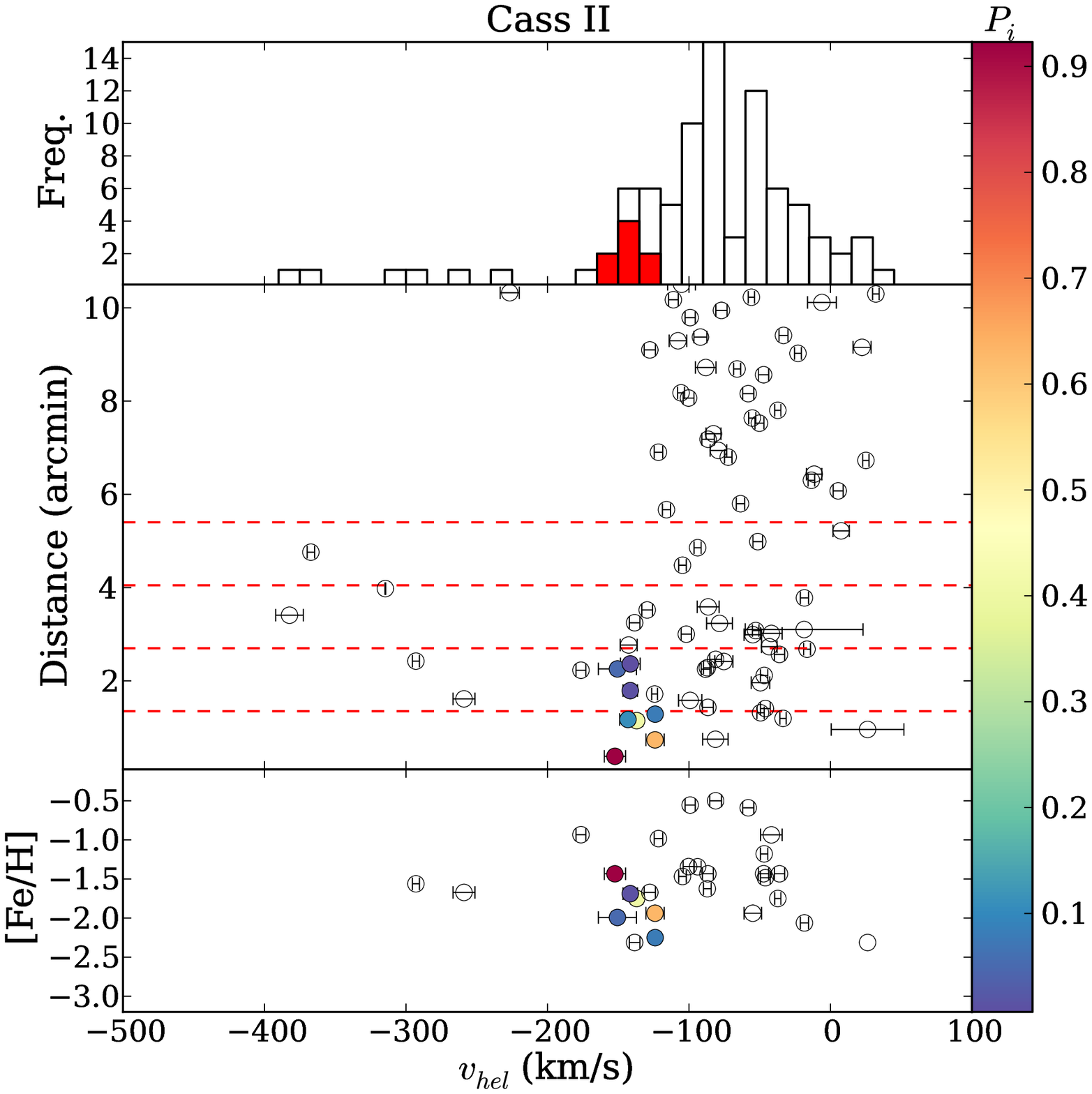}
\includegraphics[angle=0,width=0.9\hsize]{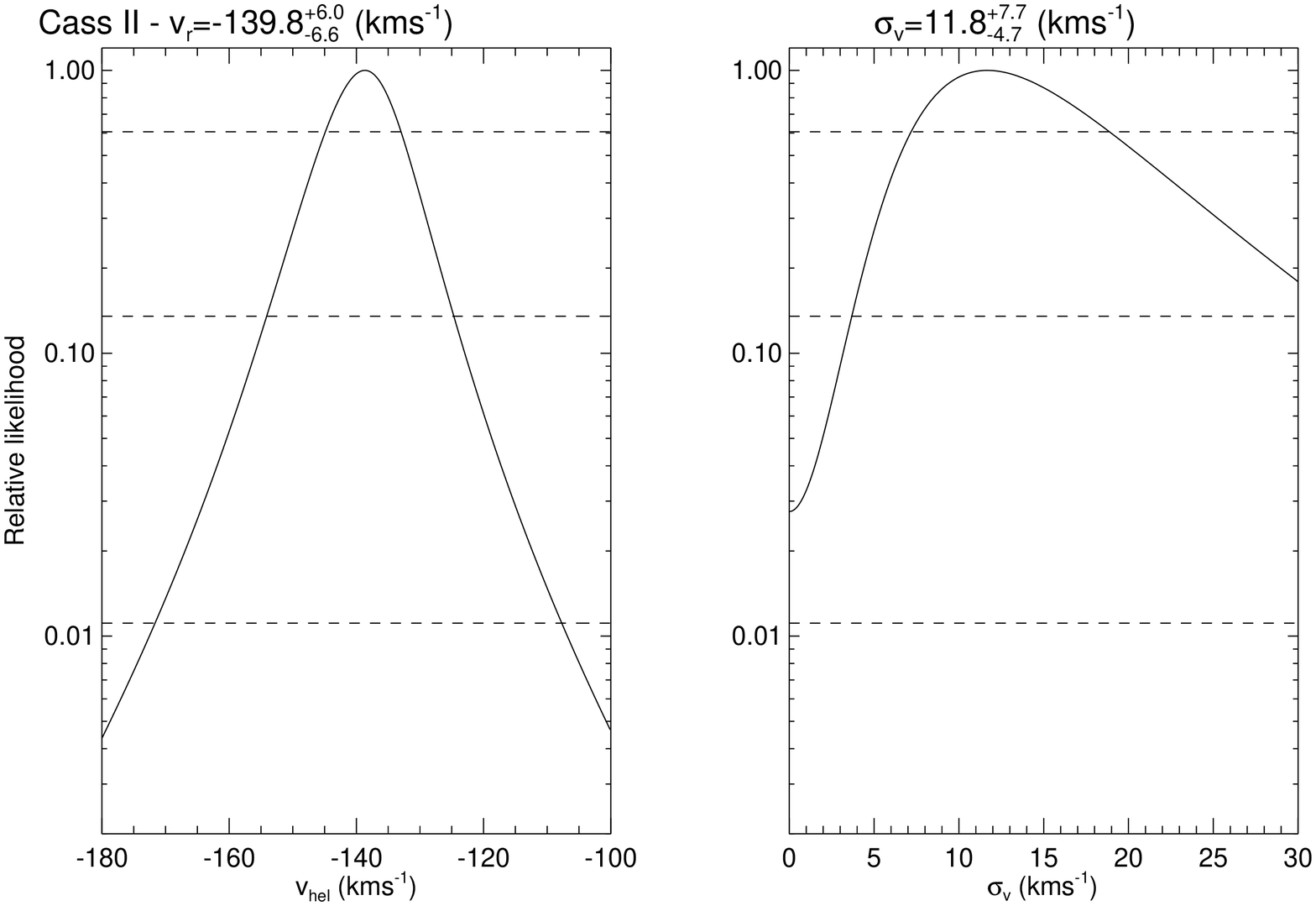}
\caption{As Fig.~\ref{fig:And17}, but for Cass II}
\label{fig:Cass2}
\end{center}
\end{figure*}

Conspiring to confound us, we find that Cass II has kinematics that place it
well within the regime of Galactic foreground, as can be seen in
Fig.~\ref{fig:Cass2}. However, our analysis is able to detect the dSph as a
cold spike consisting of 8 likely members. As for And XIX, we check the
strength of the Na I doublet in these likely members, and find no significant
absorption, making them unlikely foreground contaminants. We measure
$v_r=-139.8^{+6.0}_{-6.6}\kms$, and a fairly typical velocity dispersion of
$\sigma_v=11.8^{+7.7}_{-4.7}\kms$.

The systemic velocity of Cass II ($v_r=-139.8^{+6.0}_{-6.6}\kms$) puts it
within $\sim50\kms$ of those of NGC 147 and NGC 185 ($v_r=-193\pm3\kms$ and
$v_r=-210\pm7\kms$, \citealt{mateo98}), lending further credence to the notion
that these 3 systems are associated with one another. This will be discussed
in more detail in Irwin et al. (2013, in prep).

\subsection{A note on previous work}

  Finally, we also use our new algorithm to reanalyze all our previously
  published M31 dSph datasets. These include And V, VI \citep{collins11b}, XI,
  XII and XIII \citep{chapman07,collins10}. Details of the results of this
  reanalysis can be found in Appendix B. In summary, we find that our
  algorithm measures systemic velocities and velocity dispersions that are
  fully consistent with our previous work. We present these results in
  Table~\ref{tab:kprops}. And V, XI, XII and XIII are also analyzed by T12, so
  we compare our findings with theirs. For And XI and XII, our results are
  based on two and four times the number of stars respectively, and as such,
  supercede those presented in T12. In the case of And V and XIII, the T12
  measurements are based sample sizes with four times the number of stars as
  our datasets, making their findings more robust.

  In previous studies by our group \citep{chapman05,letarte09,collins10}, we
  also published kinematic analyses for three additional M31 dSphs; And IX,
  And XV and And XVI. In T12, it was noted that the values presented in these
  works for systemic velocities and velocity dispersions were not consistent
  with those measured in their analyses. We revisited these datasets in light
  of this discrepancy, to see if our new technique could resolve this
  issue. We found that these discrepancies remained. For And IX, we measure a
  systemic velocity of $v_r=-204.8\pm2.1\kms$ cf. $v_r=-209.4\pm2.5\kms$ in
  T12 and a velocity dispersion of $\sigma_v=2.0^{+2.7}_{-2.0}\kms$
  cf. $\sigma_v=10.9\pm2.0\kms$. Not only are their measurements determined
  from 4 times the member stars that we possess, we also experienced problems
  with our radial velocity measurements for the stars observed with this mask,
  due to the use of the minislitlet approach pioneered by
  \citet{ibata05}. This setup resulted in poor sky subtraction for many of the
  science spectra, lowering the quality of our radial velocity
  measurements. As such, the T12 results supercede those of our previous work
  \citep{chapman05,collins10}.

  For And XV and XVI, we measure a systemic velocities of
  $v_r=-354.6\pm4.9\kms$ and $v_r=-374.1\pm6.8\kms$ cf. $v_r=-323.0\pm1.4\kms$
  and $v_r=-367.3\pm2.8\kms$ from T12. We also note offsets in our velocity
  dispersions for And XV and XVI, where we measure
  $\sigma_v=9.6^{+4.1}_{-2.6}\kms$ and $\sigma_v=17.3^{+6.4}_{-4.4}\kms$
  cf. $\sigma_v=3.8\pm2.9\kms$. In this instance, the data for both And XV and
  XVI were taken in poor conditions, with variable seeing that averaged at
  $1.8^{\prime\prime}$ and patchy cirrus. These conditions significantly
  deteriorated the quality of our spectra, and made the measurement of
  reliable radial velocities extremely difficult. Again, this leads us to
  conclude that the measurements made in T12 supercede those presented by our
  group in \citet{letarte09}.

\section{The masses and dark matter content of M31 dSphs}

\subsection{Measuring the masses and mass-to-light ratios of our sample}
\label{sect:mass}

 As dSph galaxies are predominantly dispersion supported
systems, we can use their internal velocity dispersions to measure masses for
these systems, allowing us to infer how dark matter dominated they are. There
are several methods in the literature for this
(e.g., \citealt{illingworth76,richstone86}), but recent work by
\citet{walker09a} has shown that the mass contained within the half-light
radius ($M_{\rm half}$)of these objects can be reliably estimated using the
following formula:

\begin{equation}
M_{\rm half}=\mu r_{\rm half}\sigma_{v,{\rm half}}^2
\end{equation}

\noindent where $\mu=580\msun{\rm pc}^{-1}$km$^{-2}$s$^2$, $r_{\rm half}$ is
the spherical half-light radius in pc and $\sigma_{v,{\rm half}}$ is the
luminosity- averaged velocity dispersion. This mass estimator is independent
of the (unknown) velocity anisotropy of the tracer population, however, it is
sensitive to the embeddedness of the stellar component within the DM
halo. Particularly, the mass tends to be slightly over-estimated the more
embedded the stars are \citep{walker11}, especially if the dark matter halo
follows a cored density profile. 

As numerous authors have shown that the velocity dispersion profiles of dSphs
are constant with radius (e.g., \citealt{walker07,walker09b}), we assume our
measured values of $\sigma_v$ are representative of the luminosity-averaged
velocity dispersion ($\sigma_{v,{\rm half}}$) used by
\citet{walker09b}. However, if it transpired that the velocity dispersion
profiles of the Andromedean dSphs were not flat, but declined or increased
with radius, this would no longer true. We see no evidence for this behaviour
in our dataset, although low-number statistics means we are unable to
completely rule out this possibility. We calculate this for all our observed
dSphs (including those we reanalyzed from previous works, see Appendix B)
using results from the Keck LRIS and DEIMOS dataset, and report their masses
within $r_{\rm half}$ ($M_{\rm half}$) in Table~\ref{tab:kprops}.

\begin{deluxetable*}{lcccccc}
\tabletypesize{\footnotesize}
\tablecolumns{7} 
\tablewidth{0pt}
\tablecaption{Kinematic properties of Andromeda dSph galaxies
  as derived within this work, from Keck I LRIS and Keck II DEIMOS data.
\label{tab:kprops}}
\tablehead{
\colhead{Property} & $\eta$ & \colhead{$v_r$} & \colhead{$\sigma_v$}& \colhead{M$_{\rm
  half}$}  & \colhead{$[M/L]_{\rm half}$} & \colhead{$\feh_{spec}$}\\
 & & \colhead{$(\kms)$} & \colhead{$(\kms)$} & \colhead{($10^7\msun$)} & \colhead{$(\msun/\lsun)$} & }
\startdata
And V     & 2.0 & $-391.5\pm2.7$  &   $12.2^{+2.5}_{-1.9}$ & $2.6^{+0.66}_{-0.56}$& $88.4^{+22.3}_{-18.9}$ & $-2.0\pm0.1$\\
And VI    & 2.5 & $-339.8\pm1.8$      & $12.4^{+1.5}_{-1.3}$ & $4.7\pm0.7$& $27.5^{+4.2}_{-3.9}$ & $-1.5\pm0.1$\\
And XI    & 2.5 &  $-427.5^{+3.5}_{-3.4}$ & $7.6^{+4.0(*)}_{-2.8}$  & $0.53^{+0.28}_{-0.21}$& $216^{+115}_{-87}$ & $-1.8\pm0.1$\\
And XII   & 2.5 & $-557.1\pm1.7$       & $0.0^{+4.0}$        &   $0.0^{+0.3}$     & $0.0^{+194}$ & $-2.2\pm0.2$\\
And XIII  & 2.5 & $-204.8\pm4.9$     & $0.0^{+8.1(*)}$       &  $0.0^{+0.7}$      & $0.0^{+330}$ & $-1.7\pm0.3$\\
And XVII  & 2.5 & $-251.6^{+1.8}_{-2.0}$  & $2.9^{+2.2}_{-1.9}$  & $0.13^{+0.22}_{-0.13}$& $12^{+22}_{-12}$& $-1.7\pm0.2$\\
And XVIII & 2.5 & $-346.8\pm2.0$            & $0.0^{+2.7}$       & $0.0^{+0.14}$ & $0^{+5}$&$-1.4\pm0.3$  \\            
And XIX   & 2.0 & $-111.6^{+1.6}_{-1.4}$  & $4.7^{+1.6}_{-1.4}$  &$1.9^{+0.65}_{-0.66}$& $84.3^{+37}_{-38}$   & $-1.8\pm0.3$\\
And XX    & 2.5 & $-456.2^{+3.1}_{-3.6}$  & $7.1^{+3.9(*)}_{-2.5}$ & $0.33^{+0.20}_{-0.12}$& $238.1^{+147.6}_{-90.2}$& $-2.2\pm0.4$\\
And XXI   & 5.0 & $-362.5\pm0.9$      & $4.5^{+1.2}_{-1.0}$       & $0.99^{+0.28}_{-0.24}$     & $25.4^{+9.4}_{-8.7}$& $-1.8\pm0.1$\\  
And XXII &  2.0 & $-129.8\pm2.0$       & $2.8^{+1.9}_{-1.4}$  & $0.11^{+0.08}_{-0.06}$& $76.4^{+58.4}_{-48.1}$& $-1.8\pm0.6$\\   
And XXIII & 4.0 & $-237.7\pm1.2$       & $7.1\pm1.0$  & $2.9\pm4.4$& $58.5\pm36.2$& $-2.2\pm0.3$\\
And XXIV  & 1.5 & $-128.2\pm5.2$  & $0.0^{+7.3(*)}$       & $0.4^{+0.7}_{-0.4}$ & $82^{+157}_{-82}$ & $-1.8\pm0.3$\\
And XXV   & 2.5 & $-107.8\pm1.0$  & $3.0^{+1.2}_{-1.1}$  & $0.34^{+0.14}_{-0.12}$& $10.3^{+7.0}_{-6.7}$ & $-1.9\pm0.1$\\
And XXVI  & 3.0 & $-261.6^{+3.0}_{-2.8}$   & $8.6^{+2.8(*)}_{-2.2}$  & $0.96^{+0.43}_{-0.34}$ & $325^{+243}_{-225}$ & $-1.8\pm0.5$\\
And XXVII & 1.5 & $-539.6^{+4.7}_{-4.5}$      & $14.8^{+4.3}_{-3.1}$ & $8.3^{+2.8}_{-3.9}$& $1391^{+1039}_{-1128}$ & $-2.1\pm0.5$ \\
And XXVIII & 2.5 &  $-326.2\pm2.7$      & $6.6^{+2.9}_{-2.1}$ & $0.53^{+0.28}_{-0.21}$& $51^{+30}_{-25}$ & $-2.1\pm0.3$ \\
And XXX (Cass II) & 2.0  & $-139.8^{+6.0}_{-6.6}$    & $11.8^{+7.7}_{-4.7}$  & $2.2^{+1.4}_{-0.9}$ &  $308^{+269}_{-219}$ & $-1.7\pm0.4$ \\
\enddata
\tablecomments{$^{(*)}$ - indicates velocity dispersions derived from fewer than 8
  members stars, and require confirmation from further follow-up. }
\end{deluxetable*}

From these masses, it is trivial to estimate the dynamical central
mass-to-light ratios for the objects, $[M/L]_{\rm half}$. We list these values
for each dSph in Table~\ref{tab:kprops}, where the associated uncertainties also take into
account those from the measured luminosities and distances to these dSphs
\citep{mcconnachie12,conn12}, as well as those on the masses measured in this
work. 

\subsection{Comparing the mass-to-light ratios of M31 and Milky Way dSphs}

\begin{figure}
\begin{center}
\includegraphics[angle=0,width=0.95\hsize]{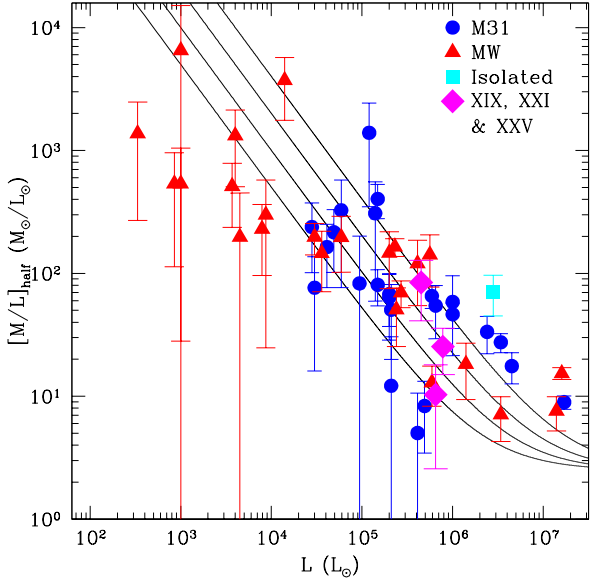}
\caption{Dynamical mass-to-light ratio within the half-light radius,
  $[M/L]_{\rm half}$, as a function of half-light radius for all M31 (blue
  circles), MW (red triangles) and
  isolated dSphs (cyan squares).}
\label{fig:ml}
\end{center}
\end{figure}

By combining our measurements of the kinematic of M31 dSphs in this work with
those from T12 and \citet{tollerud13}, we find ourselves with a set of
kinematic properties as measured for 27 of the 28 Andromeda dSphs (owing to
the difficulties experienced with the And XXIV dataset, we do not include this
object in our subsequent analysis). This near-complete sample allows us to
fully compare the masses and mass-to-light ratios for the M31 satellite system
with those measured in the Milky Way satellites. Before beginning this
analysis, we compile Table~\ref{tab:summary} which presents the kinematics for
the full M31 satellite system, which combines the results from this work, T12,
\citet{kalirai10}, and \citet{tollerud13}. In cases where two measurements for
a dSph exist, we use those that were calculated from larger numbers of likely
members, as these are the more robust. We begin by comparing the mass-to-light
ratios (which indicate the relative dark matter dominance of these objects) of
the two populations as a function of luminosity. In Fig~\ref{fig:ml} we show
these values for all MW (red triangles, with values taken from
\citealt{walker09b}), and M31 (blue circles) dSphs as a function of their
luminosity. We can see that all these objects are clearly dark matter
dominated, excluding And XII and And XVII where we are unable to resolve the
mass with current datasets. We also see that they follow the trend of
increasing $[M/L]_{\rm half}$ with decreasing luminosity, as is seen in their
MW counterparts.

The one tentative exception to this is the And XXV dSph. From our dataset, we
measure a value of $[M/L]_{\rm half}=10.3^{+7.0}_{-6.7}$ for this object,
making it consistent with a stellar population with no dark matter within its
$1\sigma$ uncertainties. This result is surprising and would be of enormous
importance if confirmed with a larger dataset than our catalog of 26 likely
members as it would be the first dSph to be observed with a negligible dark
matter component. And XXV is also one of the members of the recently
discovered thin plane of satellites in Andromeda \citep{ibata13}, and so the
presence or absence of dark matter in And XXV might tell us more about the
origins of this plane which are currently poorly understood.

\subsection{Comparing the masses of M31 and Milky Way dSphs}

Finally, we discuss how the masses for the full sample of Andromeda dSphs for
which kinematic data are available compare with those of the MW dSphs. For the
M31 dSph population, we again use our compilation of kinematic properties
assembled in Table~\ref{tab:summary}. We plot the velocity dispersions, mass
within the half-light radius, and central densities for all M31 (blue circles)
and MW (red triangles, \citealt{walker09b,aden09,koposov11,simon11}) dSphs as
a function of radius. We then overplot the best-fit NFW and cored mass
profiles for the MW, taken from \citet{walker09b}. In general, we see that the
M31 and MW objects are similarly consistent with these profiles, an agreement
that was also noted by T12. However, there are 3 objects which are clear
outliers to these relations. These are And XIX, XXI and XXV, with velocity
dispersions of $\sigma_v=4.7^{+1.6}_{-1.4}\kms,
\sigma_v=4.5^{+1.2}_{-1.0}\kms$ and $\sigma_v=3.0^{+1.2}_{-1.1}\kms$, as
derived in this work. Given their half-light radii ($r_{\rm
  half}=$1481$^{+62}_{-268}$~pc, $r_{\rm half}=$842$\pm77$~pc and $r_{\rm
  half}=$642$^{+47}_{-74}$~pc), one would expect them to have dispersions of
closer to $9\kms$ in order to be consistent with the MW mass profile.  As they
stand, these three objects are outliers at a statistical significance of
$2.5\sigma$, $3.0\sigma$ and $3.4\sigma$ (calculated directly from their
likelihood distributions as presented in
Figs.~\ref{fig:And19},~\ref{fig:And21} and~\ref{fig:And25}). Similarly, in T12
they noted that And XXII and And XIV were outliers in the same respect as And
XIX, XXI and XXV, albeit at a lower significance. These difference can also be
observed in terms of the enclosed masses and densities within $r_{\rm half}$.

\begin{figure*}
  \begin{center}
    \includegraphics[angle=0,width=0.9\hsize]{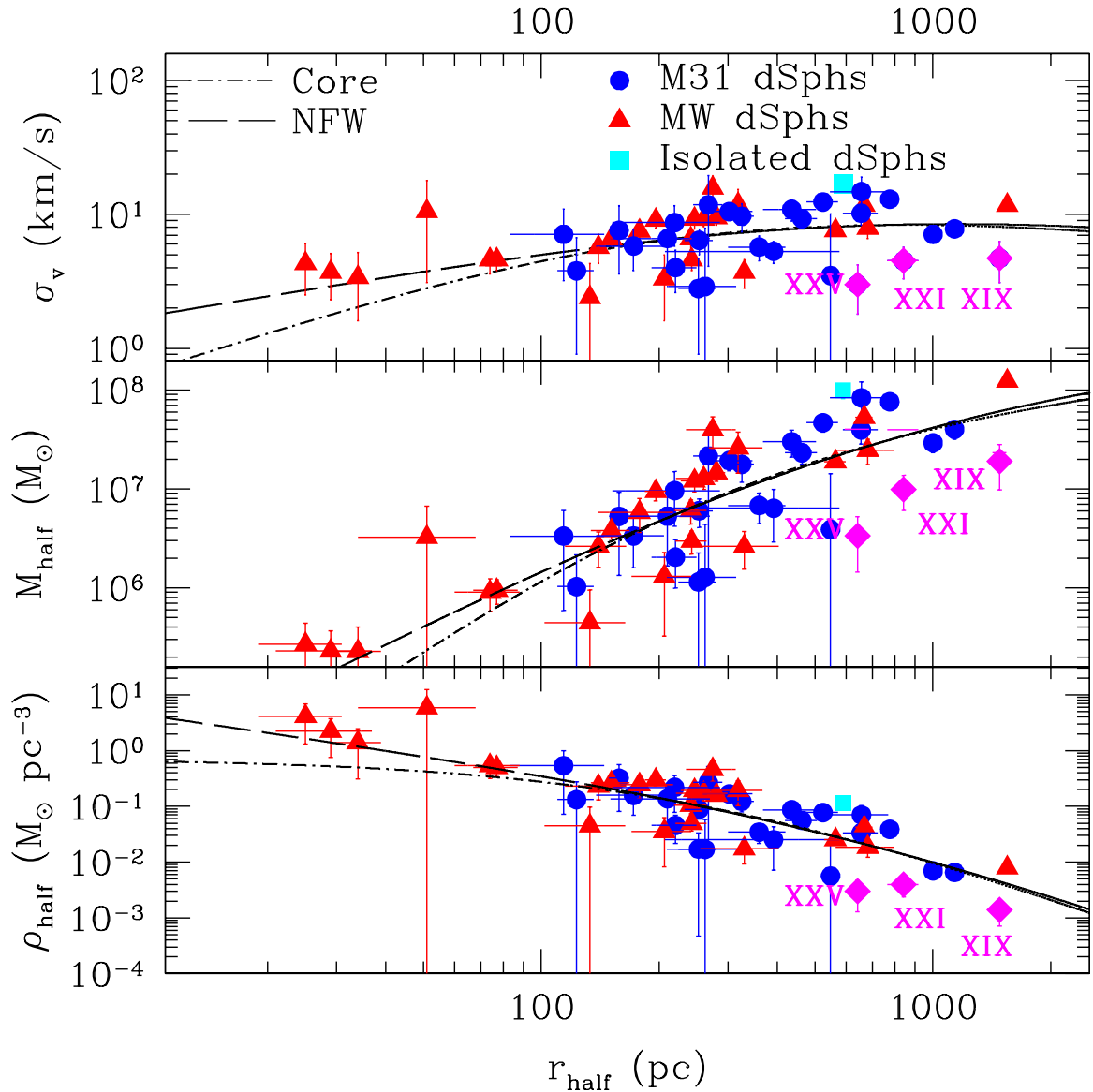}
    \caption{{\bf Top: }Half-light radius ($r_{\rm half}$) vs. velocity dispersion
      for the MW (red triangles) and M31 (blue circles) dSphs. The best fit
      NFW and cored mass profiles to the whole Local Group dSph population are
      over-plotted as dashed and dot-dashed lines respectively. {\bf Middle:} As
      above, but in the $M_{\rm half}-r_{\rm half}$ plane. {\bf Lower:} As above but
      in the $\rho_{\rm half}-r_{\rm half}$ plane.}
   \label{fig:summary}
   \end{center}
 \end{figure*}

In \citet{collins11b} we argued that the low velocity dispersion seen in some
Andromeda dwarfs were a result of tidal forces exerted on their halos by the
host over the course of their evolution, and that this effect was
predominantly seen in dSphs where their half-light radii were more extended
for a given luminosity than expected, such is the case our three outliers, And
XIX, XXI and XXV. This result therefore adds weight to the trend presented in
that work. A number of recent works trying to account for the lower than
predicted central masses of dSph galaxies within the Local Group also support
this notion. For example, \citet{penarrubia10} demonstrated that the presence
of a massive stellar disk in the host galaxy (such as those of the MW and M31)
can significantly reduce the total masses of its associated satellites. In
addition, recent, papers by \citet{zolotov12} and \citet{brooks12}, where the
effect of baryons within dark matter only simulations was measured also find
that tidal forces exerted by host galaxies where a massive disk is present
will serve to reduce the masses of its satellite population at a far greater
rate than hosts without baryons. And XIX, XXI and XXV may thus represent a
population of dSph satellites whose orbital histories about M31 have resulted
in substantial fractions of their central mass being removed by tides. It
should be noted, however, that tides not only reduce the central masses and
densities of dSph halos, they also reduce the spatial size of the luminous
component \citep{penarrubia08b,penarrubia10}, albeit at a slower rate. The
tidal scenario is therefore slightly difficult to reconcile with these
outlying M31 dSphs having the largest sizes, unless they were both more
massive and spatially larger in the past. 

Other recent theoretical works have also shown that the removal of baryons
from the very centres of dark matter halos by baryonic feedback (from star
formation and supernovae, for example) can also help to lower the central
masses and densities of satellite galaxies
(e.g., \citealt{pontzen12,zolotov12,brooks12}). For this method to work
effectively, however, very large `blow outs' of gas are required, of the order
$\sim10^8-10^9\msun$, equivalent to $\sim40000$ SNe. This would require a
minimum initial satellite luminosity of $M_V<-12$
\citep{zolotov12,garrison13}, significantly brighter than the current
luminosities of our outliers ($M_V\sim-10$). Therefore, if feedback has indeed
played a role in the shaping of the dark matter halos of And XIX, XXI and XXV,
one assumes it would have to have operated in tandem with tidal stripping. We
will discuss the implications and interpretation of these result further in a
companion paper (Collins et al. in prep).

\section{Metallicities}
\label{sect:metals}

Our observational setup was such that we cover the calcium triplet
region (Ca II) of all our observed stars. This strong, absorption
feature is useful not only for calculating velocities for each star,
but also metallicities. For RGB stars, such as we have observed, there
is a well known relation between the equivalent widths (EWs) of the Ca
II lines, and the iron abundance, $\feh$, of the object. The
calibration between these two values has been studied and tested by
numerous authors, using both globular clusters and dSphs, and is valid
down to metallicities as low as $\feh\sim-4$ (see
e.g., \citealt{battaglia08,starkenburg10}). Following the
\citet{starkenburg10} method, which extends the sensitivity of this method down
to as low as $\feh\sim-4$, we fit Gaussian functions to the three Ca
II peaks to estimate their equivalent widths (EWs), and calculate
[Fe/H] using equation~\ref{eqn:cat}:

\begin{equation}
\begin{aligned}
  \feh=-2.87+0.195(V_{RGB}-V_{HB})+0.48\Sigma \rm{Ca}\\-0.913\Sigma
  \rm{Ca}^{-1.5}+0.0155\Sigma \rm{Ca}(V_{RGB}-V_{HB})
\end{aligned}
\label{eqn:cat}
\end{equation}

\noindent where $\Sigma$Ca=0.5EW$_{8498}$+EW$_{8542}$+0.6EW$_{8662}$,
$V_{RGB}$ is the magnitude (or, if using a composite spectrum, the average,
S:N weighted magnitude) of the RGB star, and $V_{HB}$ is the mean
$V$-magnitude of the horizontal branch (HB). Using $V_{HB}-V_{RGB}$ removes
any strong dependence on distance or reddening in the calculated value of
[Fe/H], and gives the Ca II line strength at the level of the HB. For M31, we
set this value to be $V_{HB}=$25.17 \citep{holland96}\footnote{This assumed
  value is sensitive to age and metallicity effects, see \citealt{chen09} for
  a discussion, however owing to the large distance of M31, small differences
  in this value within the M31 system will have a negligible effect on
  metallicity calculations}. As the dSphs do not all sit at the same distance
as M31, assuming this introduces a small error into our calculations, but it
is at a far lower significance than the dominant uncertainty introduced by the
noise within the spectra themselves. For individual stars, these measurements
carry large uncertainties ($\gta0.4$ dex), but these are significantly reduced
when stacking the spectra into a composite in order to measure an average
metallicity for a given population.

\begin{figure*}
\begin{center}
\includegraphics[angle=0,width=0.45\hsize]{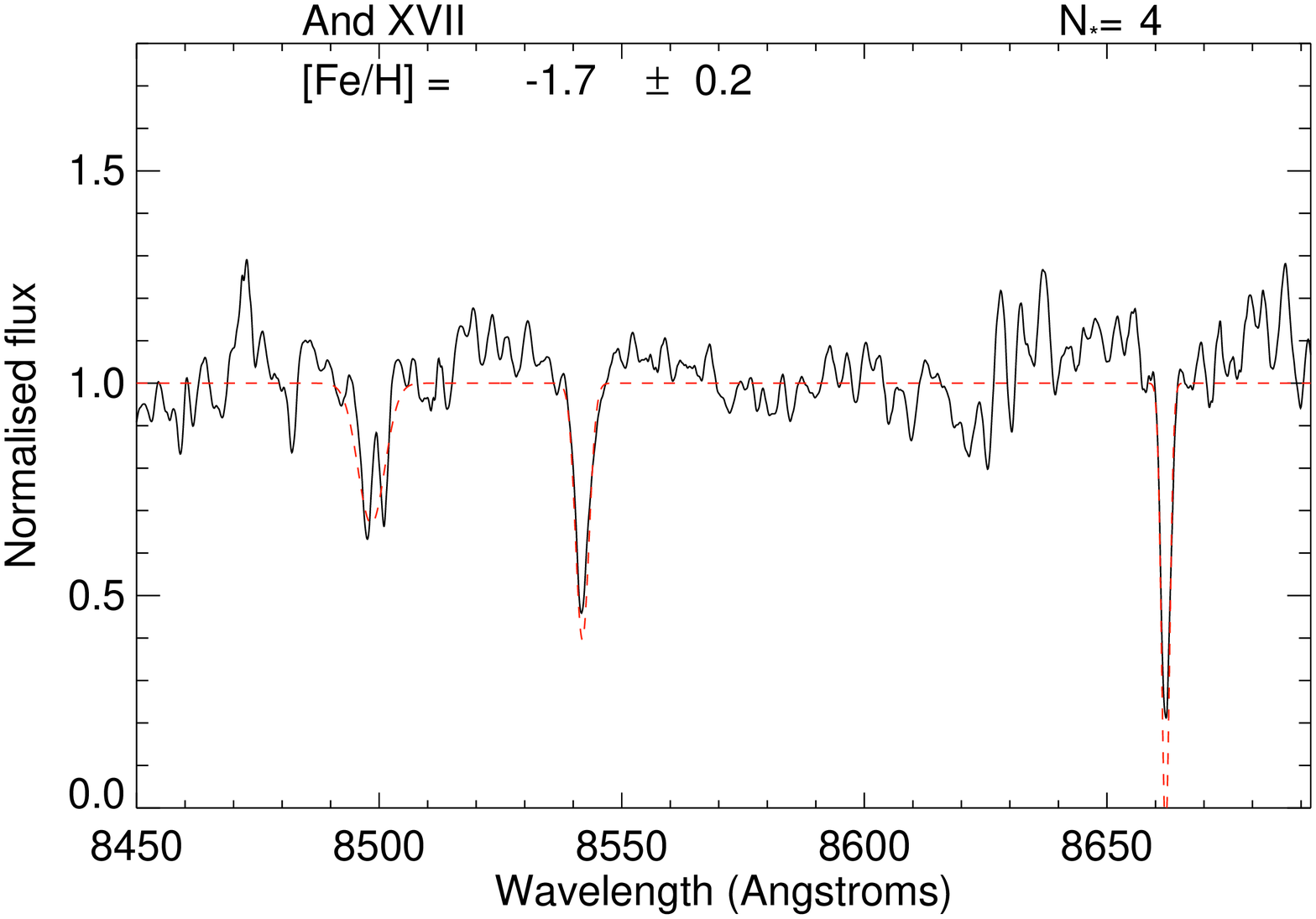}
\includegraphics[angle=0,width=0.45\hsize]{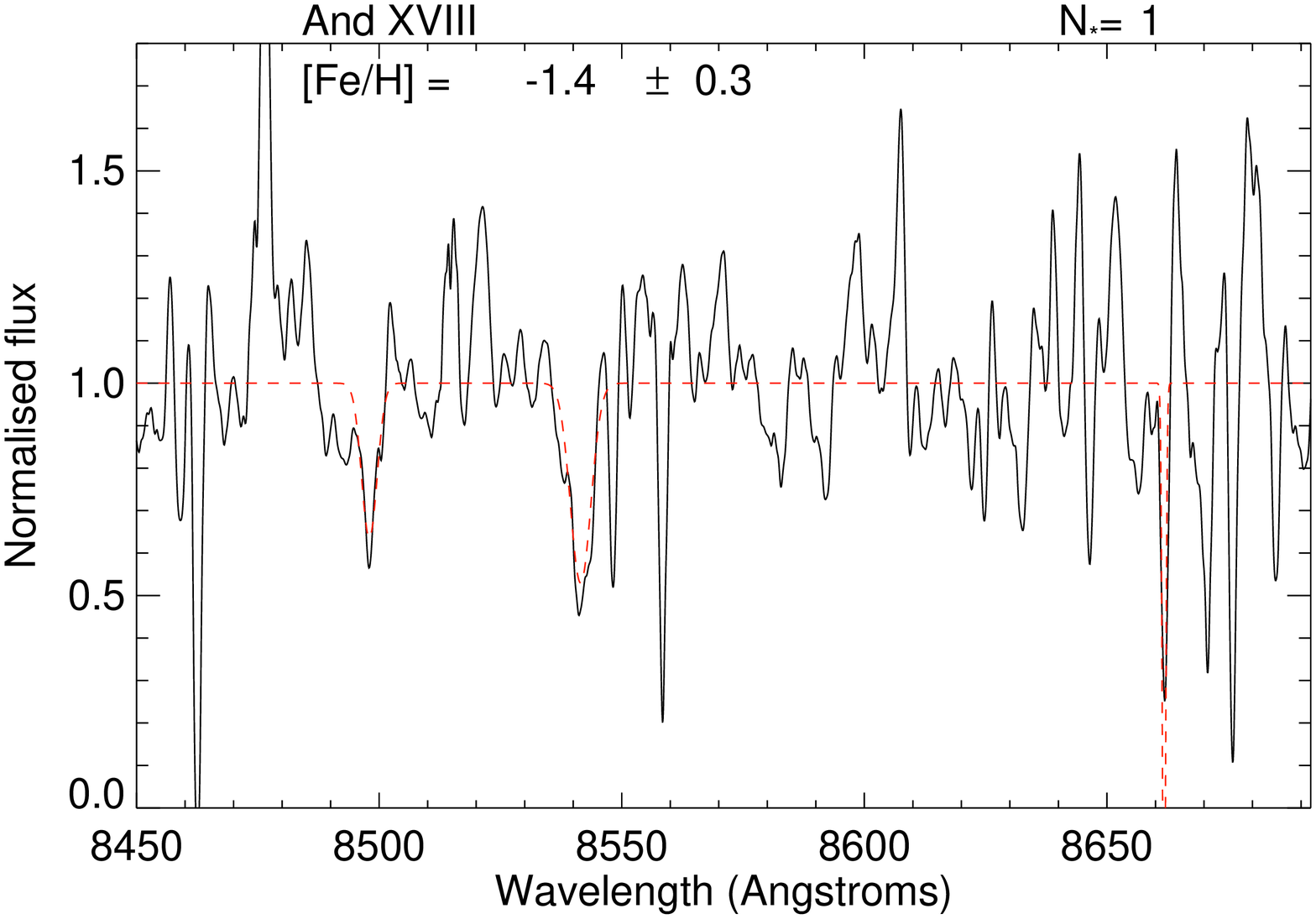}
\includegraphics[angle=0,width=0.45\hsize]{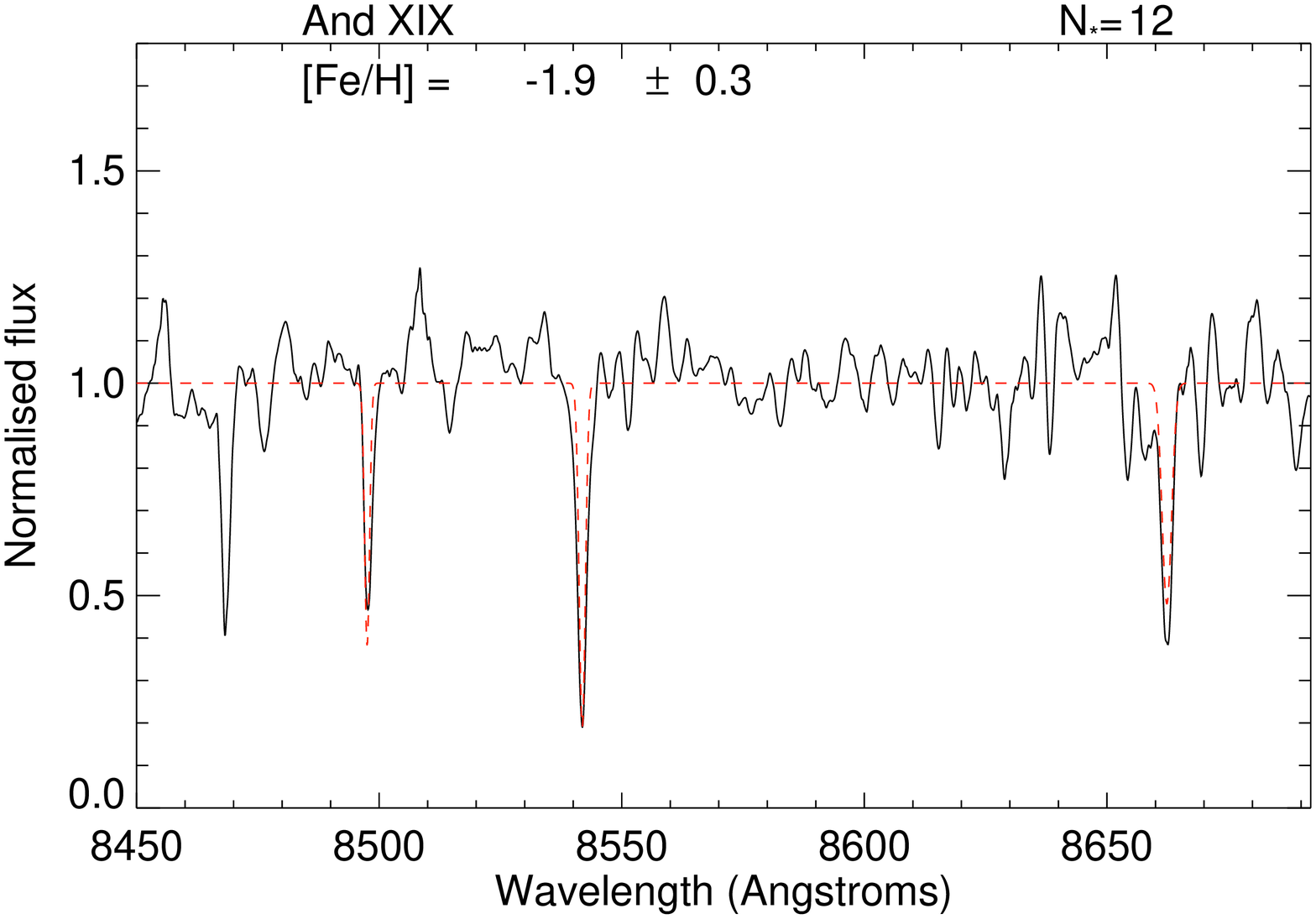}
\includegraphics[angle=0,width=0.45\hsize]{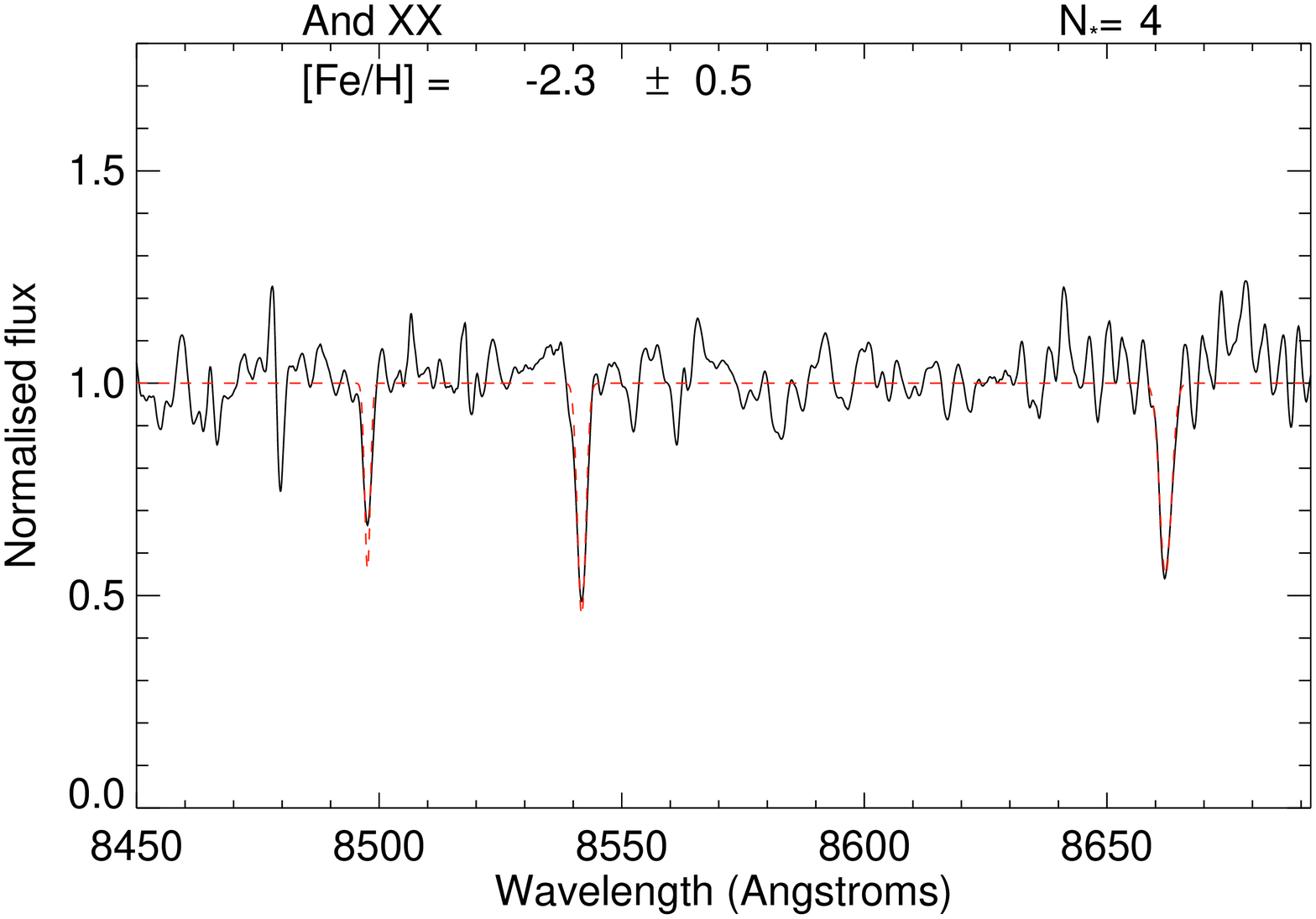}
\includegraphics[angle=0,width=0.45\hsize]{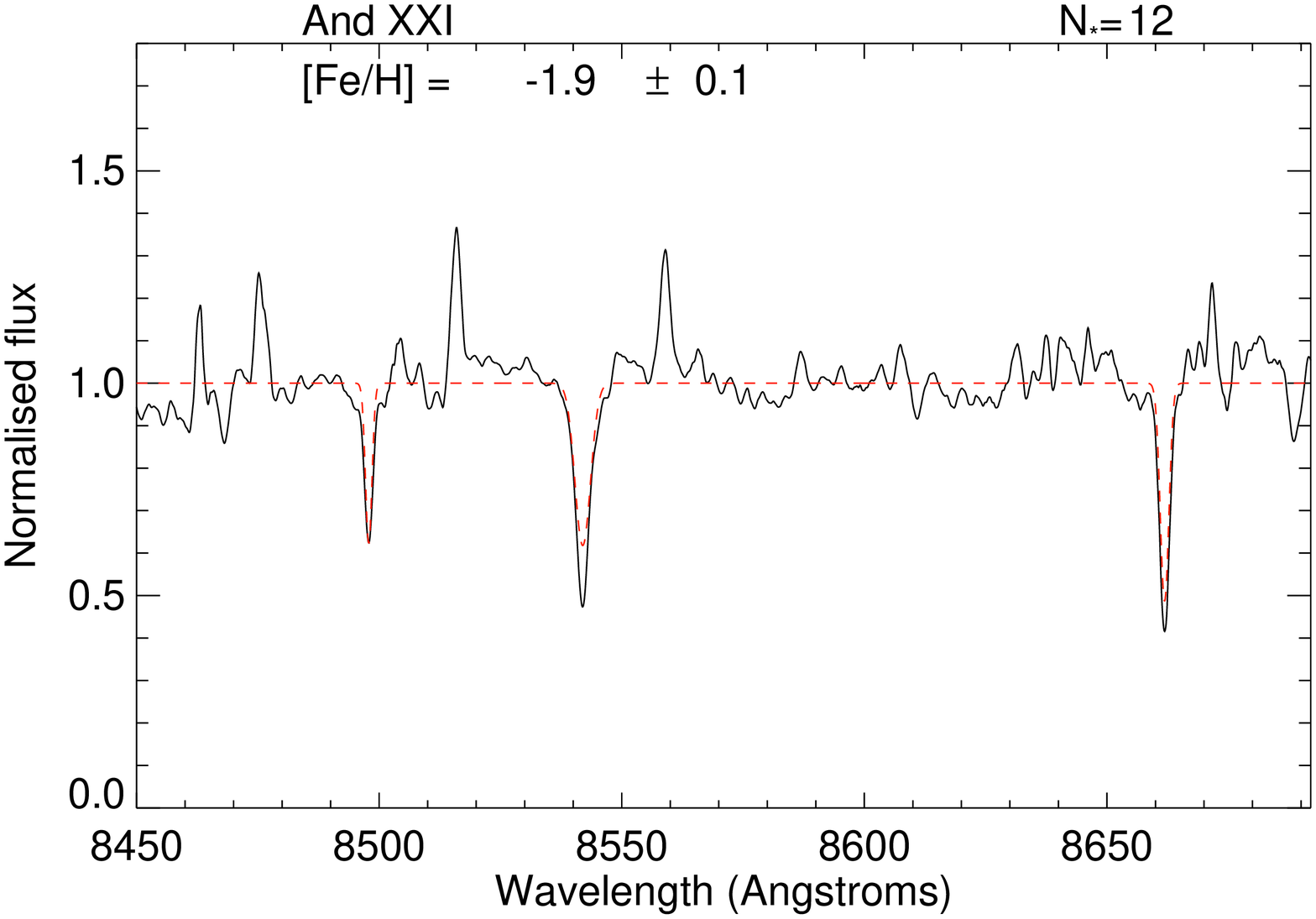}
\includegraphics[angle=0,width=0.45\hsize]{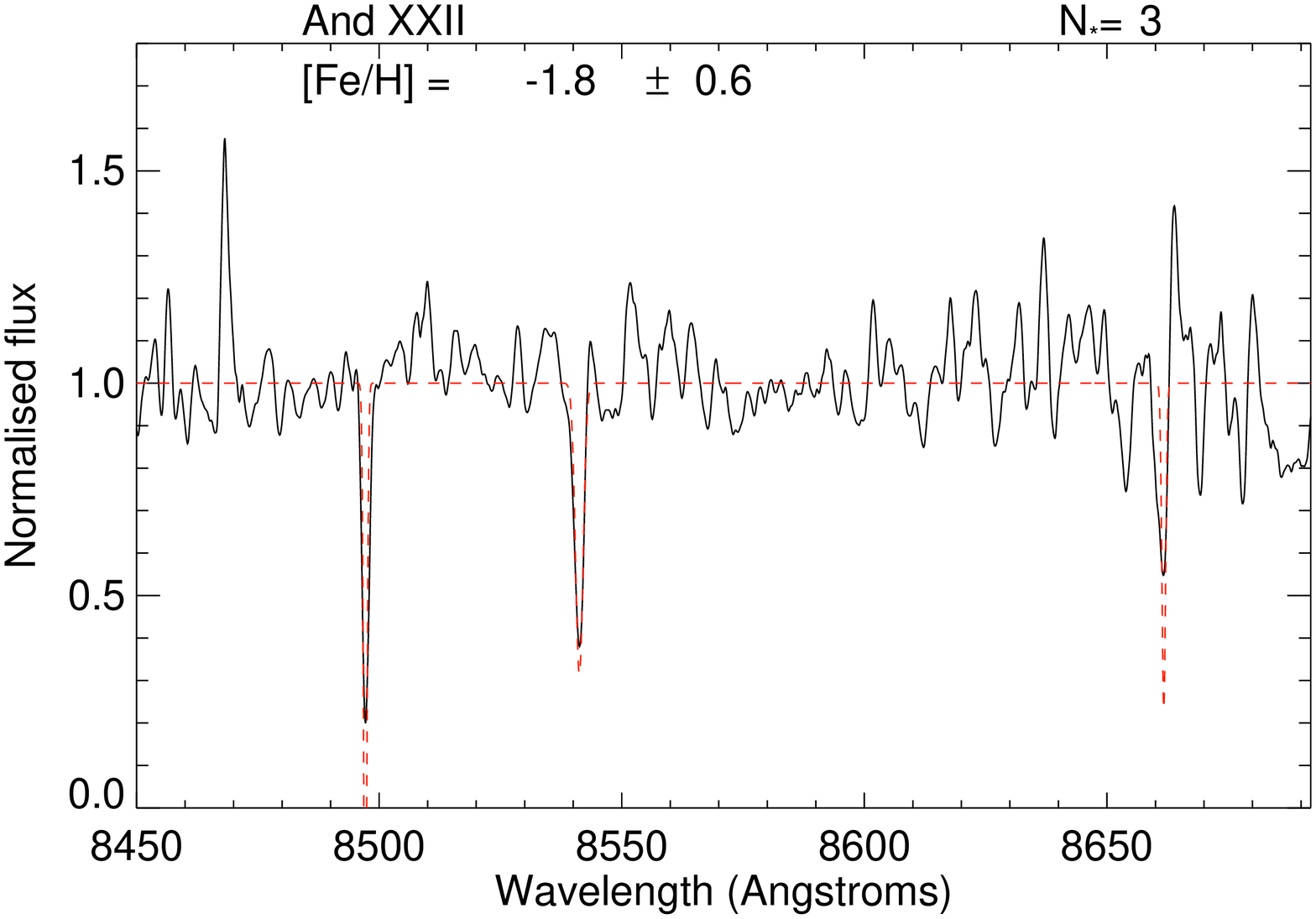}
\includegraphics[angle=0,width=0.45\hsize]{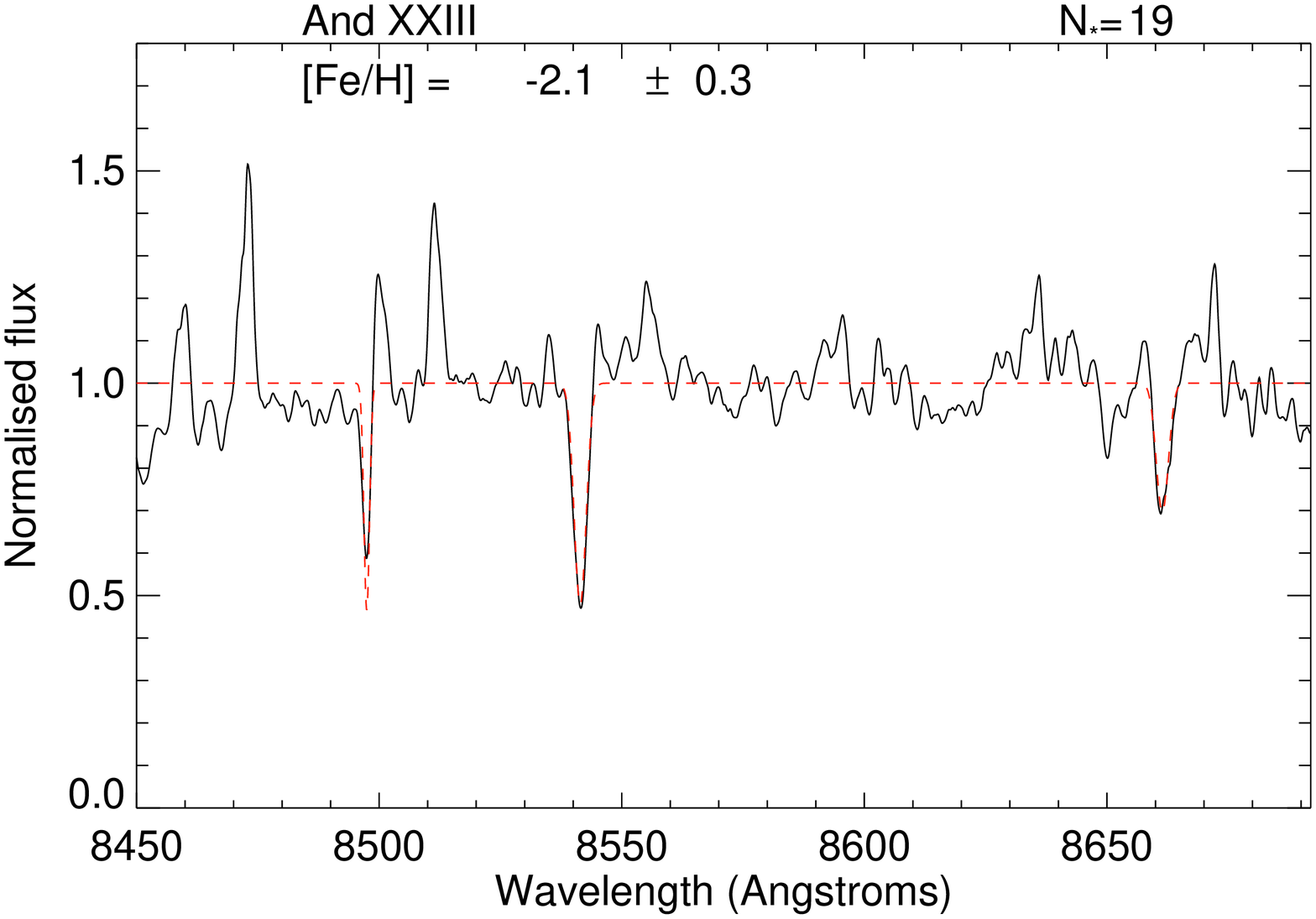}
\includegraphics[angle=0,width=0.45\hsize]{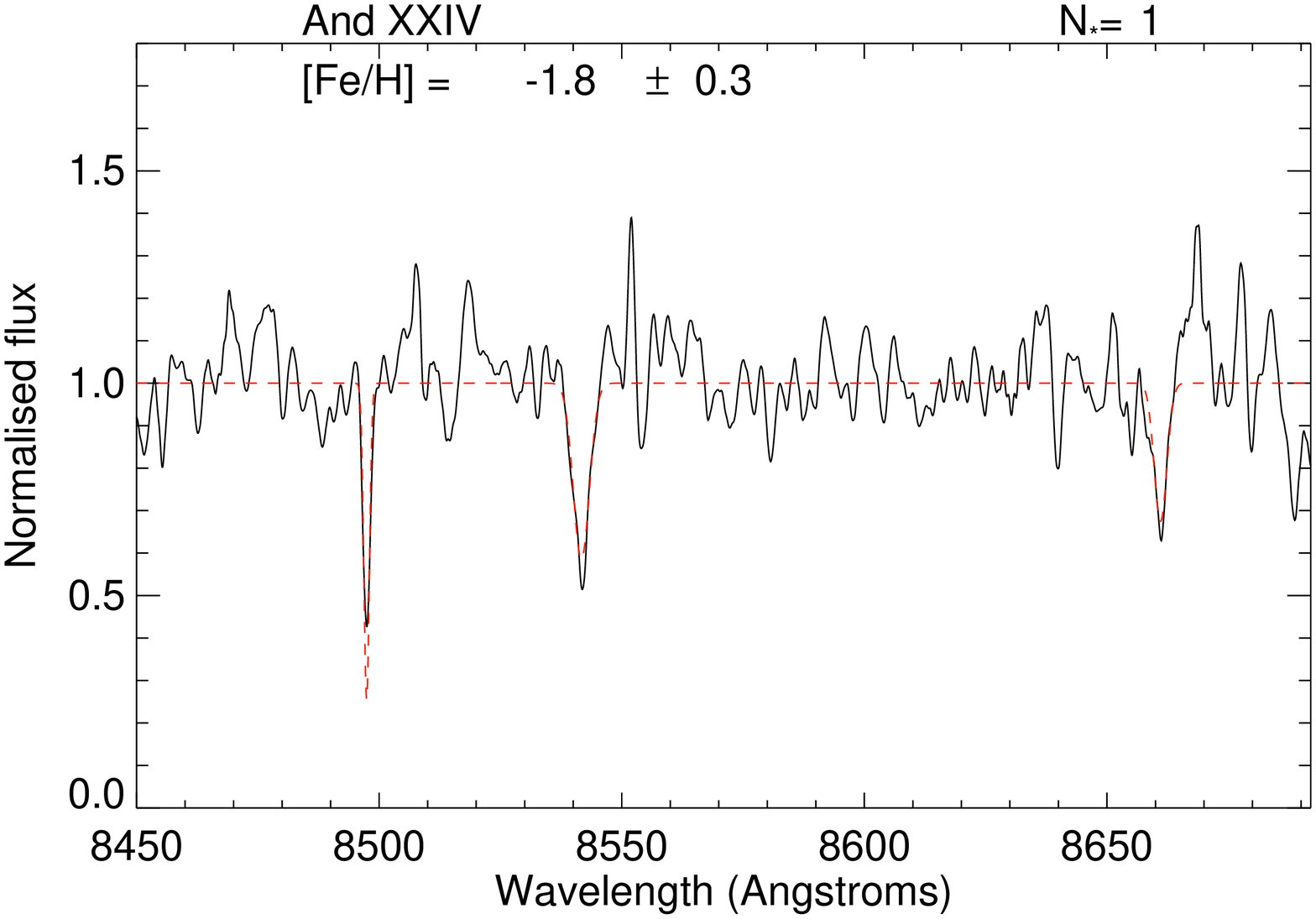}
\caption{Figures of the S:N weighted composite spectra for each dSph. The average
  metallicity for each dSph, plus the associated uncertainty is printed on
  each panel (continued over page)}
\label{fig:sumspec}
\end{center}
\end{figure*}

\begin{figure*}
\begin{center}
\includegraphics[angle=0,width=0.45\hsize]{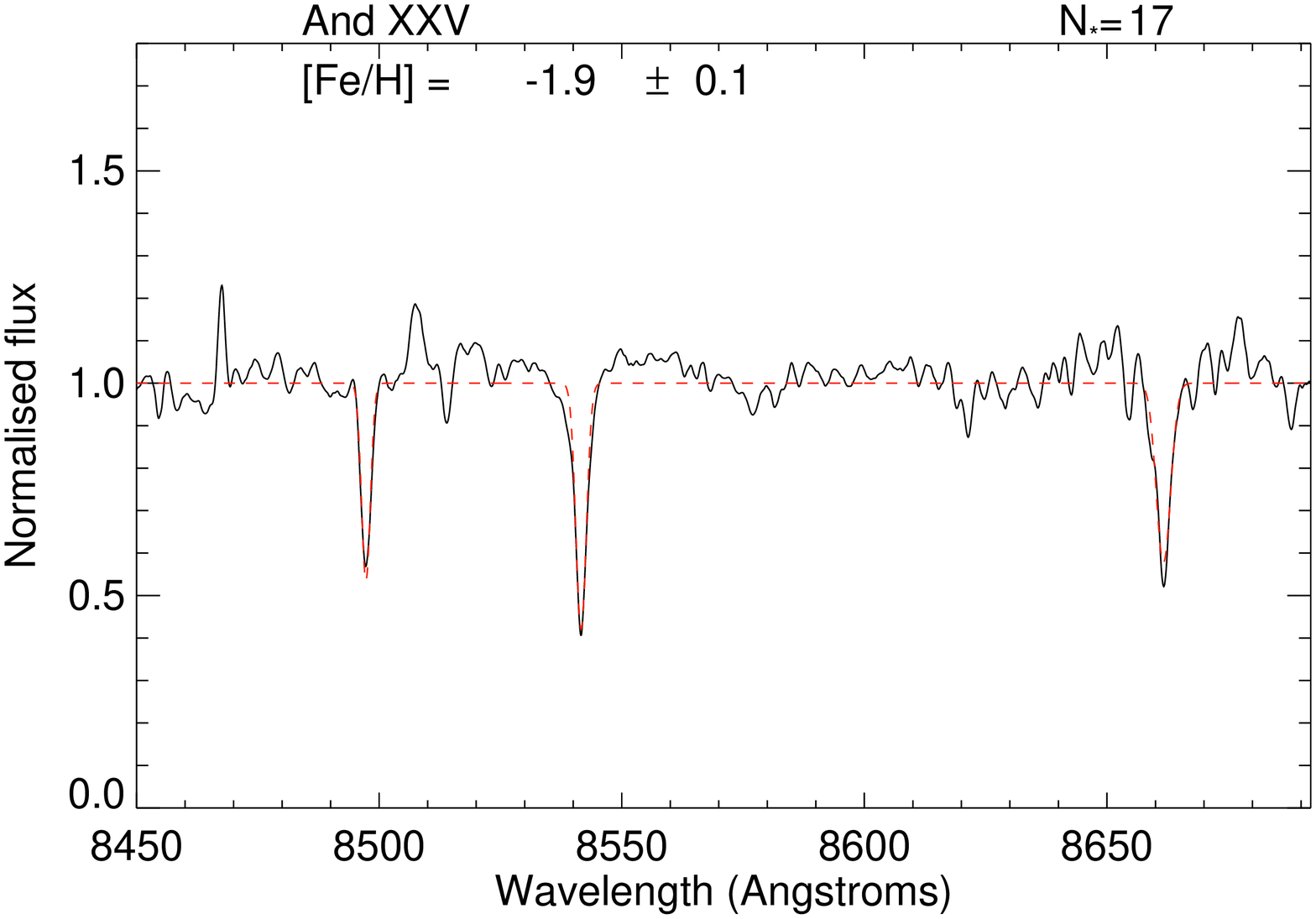}
\includegraphics[angle=0,width=0.45\hsize]{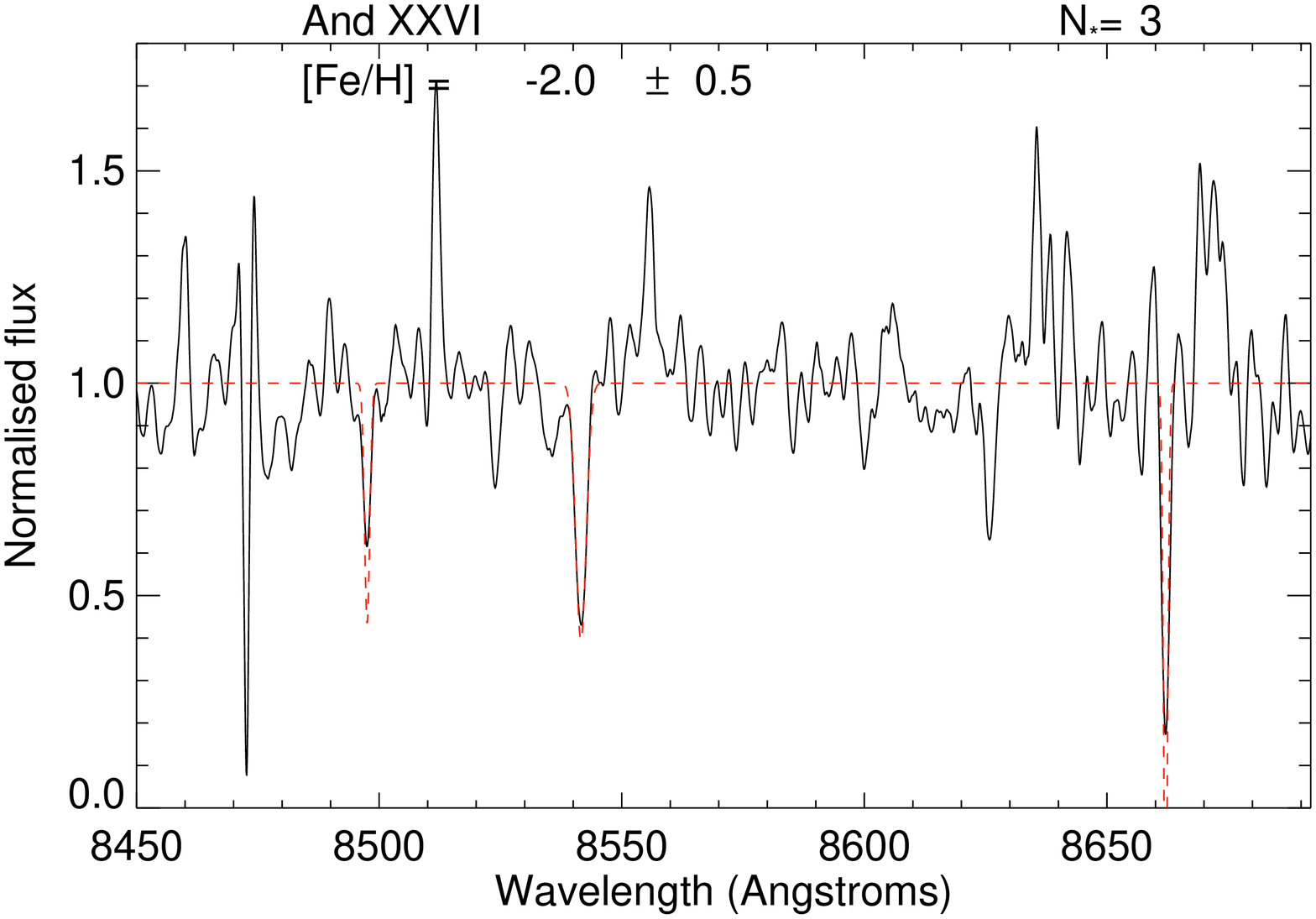}
\includegraphics[angle=0,width=0.45\hsize]{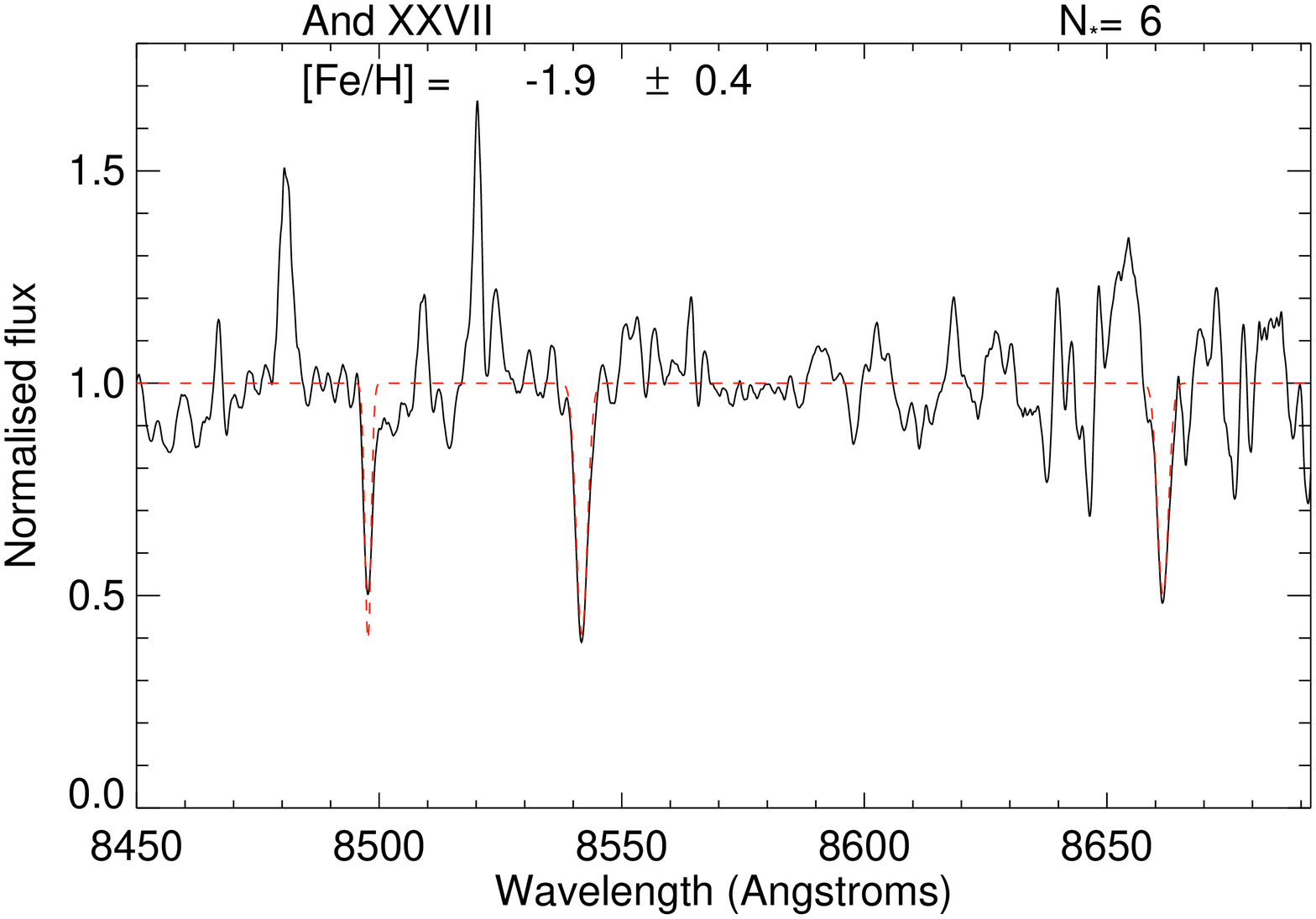}
\includegraphics[angle=0,width=0.45\hsize]{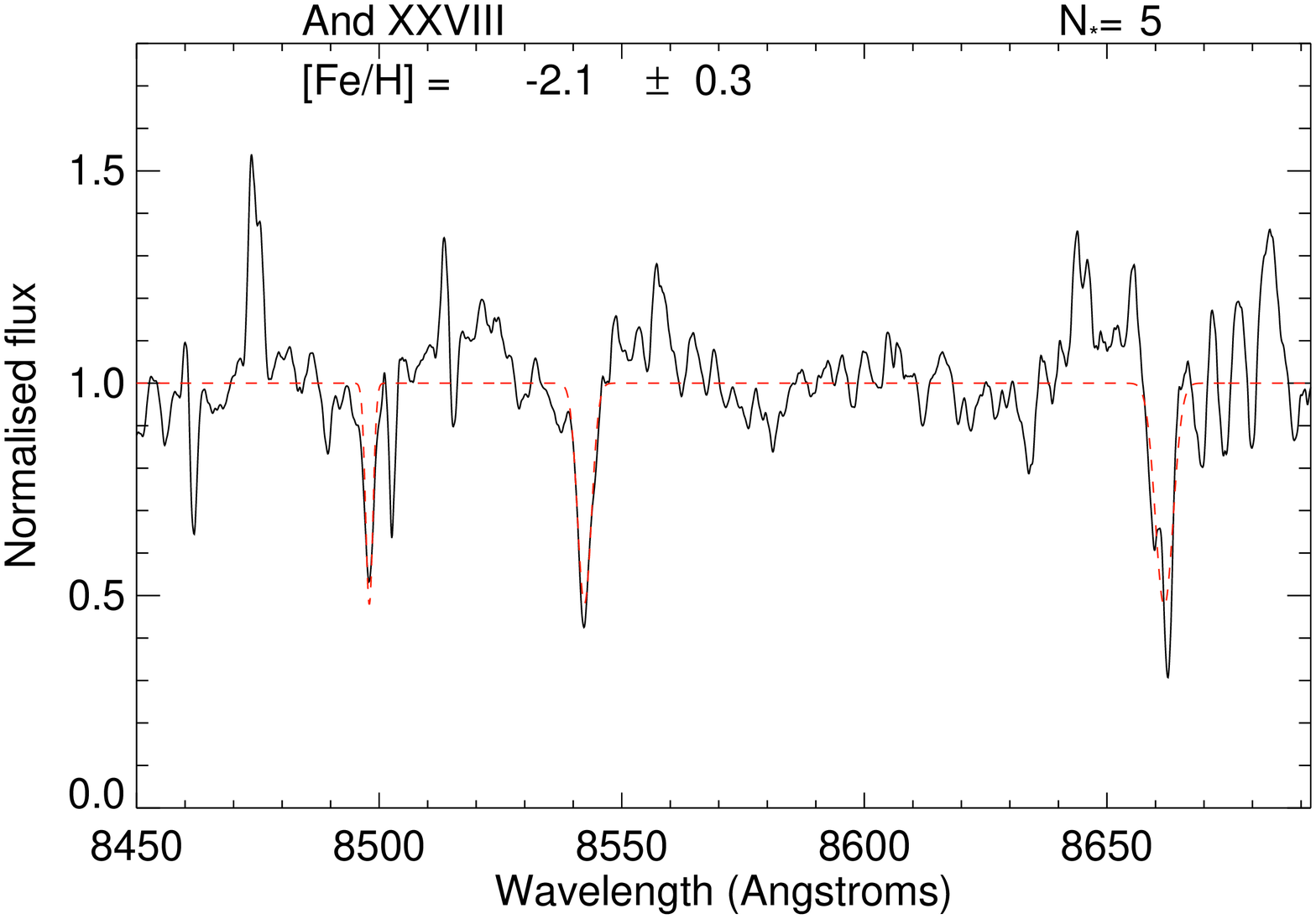}
\includegraphics[angle=0,width=0.45\hsize]{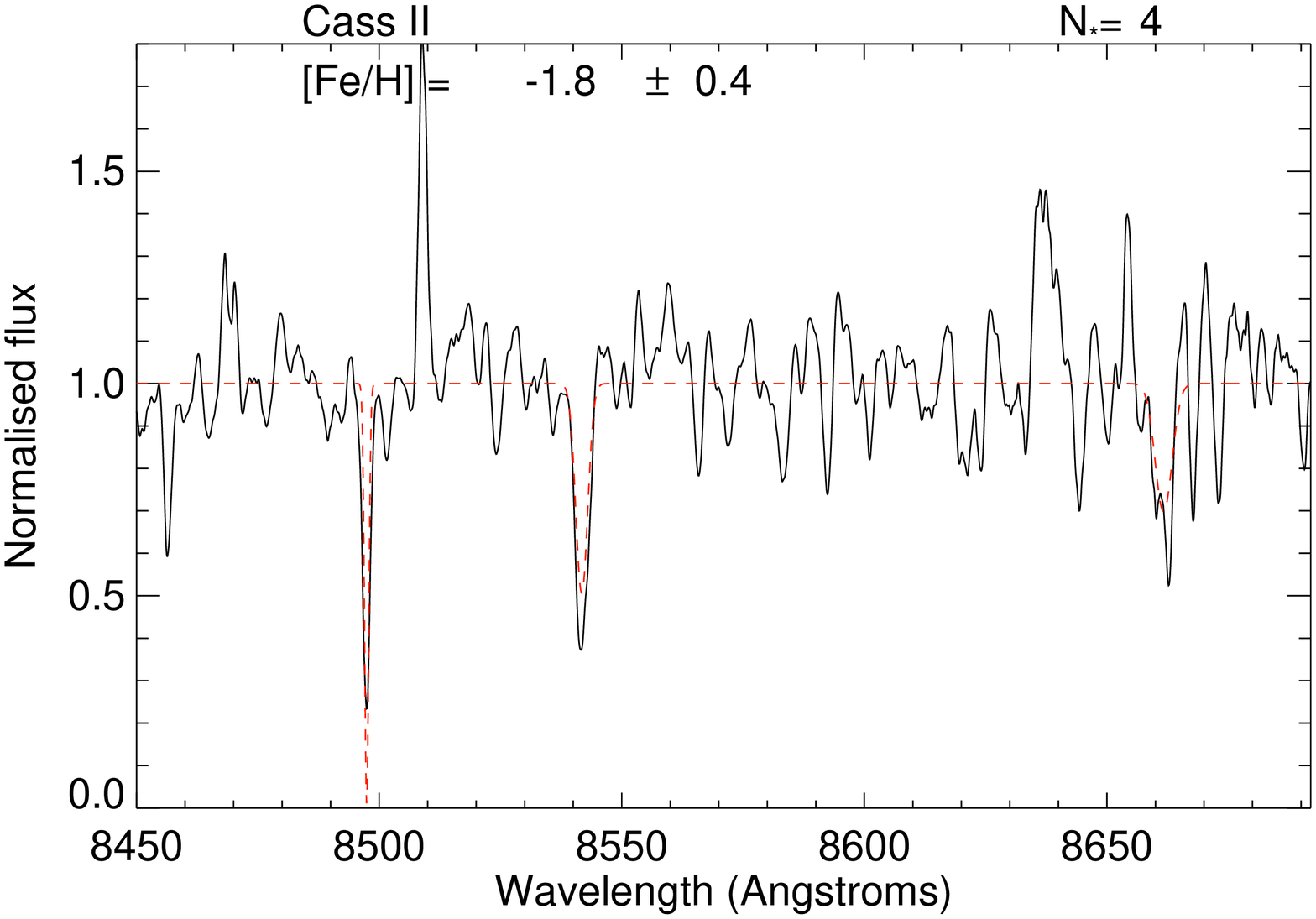}

\caption{Figures of the S:N weighted composite spectra for each dSph (excluding And XV and And XVI). The average
  metallicity for each dSph, plus the associated uncertainty is printed on
  each panel (continued over page)}
\label{fig:sumspec2}
\end{center}
\end{figure*}

Uncertainties on the individual measurements of [Fe/H] from our stellar
spectra are typically large ($\ge0.5$~dex), so for a more robust determination
of the average metallicities we co-add the spectra for each dSph (weighting by
the S:N of each individual stellar spectrum, which is required to be a minimum
of 2.5\AA$^{-1}$) and measure the resulting EWs. In a few cases, not all 3 Ca
II lines are well resolved. For And V, IX, XVII, XVIII XXVI and XXVIII, the
third Ca II line is significantly affected by skylines, whilst for And XXIV,
the first Ca II line is distorted. In the case of And XIII, only the second Ca
II line appears well resolved. In these cases, we neglect the affected lines
in our estimate of [Fe/H], and derive reduced equivalent widths from the
unaffected lines. Where the third line is affected, this gives
$\Sigma$Ca=1.5EW$_{8498}$+EW$_{8542}$. Where the second line is affected, we
find $\Sigma$Ca=EW$_{8542}$+EW$_{8662}$. Finally, where only the second line
seems reliable we use $\Sigma$Ca=1.7EW$_{8542}$. These coefficients are
derived empirically from high S:N spectra where the absolute values of [Fe/H]
are well known. We test these variations of $\Sigma$Ca by applying them to our
high S:N co-added spectra where all three lines are well resolved, such as And
XXI and XXV, and we find that all three formulae produce consistent values of
[Fe/H]. The composite spectra for each satellite are shown in
Figs.~\ref{fig:sumspec} and ~\ref{fig:sumspec2}. In all cases, we find that
our results are consistent with photometric metallicities derived in previous
works.


\begin{figure}
\begin{center}
\includegraphics[angle=0,width=0.95\hsize]{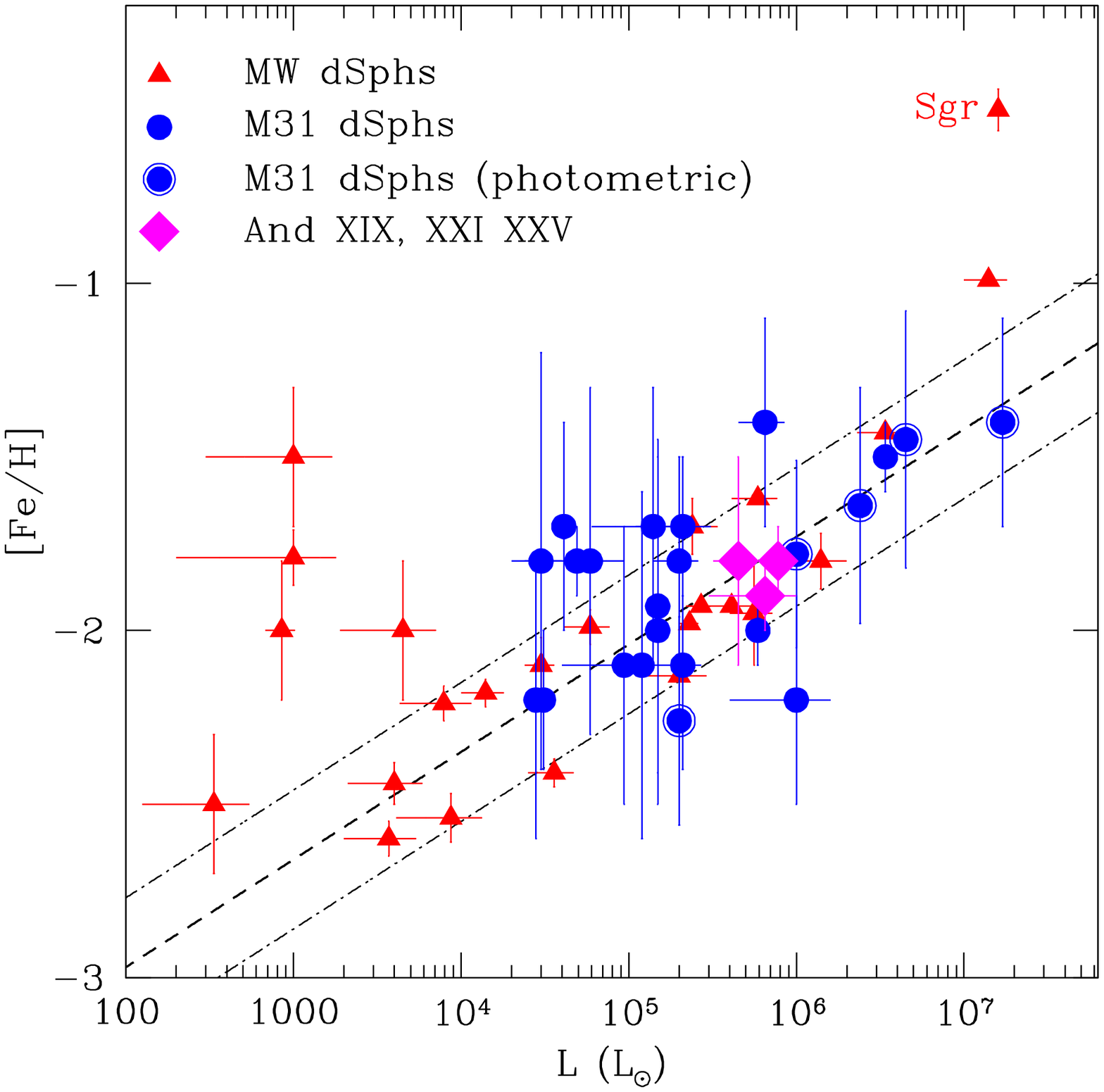}
\caption{Spectroscopically derived [Fe/H] vs. luminosity for all MW (red
  triangles, taken from \citealt{kirby11}, with additional measurements taken
  from \citealt{martin07,belokurov09,koch09}) and M31 dSphs (blue circles,
  this work). The solid line represents the best fit relationship between
  these two parameters as taken from \citet{kirby11}. The dashed lines
  represent the $1\sigma$ scatter about this relationship. We see that the M31
  dSphs follow this relationship very well within their associated
  uncertainties. As we discuss in \S~\ref{sect:metals}, those galaxies with
  $L<10^6\lsun$ are also consistent with having a constant metallicity, which
  could indicate a metallicity floor in these fainter systems.}
\label{fig:mvfeh}
\end{center}
\end{figure}

In the MW, it has been observed that the average metallicities of the dSph
population decrease with decreasing luminosity
(e.g., \citealt{kirby08,kirby11}). In Fig.~\ref{fig:mvfeh}, we plot the
spectroscopic metallicities of the M31 dSphs (blue dSphs) as a function of
absolute magnitude. We plot the MW dSphs as red triangles
(\citealt{martin07,kirby08,kirby11,belokurov09,koch09}). We also include those
M31 dSphs for which only photometric measurements of [Fe/H] are available (And
I, II, III, VII and XIV, \citealt{kalirai10,tollerud12}), and these are
highlighted as encircled blue points. The dashed line represents the best-fit
to the MW dSph population from \citet{kirby11}. The three relatively metal
rich ([Fe/H]$\sim-1.5$ to $-2.0$), faint ($L\sim1000\lsun$) MW points are the
three ultra faint dSphs Willman I, Bo\"otes II and Segue 2, and these three
were not included in the \citet{kirby11} analysis, where the best fit MW
relation was determined. We see that the metallicities for a given luminosity
in the M31 dSphs also loosely define a relationship of decreasing metallicity
with decreasing luminosity, and they agree with that defined by their MW
counterparts within their associated uncertainties. However, it is also
noteworthy that for dSphs with $L<10^6\lsun$, the Andromeda satellites are
also consistent with having a constant metallicity of $\sim-1.8$. The same
levelling off of average metallicity at lower luminosities was noted by
\citet{mcconnachie12}, where they note that this break occurs at the same
luminosity as a break in the luminosity-surface brightness relation for faint
galaxies. As such, it could imply that the denisty of baryons in these
systems, rather than the total number of baryons, could be the most important
facor in determining their chemical evolution. The error bars we present here
are still significant, so it is hard to fully interpret this result, but the
hint of a metallicity floor in these lower luminosity systems is intriguing.

In Fig.~\ref{fig:mvfeh}, we highlight the positions of our three kinematic
outliers, And XIX, XXI and XXV, and we see that they fall almost exactly on
the MW relation. In this figure, systems that have experienced extreme tidal
stripping would move horizontally to the left, as their luminosity would
gradually decrease as stars are stripped, but their chemistry would remain
unaffected. One would expect to see such behaviour only after the stellar
component began to be removed in earnest, after the majority of the dark
matter halo had been removed. If their central densities were lowered by some
active feedback mechanism, such as SNII explosions (e.g., \citealt{zolotov12}),
one would expect the objects to become more enriched and perhaps brighter,
moving them up and to the right, potentially allowing them to remain on the MW
relation. To confirm that this was the case for And XIX, XXI and XXV, we would
require more information on the abundances of these objects and their star
formation histories, which we do not currently possess.

\begin{deluxetable*}{lccccccccccc}
\tabletypesize{\footnotesize}
\tablecolumns{10} 
\tablewidth{0pt}
\tablecaption{Compilation of kinematic and structural properties of all known
Andromeda dSphs. Half-light radii are taken from \citet{mcconnachie12}, and
updated to the distances provided in \citet{conn12}.
\label{tab:summary}} \tablehead{ \colhead{Name} & \colhead{$M_V$} &
\colhead{$N_*^{(a)}$} & \colhead{[Fe/H]$_{phot}$} & \colhead{[Fe/H]$_{spec}$}&
\colhead{$v_r$} & \colhead{$\sigma_v$} & \colhead{$r_{\rm half}$} &
\colhead{D} & \colhead{Ref.}\\ & & & \colhead{(dex)} &
\colhead{(dex)} & \colhead{($\kms$)} &\colhead{ ($\kms$) } & \colhead{(pc)} &
\colhead{ (kpc)} &} \startdata 
And I& -11.8 & 80 & -1.45$\pm$0.37 & N/A &-376.3$\pm2.2$ & 10.2$\pm1.9$&$656^{+68}_{-67}$ & 727$^{+18}_{-17}$ &(1),(2),(3)\\  
And II& -12.6 & 95 &-1.64$\pm$0.34 & N/A &-193.6$\pm1.0$ & 7.3$\pm$ 0.8&
1136$\pm46$ & 630$\pm15$ &(1),(3),(4)\\ 
And III& -10.2 & 43 & -1.78$\pm$0.27 & N/A &-344.3$\pm1.7$ & 9.3$\pm1.4$&
$463^{+44}_{-45}$ & 723$^{+18}_{-24}$&(1),(2),(3)\\  
And V& -9.6 & 85 & -1.6$\pm0.3$ & -1.8$\pm$ 0.2& -397.3$\pm1.5^{(b)}$ &
10.5$\pm1.1^{(b)}$& $302\pm44$ & 742$^{+21}_{-22}$ &(2),(3),(5)\\ 
And VI& -11.5 & 38 & -1.3$\pm0.3$ & -1.5$\pm$ 0.3& -339.8$\pm1.9$ & 12.4$^{+1.5}_{-1.3}$& $524\pm49$ & 783$\pm28$ &
(3),(5),(6)\\ 
And VII& -13.3 & 18 & -1.4$\pm$0.3 & N/A &-307.2$\pm1.3$ & 13.0$\pm$ 1.0&
$776\pm42$ & $762\pm35$ &(1),(2),(3)\\ 
And IX& -8.1 & 32 & -2.2$\pm0.2$ & -1.9$\pm$ 0.6&-209.4$\pm2.5^{(b)}$ &
10.9$\pm2.0^{(b)}$& $436^{+68}_{-24}$ &$600^{+91}_{-23}$ & (1),(2),(3)\\ 
And X& -8.1 & 22 &-1.93$\pm$0.48 & N/A & -164.1$\pm1.7$ & 6.4$\pm1.4$&
$253^{+21}_{-65}$ &$670^{+24}_{-39}$ & (1),(2),(3)\\ 
And XI& -6.9 & 5 &-2.0$\pm0.2$ & -2.0$\pm$ 0.3 & -427.5$^{+3.4}_{-3.5}$ &
7.6$^{+4.0}_{-2.8}$&$158^{+9}_{-23}$ & $763^{+29}_{-106}$& (3),(6),(7)\\ 
And XII&-6.4 & 8 & -1.9$\pm0.2$ & -2.0$\pm$ 0.3 & -557.1$\pm1.7$ &
0.0$^{+4.0}$&$324^{+56}_{-72}$ & $928^{+40}_{-136}$ & (3),(6),(7)\\ 
And XIII& -6.7 & 12 & -2.0$\pm0.2$ & -1.9$\pm$ 0.7 & -185.4$\pm2.4^{(b)}$ &5.8$\pm2.0^{(b)}$ & $172^{+34}_{-39}$ & $760^{+126}_{-154}$ &(2),(3),(7)\\ 
And XIV & -8.3 & 38 & -2.26$\pm$0.3 & N/A & -480.6$\pm1.2$ &5.3$\pm$ 1.0 &
$392^{+185}_{-205}$ & $793^{+23}_{-179}$ &(1),(2),(3)\\ 
And XV & -9.4 & 29 & -1.1 & N/A &-323$\pm1.4^{(b)}$ & 4.0$\pm1.4^{(b)}$ &
$220^{+29}_{-15}$ & $626^{+79}_{-35}$& (1),(2),(3) \\ 
And XVI & -9.4 & 7 & -1.7 & -2.0$\pm$0.5 &-367.3$\pm2.8^{(b)}$ &
3.8$\pm2.9^{(b)}$& $123^{+13}_{-10}$ &$476^{+44}_{-29}$ & (1),(2),(3)\\ 
And XVII& -8.5 & 7 & -1.9 &-1.7$\pm0.3$ & -251.6$^{+1.8}_{-2.0}$ &
2.9$^{+2.2}_{-1.9}$ &$262^{+53}_{-46}$ &$727^{+39}_{-25}$ & (1),(3),(6)\\ 
And XVIII&-9.7 & 22 & -1.8$\pm0.5$ & N/A &-332.1$\pm2.7^{(b)}$ &
9.7$\pm2.3^{(b)}$ &$325\pm24$ & $1214^{+40}_{-43}$ & (1),(2),(3)\\ 
And XIX & -9.3& 27 & -1.9$\pm0.4$ & -1.9$\pm0.6$ & -111.6$^{+1.6}_{-1.4}$ &
4.7$^{+1.6}_{-1.4}$ &$1481^{+62}_{-268}$ & $821^{+32}_{-148}$ & (1),(3),(6)\\ 
And XX& -6.3 & 4 & -1.5$\pm0.5$ & -2.3$\pm0.8$ & -456.2$^{+3.1}_{-3.6}$
&7.1$^{+3.9}_{-2.5}$ &$114^{+31}_{-12}$ & $741^{+42}_{-52}$&(1),(3),(6)\\ 
And XXI & -9.9 & 32 & -1.8 & -1.8$\pm0.4$ &-362.5$\pm0.9$ & 4.5$^{+1.2}_{-1.0}$ &
$842\pm77$ & $827^{+23}_{-25}$ &(1),(3),(6)\\ 
And XXII& -6.5 & 12 & -1.8 &-1.85$\pm$ 0.1&-129.8$\pm2.0$& 2.8$^{+1.9}_{-1.4}$
& $252^{+28}_{-47}$ &$920^{+32}_{-139}$ & (1),(2),(3),(8)\\ 
And XXIII &-10.2 & 42&-1.8$\pm0.2$ & -2.3$\pm0.7$ & -237.7$\pm1.2$ &
7.1$\pm1.0$&$1001^{+53}_{-52}$ & $748^{+31}_{-21}$ & (1),(3),(6)\\ 
AndXXIV& -7.6 & 3 &-1.8$\pm0.2$ & -1.8$\pm0.3$ & $-128.2\pm5.2^{(c)}$
&$0.0^{+7.3(c)}$ & $548^{+31}_{-37}$ & $898^{+28}_{42}$ &(1),(3),(6)\\ 
And XXV & -9.7 & 25 &-1.8$\pm0.2$ &-2.1$\pm0.2$ &-107.8$\pm1.0$ &
3.0$^{+1.2}_{-1.1}$& $642^{+47}_{-74}$ &$736^{+23}_{-69}$ & (1),(3),(6)\\ 
And XXVI& -7.1 & 6&-1.9$\pm0.2$ & -1.8$\pm0.5$ & -261.6$^{+3.0}_{-2.8}$ & 8.6$^{+2.8}_{-2.2}$&$219^{+67}_{-52}$ & $754^{+218}_{-164}$ & (1),(3),(6)\\
AndXXVII & -7.9 & 11 &-1.7$\pm0.2$ & -1.5$\pm0.28$ & -539.6$^{+4.7}_{-4.5}$ &
14.8$^{+4.3}_{-3.1}$& $657^{+112}_{-271}$ &$1255^{+42}_{-474}$ &
(1),(3),(6)\\ 
AndXXVIII & -8.5 & 17 &-2.0$\pm0.2$ & -2.1$\pm0.3$ & -326.2$\pm2.7$ &
6.6$^{+2.9}_{-2.1}$& $210^{+60}_{-50}$ &$650^{+150}_{-80}$ &
(6),(9)\\ 
AndXXIX & -8.3 & 24 &-1.8$\pm0.2$ & N/A & $-194.4\pm1.5$ &
$5.7\pm1.2$& $360\pm60$ &$730\pm75$ &
(10),(11)\\ 
And XXX (Cass II)& -8.0 & 8 &-1.6$\pm0.4$ & -2.2$\pm0.4$
&-139.8$^{+6.0}_{-6.6}$ & 11.8$^{+7.7}_{-4.7}$& $267^{+23}_{-36}$ &
$681^{+32}_{-78}$ & (3),(6),(12)\\
\enddata
\tablecomments{$^{(a)}$ Number of spectroscopically confirmed member
  $^{(b)}$ Kinematics taken from T12 rather than this work, as they possess
  greater numbers of member stars.
  stars. $^{(c)}$ Owing to shorter than typical exposure time, lower resolution
  data and difficult observing conditions, this kinematic identification of And
  XXIV remains tentative, and needs to be confirmed with further follow-up. 
  References: (1) \citet{mcconnachie12}, (2)
  \citet{tollerud12}, (3) \citet{conn12},
  (4)\citet{kalirai10} (5)\citet{collins11b}, (6) This work, (7)
  \citet{collins10}, (8) \citet{chapman12}, (9) \citet{slater11}, (10)
  \citet{tollerud13}, (11) \citet{bell11}, (12) Irwin et al. (in prep).}
\end{deluxetable*}

\section{Conclusions}
\label{sect:conc}

Using new and existing spectroscopic data from the Keck I LRIS and Keck II
DEIMOS spectrographs, we have homogeneously derived kinematic properties for
18 of the 28 known Andromeda dSph galaxies. Using a combination of their $g-i$
colors, positions on the sky and radial velocities, we determine the
likelihood of each observed star belonging to a given dSph, thus filtering out
MW foreground or M31 halo contaminants. We have measured both their systemic
velocities and their velocity dispersions, with the latter allowing us to
constrain the mass and densities within their half-light radii. For the first
time, we confirm that And XVII, XIX, XX, XXIII, XXVI and Cass II are dark
matter dominated objects, with dynamical mass-to-light ratios within the
half-light radius of $[M/L]_{\rm half}>10\msun/\lsun$. 

For And XXV, a bright M31 dSph ($M_V=-9.7$) we measure a mass-to-light ratio
of only $[M/L]_{\rm half}=10.3^{+7.0}_{-6.7}\msun/\lsun$ from a sample of 26
stars, meaning that it is consistent with a simple stellar system with no
appreciable dark matter component within its $1\sigma$ uncertainties. If this
were confirmed with larger datasets, it would prove to be a very important
object for our understanding of the formation and evolution of galaxies.

We compare our computed velocity dispersions and mass estimates with those
measured for MW dSphs, and find that the majority of the M31 dSphs have very
similar mass-size scalings to those of the MW. However, we note 3 significant
outliers to these scalings, namely And XIX, XXI and XXV, who possess
significantly lower velocity dispersions than expected for their size. These
results builds on the identification of three potential outliers in the
\citet{tollerud12} dataset (And XIV, XV and XVI). We suggest that the lower
densities of the dark matter halos for these outliers could be an indication
that they have encountered greater tidal stresses from their host over the
course of their evolution, decreasing their masses. However, these bright
systems still fall on the luminosity-metallicity relation established for the
dSph galaxies of the Local Group. If these objects had undergone significant
tidal disruption, we would expect them to lie above this relation. As such,
this remains puzzling, and requires dedicated follow up studies to
fully map out the kinematics of these unusual systems.

We measure the metallicities of all 18 dSphs from their co-added
spectra and find that they are consistent with the established MW
trend of decreasing metallicity with decreasing luminosity.

This work represents a significant step forward in understanding the mass
profiles of dwarf spheroidal galaxies. Far from residing in dark matter halos
with identical mass profiles, we show that the halos of these objects are
complex, and differ from one to the next, with their environment and tidal
evolution imprinting themselves upon the dynamics of their stellar
populations. The Andromeda system of dSphs presents us with an opportunity to
better understand these processes, and our future work will further illuminate
the evolutionary paths taken by these smallest of galaxies.

\section*{Acknowledgments}
We would like to thank Hans-Walter Rix for helpful discussions regarding this
manuscript. We are also grateful to the referee for their helpful and detailed
suggestions for improving this work. We thank the SPLASH collaboration for
providing us with details of their observations of dSphs as presented in T12.

Most of the data presented herein were obtained at the W.M. Keck Observatory,
which is operated as a scientific partnership among the California Institute
of Technology, the University of California and the National Aeronautics and
Space Administration. The Observatory was made possible by the generous
financial support of the W.M. Keck Foundation.

Based in part on observations obtained with MegaPrime/MegaCam, a joint project
of CFHT and CEA/DAPNIA, at the Canada-France-Hawaii Telescope (CFHT) which is
operated by the National Research Council (NRC) of Canada, the Institute
National des Sciences de l'Univers of the Centre National de la Recherche
Scientifique of France, and the University of Hawaii.

Based in part on data collected at Subaru Telescope, which is operated by the
National Astronomical Observatory of Japan.

The authors wish to recognize and acknowledge the very significant cultural
role and reverence that the summit of Mauna Kea has always had within the
indigenous Hawaiian community.  We are most fortunate to have the opportunity
to conduct observations from this mountain.  Funding for SDSS-III has been
provided by the Alfred P. Sloan Foundation, the Participating Institutions,
the National Science Foundation, and the U.S. Department of Energy Office of
Science. The SDSS-III web site is http://www.sdss3.org/.

SDSS-III is managed by the Astrophysical Research Consortium for the
Participating Institutions of the SDSS-III Collaboration including the
University of Arizona, the Brazilian Participation Group, Brookhaven National
Laboratory, University of Cambridge, Carnegie Mellon University, University of
Florida, the French Participation Group, the German Participation Group,
Harvard University, the Instituto de Astrofisica de Canarias, the Michigan
State/Notre Dame/JINA Participation Group, Johns Hopkins University, Lawrence
Berkeley National Laboratory, Max Planck Institute for Astrophysics, Max
Planck Institute for Extraterrestrial Physics, New Mexico State University,
New York University, Ohio State University, Pennsylvania State University,
University of Portsmouth, Princeton University, the Spanish Participation
Group, University of Tokyo, University of Utah, Vanderbilt University,
University of Virginia, University of Washington, and Yale University.

R.I. gratefully acknowledges support from the Agence Nationale de la Recherche
though the grant POMMME (ANR 09-BLAN-0228).

G.F.L. gratefully acknowledges financial support for his
ARC Future Fellowship (FT100100268) and through the
award of an ARC Discovery Project (DP110100678).

N.B. gratefully acknowledges financial support through the
award of an ARC Discovery Project (DP110100678).

A.K. thanks the Deutsche Forschungsgemeinschaft for funding from Emmy-Noether
grant Ko 4161/1.


\appendix
\section{Testing the membership probability algorithm}
\subsection{The inclusion of a velocity term in the calculation of $P_i$}
\label{sect:vptest}

In T12, the authors do not impose a velocity probability criterion for their
membership calculations. Instead they require all member stars to have a total
probability, based on their positions and colors, of $P_{member}>0.1$, and
then apply a $3\sigma$ clipping to this final sample to prevent any outliers
from significantly inflating their calculated velocity dispersions. In our
analysis, we have avoided making any hard cuts to our sample by also utilising
prior information on the velocities of our expected contaminant populations
and member stars. In \S~\ref{sect:velprob}, we tested our velocity probability
criteria was not overly biasing our final measurements of velocity dispersion
to artificially lower values with the introduction of an extra parameter,
$\eta$, that allows us to add additional weight to stars in the tails of the
Gaussian velocity distribution. However, we can further test our velocity
criterion by removing it entirely from the probabilistic determination, and
instead implementing the same cuts presented by T12. This involves cutting
stars where $P_i<0.1$ (as determined from $P_{CMD}$ and $P_{dist}$), and also
by iteratively removing all stars that have velocities that do not lie within
$3\sigma$ of the mean of the remaining sample. In Table~\ref{tab:vpcuts}, we
present the results of this on our measured values of $v_r$ and $\sigma_v$ for
our full sample of dSphs. For all objects (bar And XXVII, which is a unique
case, as described in \S~\ref{sect:and27}) the systemic velocities derived are
within $\sim2-3\kms$ of one another. The velocity dispersions we measure from
our full algorithm tend to be slightly higher on average, and this is to be
expected as we do not cut any stars from our analysis, and therefore outliers
in the velocity profile may be assigned non-negligible membership
probabilities that will allow them to increase this measurement. By and large,
these differences are not significant, with the final values agreeing to well
within their $1\sigma$ uncertainties.

\begin{figure*}
 \begin{center}
 \includegraphics[angle=0,width=0.45\hsize]{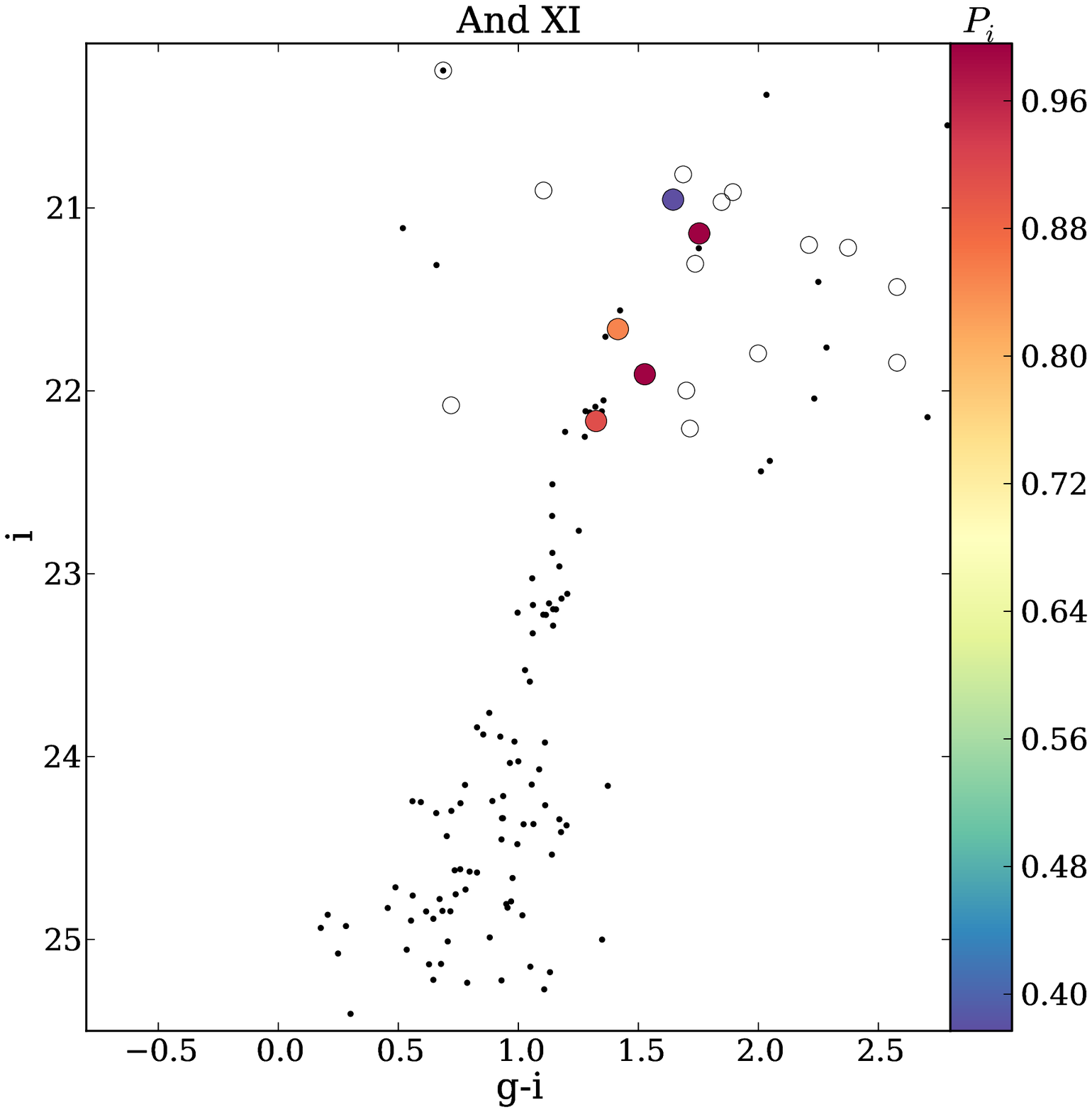}
 \includegraphics[angle=0,width=0.45\hsize]{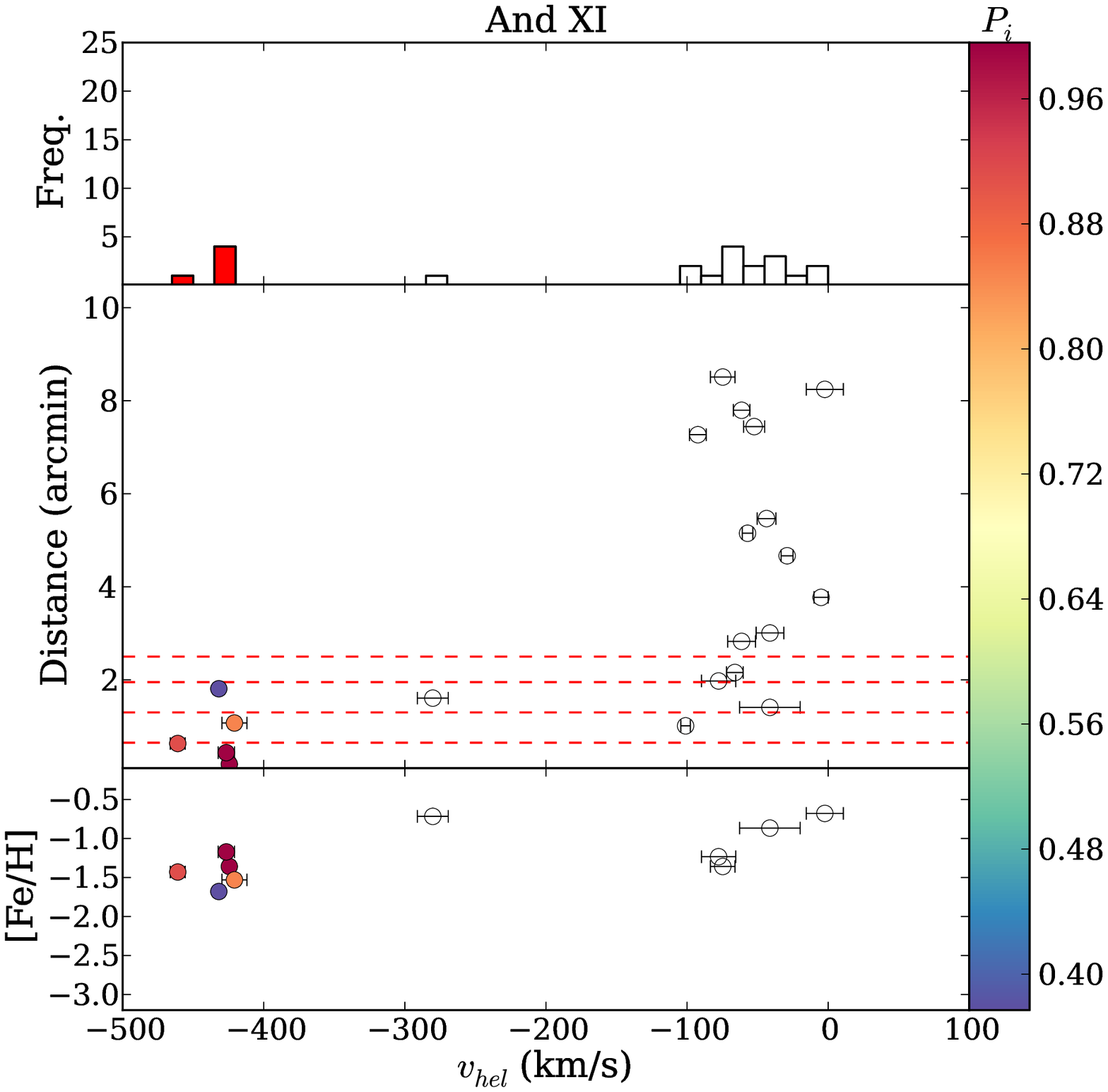}
 \includegraphics[angle=0,width=0.9\hsize]{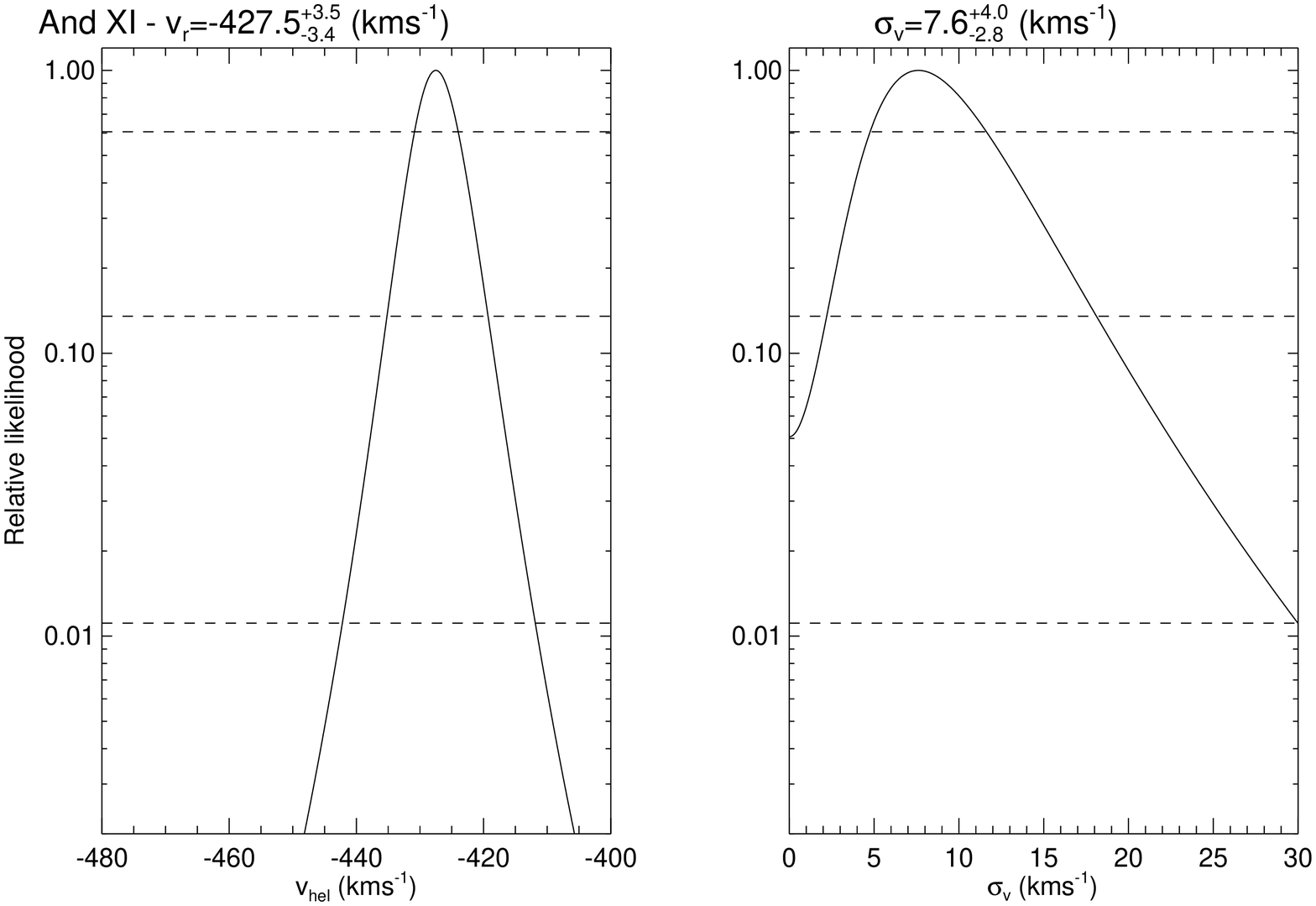}
 \caption{As Fig.~\ref{fig:And17}, but for And XI. Using our full
   probabilistic algorithm (including our $P_{vel}$ term), 5 stars are
   identified as likely members. When implementing the $3\sigma$ clipping of
   T12, the star with $v_r\sim-450\kms$ is removed from the analysis, despite
   having a high probability of membership based on its color ($P_{CMD}$) and
   distance from the centre of the dSph ($P_{dist}$). This has a substantial
   effect of the calculated velocity dispersion, as shown in Table~\ref{tab:vpcuts}.}
 \label{fig:And11}
 \end{center}
 \end{figure*}

It is interesting that our algorithm appears to perform better when dealing
with dSphs where the number of member stars is low. This is best demonstrated
by And XI (see Fig.~\ref{fig:And11}). Our algorithm identifies 5 stars with
non-negligible probabilities of membership, clustered around
$v_r\sim-430\kms$. Our full algorithm measures a systemic velocity of
$v_r=-427.5^{+3.5}_{-3.4}\kms$ and a velocity dispersion of
$\sigma_v=7.5^{+4.0}_{-2.8}\kms$. One of these stars is slightly offset from
the other 4 with a more negative velocity of $v_r=-456.8\kms$. Although this
star has a reasonably high probability of being a member based on its
distance from the centre of And XI, and its position in the CMD, it does not
survive the $3\sigma$ velocity clipping procedure of T12. As the number of
member stars is so low, cutting one star from the sample can have a
significant effect, and as such, while the T12 procedure determines a very
similar systemic velocity of $v_r=-425.0\pm3.1\kms$ it is unable to resolve a
velocity dispersion. This effect is also seen other systems (such as And XIII,
XVII, XXII and XXVI), although it is typically less pronounced.

Another regime where our algorithm performs better than that of T12 is where
the systemic velocity of the system in question is within the regime of the
contaminating Milky Way K-dwarfs. An example of this is the unusual system,
And XIX, where our algorithm measures a systemic velocity of
$v_r=-111.6^{+1.6}_{-1.4}\kms$ and a velocity dispersion of
$\sigma_v=4.7^{+1.6}_{-1.4}$. However, the procedure of T12 is less able to
resolve the kinematics of the system, measuring $v_r=-109.3\pm5.3\kms$ and a
velocity dispersion of $\sigma_v=1.8^{+9.1}_{-1.8}$. The much larger
uncertainty on the dispersion is a result of including Milky Way contaminants
in the sample which can be difficult to cut out without applying prior
knowledge of the velocity profile of this population. $3\sigma$ clipping
allows outliers to contribute more significantly to the measured profile in
this instance, increasing the uncertainty. A similar effect is seen in the And
XXIV and And XXX (Cass II) objects, which also have systemic velocities in the
Milky Way contamination regime.

These results lead us to conclude that the inclusion of a $P_{vel}$ term in
our analysis allows us to more effectively determine the true kinematics of
the systems we are studying. Further, as no cuts to the sample are required
using this method, it allows for a more unbiased study of the kinematics of
dSphs than that of T12.

\begin{deluxetable*}{lcccc}
\tabletypesize{\footnotesize}
\tablecolumns{5} \tablewidth{0pt} \tablecaption{\label{tab:vpcuts}Measured
  velocities ($v_r$) and dispersions ($\sigma_v$) for our full sample of dSphs
  as calculated using cuts on probability of membership calculated from only
  color and position of stars, combined with $3\sigma$ clipping on velocities
  vs. those calculated using our full probabilistic algorithm.} \tablehead{
  \colhead{Object} & \colhead{$v$ \&\ $P_i$ cuts } & & \colhead{Full
    algorithm} & \\ & \colhead{$v_r (\kms)$} & \colhead{$\sigma_v(\kms)$}
  &\colhead{$v_r (\kms)$} & \colhead{$\sigma_v(\kms)$}} \startdata
And V & $-391.1\pm2.9$ & $10.8^{+3.0}_{-2.3}$ & $-391.5\pm2.7$ & $12.2^{+2.5}_{-1.9}$\\
And VI & $-339.0\pm3.0$ & $11.9^{+2.9}_{-2.3}$ & $-339.8\pm1.8$ & $ 12.4^{+1.5}_{-1.3}$\\
And XI & $-425.0\pm3.1$ & $0^{+3.5}$ & $-427.5^{+3.5}_{-3.4}$ & $7.6^{+4.0}_{-2.8}$\\
And XII & $-558.8\pm3.7$ & $0^{+6.8}$ & $-557.1\pm1.7$ & $0^{+4.0}$ \\
And XIII & $-203.8\pm8.4$ & $0^{+16.2}$ & $-204.8\pm4.9$& $0.0^{+8.1}$ \\
And XVII & $-260.0^{+8.0}_{-7.8}$ & $1.8^{+9.1}_{-1.8}$ &$-254.3^{+3.3}_{-3.7}$ &$2.9^{+5.0}_{-2.9}$ \\
And XVIII &$-345.1\pm3.3$ & $0^{+4.4}$ & $-346.8\pm2.0$& $0^{+2.7}$\\
And XIX & $-109.3\pm5.3$ & $1.5^{+6.8}_{-1.5}$ & $-111.6^{+1.6}_{-1.4}$& $4.7^{+1.6}_{-1.4}$\\
And XX & $-454.6^{+4.6}_{-5.7}$ & $7.7^{+8.4}_{-3.9}$ &$-456.2^{+3.1}_{-3.6}$ &$7.1^{+3.9}_{-2.5}$ \\
And XXI &$-363.4^{+2.0}_{-1.8}$ & $3.2^{+2.3}_{-2.1}$ & $-362.5\pm0.9$&$4.5^{+1.2}_{-1.0}$ \\
And XXII &$-131.4\pm2.7$ & $0^{+3.1}$ &$-129.8\pm2.0$ &$2.8^{+1.9}_{-1.4}$ \\
And XXIII & $-236.9\pm2.1$ & $8.4^{+1.9}_{-1.5}$ &$-237.7\pm1.2$ &$7.1\pm1.0$ \\
And XXIV &$-129.2\pm3.6$ & $0^{+6.1}$ & $-129.9^{+4.3}_{-4.4}$& $3.5^{+6.6}_{-3.5}$ \\
And XXV & $-107.7^{+1.9}_{-1.8}$ & $3.3^{+2.2}_{-1.8}$ &$-107.8\pm1.0$ & $3.0^{+1.2}_{-1.1}$\\
And XXVI & $-264.1\pm4.5$ & $0^{+4.8}$ & $-261.7^{+3.1}_{-2.8}$& $8.7^{+2.9}_{-2.3}$\\
And XXVII & $-517.6^{+42.8}_{-43.2}$ & $19.3^{+17}_{-19}$ &
$-539.6^{+4.7}_{-4.5}$& $14.8^{+4.3}_{-3.1}$ \\
And XXX (CassII) & $-140.1^{+8.6}_{-9.3}$ & $14.1^{+12.9}_{-6.1}$ &$-139.8^{+6.0}_{-6.6}$ &$11.8^{+7.7}_{-4.7}$ \\
\enddata
\end{deluxetable*}

\subsection{The effect of low signal-to-noise data on measuring $v_r$ and $\sigma_v$}

For our brightest targets ($i\lta22.5$) the S:N of our spectra is typically
$>3$\AA$^{-1}$. However, as our targets become fainter, so too their S:N
falls. For spectra with S:N$\gta1.5$\AA$^{-1}$, our pipeline is still able to
measure velocities based on the Ca II triplet, with reasonable measurement
uncertainties. However, it is prudent to check whether the inclusion of these
velocities, calculated from significantly noisier spectra, has a detrimental
effect on our ability to measure the kinematic properties of our dSph sample.

Such a test is straightforward to implement. We have a number of dSphs within
our sample for which our probabilistic analysis identifies $\sim30$ likely
members (such as And XXI, XXIII and XXV). We can therefore use these samples
to impose S:N cuts on our data to see the effect of this on our measurements of
$v_r$ and $\sigma_v$. We present the results of this test in
Table~\ref{tab:sntest}, and our finding is that, as the level of our imposed
S:N cut increases (and so the number of included stars decreases), the
systemic velocity remains more or less constant. The measured velocity
dispersion, however, shows some variation. In the case of And XXI and XXV, the
dispersion increases with increased S:N, however not significantly. In both
cases the dispersion calculated from the higher S:N data lies well within
$1\sigma$ of that calculated from the lower S:N data. Intuitively, this makes
sense as the spectra with higher S:N are likely to have lower velocity
uncertainties, and so our maximum-likelihood analysis will attribute more of
the spread in measured velocities to an intrinsic dispersion, rather than to
our measurement errors. In the case of And XXIII, we find the opposite to be
true. As our S:N cut increases, we find that our measured dispersion
decreases. This may be because the number of member stars in subsequent
quality cuts drops off more rapidly for And XXIII than And XXI and XXV. This
suggests that we should be extra cautious when interpreting our measured
velocity dispersions for dSphs where both the average S:N of member stars, and
the number of member stars, is low.

\begin{deluxetable*}{lccccccccc}
\tabletypesize{\footnotesize}
\tablecolumns{10} 
\tablewidth{0pt}
\tablecaption{\label{tab:sntest}The effect of including low S:N data in our
  analysis on determining $v_r$ and $\sigma_v$.} \tablehead{ \colhead{Object} & & \colhead{And
  XXI} & & &
\colhead{And XXIII}& & & \colhead{And XXV} & \\ &\colhead{$N_*$} & \colhead{$v_r
  (\kms)$} &  \colhead{$\sigma_v(\kms)$} &\colhead{$N_*$} & \colhead{$v_r
  (\kms)$} &  \colhead{$\sigma_v(\kms)$} &\colhead{$N_*$} & \colhead{$v_r
  (\kms$)} &  \colhead{$\sigma_v(\kms)$}} 
\startdata
S:N$>2$ & 20 & $-362.9\pm0.9$& $3.5^{+0.9}_{-0.7}$ & 22 & $-238.0\pm1.2$ &
$6.6\pm1.1$ & 27 & $-107.7\pm0.9$ & 2.7$\pm1.1$ \\
S:N$>3$ & 11 & $-364.2\pm0.9$ & $3.1\pm0.8$ & 10 & $-238.3\pm1.4$ &
$5.1^{+1.4}_{-1.2}$ & 24 & $-107.7\pm1.0$ & $3.0\pm1.2$\\
S:N$>4$ & 5 & $-363.9\pm1.5$ & $4.0^{+1.5}_{-1.1}$ & 5 & $-239.4\pm1.1$ &
$5.7^{+1.5}_{-1.3}$ & 16 & $-108.2\pm1.2$ & $3.0^{+1.4}_{-1.2}$\\
S:N$>5$ & 3 & $-362.6^{+2.2}_{-2.3}$ & $4.5^{+2.6}_{-1.5}$ & 2 & $-239.5\pm1.9$
& $0.0^{+4.9}$ & 13 & $-109.0\pm1.2$ & $2.8^{+1.5}_{-1.3}$\\ 
\enddata
\tablecomments{xxx}
\end{deluxetable*}

\subsection{The effect of small sample sizes on determining kinematic properties
  of dwarf galaxies}

Obtaining reliable velocities for member stars of faint and distant systems is
a difficult task that can only be achieved with the largest optical
telescopes, such as Keck. Given the demand for facilities such as this, any
observing time awarded must be used as effectively as possible, and this often
means compromising between deep pointings for a few objects, and shallower
pointings for a number of objects. With longer or multiple exposures on a
single target, one can build up impressive samples of member stars for an
individual dSph. For example, the SPLASH collaboration observed a total of 95
members in And II, one of the brightest M31 dSph companions \citep{kalirai10}
by taking 2 separate exposure fields over this large object. However, multiple
exposures such as these produce diminishing returns as you move down the
luminosity scale to fainter, more compact dSphs. This is both because of their
smaller size with respect to the DEIMOS field of view, and the fewer number of
bright stars available on the RGB to target. In this case, the only way to
identify more members is by integrating for longer, but given the paucity of
stars, the trade-off between time spent exposing and additional members
observed can be quite expensive. Such difficulties inevitably lead to the
inference of dynamical properties for an entire system from a handful of
stars. It is important for us to understand the effect this bias has on our
results, and how reliable the quoted values are. We test this using our
datasets for which we identify $>25$ member stars, namely And XIX, XXI, XXIII
and XXV using the following method. We select 4, 6, 8, 10, 15, 20, 25 and 30
stars at random from each dataset and then measure the systemic velocity and
velocity dispersion using our probability algorithm. This was repeated 1000
times for each sample size. In cases where the algorithm is unable to resolve
a velocity dispersion, we throw out the result and resimulate, as null results
here will affect our averages and will not inform us whether the instances in
which we are able to resolve a velocity dispersion from small numbers of stars
are producing valid, reliable result. We display the resulting values in
Table~\ref{tab:sampsz}, with the true value recovered from the full sample
shown in bold in the final row for comparison. We show that in all these
cases, the systemic velocity and velocity dispersion are recovered well within
the scatter of the 1000 simulations even when dealing with sample sizes as
small as 4 stars, so long as the measurement is resolved. In cases where we
are unable to resolve a dispersion, we find that our resulting uncertainties
are not meaningful. This is shown explicitly in the case of And XVIII, where
we can compare our upper limit for the velocity dispersion as determined from
our algorithm with the dispersion calculated in T12 from a much larger
dataset. We see that our uncertainty is not consistent with their result. As
such, we advise that in all cases where we calculate velocity dispersions from
small samples ($N_*<8$, And XI, XII, XX XXIV and XXVI), the dispersion
measurements should be treated as indications of the likely dispersion, and
need to be confirmed with follow-up studies.

\begin{deluxetable*}{ccccccc}
\tabletypesize{\footnotesize}
\tablecolumns{10} \tablewidth{0pt} \tablecaption{\label{tab:sampsz} The effect
  of varying sample size ($N_*$) on the measurements of systemic velocity,
  $v_r$, and velocity dispersion, $\sigma_v$. $N$ stars from And XXI, XXIII
  and XXV are randomly selected, and their properties are derived using our
  full probabilistic analysis. This was repeated 1000 times. Values reported
  below are the averages from 1000 realizations, with the uncertainties
  represent the standard deviation of the 1000 realizations. } \tablehead{
  \colhead{Object} & \colhead{And XXI} & & \colhead{And XXIII}& & \colhead{And
    XXV} & \\ \colhead{$N_*$} & \colhead{$v_r (\kms)$} &
  \colhead{$\sigma_v(\kms)$} & \colhead{$v_r (\kms)$} &
  \colhead{$\sigma_v(\kms)$} & \colhead{$v_r (\kms$)} &
  \colhead{$\sigma_v(\kms)$}} \startdata
4 & $-362.6\pm3.1$ & $5.0\pm3.3$ & $-237.2\pm3.8$ & $7.5\pm3.6$ & $-108.8\pm2.4$ & 5.4$\pm4.3$ \\
6 & $-363.0\pm2.0$ & $4.0\pm1.9$ & $-237.3\pm3.0$ & $6.9\pm2.6$& $-108.0\pm2.7$ & $4.0\pm2.3$\\
8 & $-362.9\pm2.1$ & $4.2\pm2.0$ & $-236.8\pm2.8$ & $6.9\pm2.3$ & $-107.6\pm2.0$ & $2.7\pm1.1$\\
10 & $-362.7\pm1.8$ & $4.1\pm1.8$ & $-237.3\pm2.5$ & $6.5\pm2.1$ & $-107.9\pm1.5$ & $3.1\pm1.0$\\
15 & $-362.9\pm1.5$ & $4.3\pm1.2$ & $-237.4\pm2.0$ & $6.8\pm1.4$ &
$-107.9\pm1.5$ & $3.1\pm1.0$ \\
20 & $-362.9\pm1.3$ & $4.3\pm1.2$ & $-237.6\pm1.7$ & $6.9\pm1.4$ &
$-107.8\pm1.2$ & $3.0\pm0.9$ \\
25 & $-363.0\pm1.0$ & $4.4\pm1.0$ & $-237.4\pm1.4$ & $7.1\pm1.7$ &
$-107.8\pm1.0$ & $3.1\pm0.8$\\
30 & $-362.8\pm1.1$ & $4.4\pm1.1$ & $-237.7\pm1.3$ & $7.2\pm1.4$ & -- & --\\
{\bf Full sample} &  $\boldsymbol{-362.5\pm0.9}$ &$\boldsymbol{4.5^{+1.2}_{-1.1}}$ & $\boldsymbol{-237.7\pm1.2}$ & $\boldsymbol{7.1\pm1.0}$ & $\boldsymbol{-107.8\pm1.0}$& $\boldsymbol{3.0^{+1.2}_{-1.1}}$ \\
\enddata
\end{deluxetable*}

\subsection{Testing our algorithm on the SPLASH sample of M31 dSphs}

In T12, the authors reported on the kinematic properties of 15 M31 dSphs, And
I, III, V, VII, IX, X, XI, XII, XIII, XIV, XV, XVI, XVIII, XXI and XXII, and
the positions, and measured velocities (plus uncertainties) for each likely
member stars were published as part of that work. The authors were kind enough
to also give us access to these properties for their non-member stars so that
we might run our algorithm over the full samples to see if we reproduce their
results. Our technique for assigning membership probability differs from
theirs in that we use the velocities of stars as an additional criterion for
membership, whereas they use a cut on both the resulting probability
($P({\mathrm member}>0.1$), and a $3\sigma$ clipping on the velocity. In
addition, where we use PAndAS CFHT MegaCam $g-$ and $i-$band photometry for
our membership analysis,the SPLASH team use their own Washington-DDO51 filter
photometric dataset, in membership classification. As such, small differences
might be expected, but if our technique is robust our results should well
mirror those of T12. In Table~\ref{tab:splashcomp} we compare our calculated
values of $v_r$ and $\sigma_v$ to those published in T12. As the measurements
made in T12 for And XI and XII are made from only 2 stars, we do not include
these in this test. In general, the results from both analyses agree to within
$1\sigma$ of one another, with the majority of them being well within this
bound. Typically, we find that our procedure measures slightly larger values
for $\sigma_v$ than that of T12 (with the exception of And IX, And XVIII and
And XXII). This is to be expected, as we do not cut stars from our analysis
based on their velocity, instead we down-weight their probability of
membership. As such, those stars considered as outliers would naturally
inflate our dispersions above those measured by T12, but the effect is
marginal. These results demonstrate that our technique for assigning
probability of membership of individual stars within M31 dSphs based on their
photometric properties and velocities is robust, and comparable to that of
T12. However, as discussed in \S~\ref{sect:vptest}, we find that our technique
is superior as it requires no cuts to the final dataset to be made, reducing
the bias in these measurements.

\begin{deluxetable*}{lcccc}
\tabletypesize{\footnotesize}
\tablecolumns{5} 
\tablewidth{0pt}
\tablecaption{\label{tab:splashcomp}} \tablehead{ \colhead{Object} &
  \colhead{T12 analysis} & & \colhead{Our analysis}&  \\ & \colhead{$v_r
  (\kms)$} &  \colhead{$\sigma_v(\kms)$} &\colhead{$v_r
  (\kms)$} &  \colhead{$\sigma_v(\kms)$}} 
\startdata
And I & -376.3$\pm2.2$ & 10.2$\pm1.9$ & $-376.3\pm1.3$ & $10.6\pm1.0$\\
And III & -344.3$\pm1.7$ & 9.3$\pm1.4$& $-344.2\pm1.2$ & $10.1\pm1.9$ \\
And V & -397.3$\pm1.5$ & 10.5$\pm1.1$ & $-396.0\pm1.0$ & $11.4\pm1.2$\\
And VII & -307.2$\pm1.3$ & 13.0$\pm$ 1.0& $-307.1\pm1.1$ & $13.1\pm0.9$\\
And IX & -209.4$\pm2.5$ & 10.9$\pm2.0$ & $-210.3\pm1.9$ & $10.2^{+1.9}_{-1.7}$ \\
And X &  -164.1$\pm1.7$ & 6.4$\pm1.4$ & $-165.3\pm1.5$ & $6.0^{+1.3}_{-1.2}$\\
And XIII& -185.4$\pm2.4$ & 5.8$\pm2.0$ & $-183.0^{+2.4}_{-2.3}$ & $8.6^{+2.1}_{-1.7}$\\
And XV & -323$\pm1.4$ & 4.0$\pm1.4$ & $-322.6\pm1.1$ & $6.0^{+2.0}_{-1.8}$\\
And XVI & -367.3$\pm2.8$ & 3.8$\pm2.9$& $-366.1^{+4.0}_{-3.1}$ & $4.2^{+4.8}_{-4.2}$\\
And XVIII &-332.1$\pm2.7$ & 9.7$\pm2.3$ & $-330.7^{+3.9}_{-4.1}$ & $7.5^{+4.5}_{-3.1}$\\
And XXI & $-361.4\pm5.8$ & $7.2\pm5.5$ & $-358.9^{+5.1}_{-5.6}$ & $8.5^{+6.3}_{-5.1}$\\
And XXII & $-126.8\pm3.1$ & $3.54^{+4.16}_{-2.49}$ & $-124.2^{+4.6}_{-4.5}$ & $0.0^{+5.7}$\\
\enddata
\end{deluxetable*}

\section{Reanalyzing our previously published results}

To ensure our analysis of the global properties of the M31 dSph population in
\S~\ref{sect:mass} is homogenous, we reanalyzed our previously published
datasets using our new probability algorithm. Here we briefly summarize the
results of this analysis and compare the new results to the published
works. The objects we discuss here are And V, VI (first published in
\citealt{collins11b}), XI, XII and XIII (published in
\citealt{chapman05,chapman07,collins10}). And V was observed with the LRIS
instrument, while the remaining objects were observed with the DEIMOS
instrument.

\subsection{Andromeda V}
\label{sect:and5}

Andromeda V (And V) was observed using the LRIS instrument on Keck I rather
than the DEIMOS instrument on Keck II. LRIS has a lower resolution than
DEIMOS, and a smaller field of view, which lowers the accuracy of velocity
measurement and limits us to only $\sim50$ targets within a mask compared with
$100-200$ for a DEIMOS mask. The raw data were also not reduced using our
standard pipeline, owing to problems with the arc-lamp calibrations, and were
instead analyzed using the NOAO.ONEDSPEC and NOAO.TWODSPEC packages in
IRAF. The results from this reduction, plus an analysis of the data using hard
cuts in velocity, distance and color to determine membership were first
published in \citet{collins11b}. Our full probabilistic analysis identifies 17
stars with a non-negligible probability of belonging to And V. Our technique
determines a systemic velocity of $v_r=-391.5\pm2.7\kms$ and
$\sigma_v=12.2^{+2.5}_{-1.9}\kms$. Comparing these values to our previously
published results ($v_r=-393.1\pm4.2\kms$ and
$\sigma_v=11.5^{+5.3}_{-4.3}\kms$, \citealt{collins11b}) we find them to be
consistent within the quoted uncertainties. We also compare our results to
those of T12, who measured $v_r=−397.3\pm1.5\kms$ and
$\sigma_v=10.5\pm1.1\kms$ from a larger sample of stars (85 members cf. 17)
using the higher resolution DEIMOS spectrograph. The velocity dispersions of
both are consistent within their $1\sigma$ uncertainties, as are the systemic
velocities. Given the difference of a factor of 5 in number of probable member
stars between our study and that of T12, this consistency is reassuring, and
demonstrates the ability of our technique to accurately determine the
kinematics of M31 dSph galaxies from small sample sizes.


\subsection{Andromeda VI}
\label{sect:and6}

As And VI sits at a large projected distance from the centre of M31
($\sim270$~kpc), it was not observed as part of the PAndAS survey and we were
unable to use CFHT data for our $P_{CMD}$ determination. Instead, we use
Subaru Suprime-cam data (PI. N. Arimoto, see \citealt{collins11b} for a full
discussion of this data). Using our full probabilistic analysis, we identify
45 stars with $P_i>10^{-6}$. Our technique determines a most-likely
$v_r=-339.8\pm1.8\kms$, with $\sigma_v=12.4^{+1.5}_{-1.3}\kms$. Comparing
these values with the results of \citet{collins11b} who measured
$v_r=-344.8\pm2.5\kms$ and $\sigma_v=9.4^{+3.2}_{-2.4}\kms$, we see that both
the systemic velocities and the velocity dispersions are consistent within
quoted uncertainties. The slight differences between our previous study and
this work are simply a result of the application of our new technique.


\subsection{Andromeda XI}
\label{sect:and11}

The kinematic properties for Andromeda XI as measured from this DEIMOS data
set were first published in \citet{collins10}. Here we identify 5 stars as
probable members. We determine most-likely parameters of
$v_r=-427.5^{+3.5}_{-3.4}\kms$ and $\sigma_v=7.6^{+4.0}_{-2.8}\kms$. In
\citet{collins10} we measured $v_r=-419.4^{+4.4}_{-3.8}\kms$ and we were
unable to resolve the velocity dispersion for the dSph, measuring
$\sigma_v=0.0^{+4.6}\kms$ (where the upper bound represents the formal
1$\sigma$ uncertainty on the unresolved dispersion), which implied a higher
systemic velocity and lower velocity dispersion for And XI. However, in that
analysis, one star with a velocity of $\sim-440\kms$ was considered to be an
outlier based on its velocity, and thus excluded from the kinematic
analysis. Here, our algorithm gives this star a non-zero probability of
membership, which likely decreases the systemic velocity and increases the
dispersion.  T12 also presented observations for And XI, but they were not
able to cleanly detect the galaxy. They identified 2 stars with highly
negative velocities ($\sim-460\kms$), which are offset from our systemic
velocity by $\sim30\kms$.  Given the very negative velocities of their 2
stars, the probability of them both being M31 contaminants seems low, and some
other explanation may be more suitable. Between observations, there is one
star common to both ($\alpha=$00:46:19.10,$\delta=$+33:48:4.1), for which we
measure a velocity of $-427.16\kms$ compared with $-461.6\kms$. This amounts
to a statistical difference at the level of $5\sigma$. One obvious avenue to
check is that there has been no velocity offset introduced by a rogue skyline
that falls within the region of one of the three Ca II lines. We have
carefully checked the spectra of each of our 5 probable members to see if this
has been the case. We also rederive the velocity based on cross-correlations
with each of the three lines individually, rather than with the full
triplet. We find these results, and their average to be entirely consistent
within the associated errors from the velocities derived using the technique
discussed in \S~\ref{sect:specobs}. This large discrepancy is puzzling,
particularly as the methods used to measure velocities in this work are almost
identical to those of T12. Without further data it is not possible for either
team to pin down the exact issue, or which of the datasets gives the true
systemic velocity. This argues for taking further observations within these
faint systems, in the hopes of better understanding both the systems
themselves, and any systematics introduced by the DEIMOS instrument. It is
comforting that, in all cases where this offset is observed, the velocity
dispersions measured by each team are consistent with one another, suggesting
the problem affects all observations identically.

In this case, we have identified a greater number of potential members stars
than in the work of T12, the spectra for all of which have relatively high S:N
(S:N$\sim5-10$\AA$^{-1}$). Therefore, our results should be considered as more
robust than those of T12.

\subsection{Andromeda XII}
\label{sect:and12}

Kinematic properties for Andromeda XII as determined from the DEIMOS data set
presented here were previously published in \citet{chapman07} and
\citet{collins10}, and in both cases, membership was largely determined using
hard cuts in velocity. This object has been of particular interest as it
possesses an extremely negative systemic velocity with respect to Andromeda,
suggesting that it is on its first infall into the local group (see
\citealt{chapman07} for a full discussion). Using our new algorithm, we measure
$v_r=-557.1\kms$ and an unresolved velocity dispersion of
$\sigma_v=0.0^{+4.0}\kms$, where the upper bound represents the formal
$1\sigma$ uncertainty on the measurement. As the dispersion is unresolved, the
lower error bound is undefined, as having a negative velocity dispersion is
unphysical. These values are completely consistent with the results of
\citet{chapman07} and \citet{collins10} ($v_r=-558.4\pm3.2\kms$ and
$\sigma_v=2.6^{+5.1}_{-2.6}\kms$). T12 also presented observations for And
XII, but as for And XI, they were not able to cleanly detect the galaxy. They
identified 2 stars with highly negative velocities ($\sim-530\kms$), which are
offset from our systemic velocity by $\sim30\kms$. In this instance, both the
stars they observed overlap with two of our likely members, situated at
$\alpha$=00:47:28.63,$\delta$=+34:22:43.1 and
$\alpha=$00:47:24.69,$\delta=$+34:22:23.9, and these velocities are offset
from those that we measure at a statistical level of $3.8\sigma$. This
suggests that the self-same calibration effect that causes an offset between
our results for And XI and those of T12, is present here also. In the case of
And XII, our mask was observed over two separate nights, giving us two
velocity measurements for each star (as discussed in \citealt{chapman07} and
\citealt{collins10}), and we saw no evidence for systematic offsets of this
magnitude in the night to night velocities, making a calibration error within
our dataset seem unlikely, though not impossible. We therefore conclude that,
owing to our larger sample of members and repeat observations, our
measurements for the kinematic properties of And XII are more robust than
those of T12.


\subsection{Andromeda XIII}
\label{sect:and13}

The kinematic properties of Andromeda XIII were also presented in
\citet{collins10}, And XIII sits at a large projected distance from Andromeda
$(~\sim120$ kpc) in the southern M31 halo, so we expect contamination from the
Milky Way and Andromeda halo to be low. It is surprising then, that we see
significant structure within our DEIMOS field. This is also seen within the 3
fields observed by T12, who attribute this over-density of stars located at
$v_r\sim-120\kms$ to an association with the TriAnd over-density within the
Galactic halo. We too see a number of stars between $-140\kms$ and
$-100\kms$. From their positions within the CMD of And XIII, they appear more
consistent with MW foreground K-dwarfs than M31 RGB stars. As such, these are
also likely associated to this MW substructure.

Using our full probabilistic analysis, we identify the most probable And XIII
stars as those 4 that cluster around $v\sim-200\kms$, and we determine
$v_r=-204.8\pm4.9\kms$ and are unable to resolve a velocity dispersion, with
$\sigma_v=0.0^{+8.1}$, where the upper limit indicates the formal $1\sigma$
uncertainty on the likelihood distribution. Given the very large uncertainties
on these values (mostly a factor of the low number of member stars) it comes
as no surprise perhaps that these results are consistent with the results in
\citet{collins10} ($v_r=-195.0^{+7.4}_{-8.4}\kms$ and
$\sigma_v=9.7^{+8.9}_{-4.5}$), although not with those of T12
($v_r=-185.4\pm2.4\kms$ and $\sigma_v=5.6\pm2.0\kms$), where they derived
parameters from three times the number of member stars that we present
here. Given the highly substructured nature of the And XIII field, and the
fact that our detection is at a very low significance (only 4 stars), we find
that the T12 measurements for the kinematics of And XIII are more
robust than ours.



\end{document}